\documentclass{aa}  

\usepackage{epsfig}
\usepackage{graphics,graphicx}
\usepackage[varg]{txfonts}
\def\NGC{NGC7538}
\def\NGC1{NGC7538~IRS1}

\def\meth{CH$_{3}$OH}
\def\nh3{NH$_{3}$}
\def\kms{km~s$^{-1}$}
\def\kmsy{km~s$^{-1}$~yr$^{-1}$}
\def\kmo{km~s$^{-1}$~mas$^{-1}$}

\def\Vlsr{$V_{\rm LSR}$}
\def\Jyb{Jy~beam$^{-1}$}

\def\HII{H{\sc ii}}
\newcommand{\ms}{$M_{\odot}$}

\newcommand{\pas}{$\rlap{.}^{\prime\prime}$}
\newcommand{\degree}{$^{\circ}$}

\begin{document}
   \title{A multiple system of high-mass YSOs surrounded by disks in \NGC1}
   
   \subtitle{Gas Dynamics on Scales 10--700 AU from CH$_{3}$OH Maser and \nh3\ Thermal Lines}

   \author{L. Moscadelli\inst{1}
          \and
          C. Goddi\inst{2}
          }

   \institute{INAF-Osservatorio Astrofisico di Arcetri, Largo E. Fermi 5, 50125 Firenze, Italy \\
             \email{mosca@arcetri.astro.it}
             \and
             Joint Institute for VLBI in Europe, Postbus 2, NL-7990 AA Dwingeloo, the Netherlands \\
             \email{goddi@jive.nl}
             }

   \date{}
 
\abstract{
NGC7538 IRS1 is claimed to be a high-mass young stellar object (YSO) with $30~M_{\odot}$, surrounded by a rotating  Keplerian-disk, probed by a linear distribution of methanol masers. The YSO is also powering a strong compact H II region or ionized wind, and is driving at least one molecular outflow. The axes orientations of the different structures (ionized gas, outflow, and disk) are however misaligned with each other, which led to different competing models proposed to explain individual structures. 
   We investigate the 3D kinematics and dynamics of circumstellar gas with very high linear resolution, from tens to 1500 AU, with the ultimate goal of building a comprehensive dynamical model for what is considered the best high-mass accretion disk candidate around an O-type young star in the northern hemisphere.
   We use high-angular resolution observations of 6.7~GHz \meth\ masers with the EVN, \nh3\ inversion lines with the JVLA B-Array, and  radio continuum with the VLA A-Array. In particular, we employed four different observing epochs of EVN data at 6.7 GHz, spanning almost eight years, which enabled us to measure, besides line-of-sight (l.o.s.) velocities and positions (as done in previous works), also l.o.s. accelerations and proper motions of \meth\ masers. 
   In addition, we  imaged highly-excited \nh3\  inversion lines, from (6,6) to (13,13), which enabled us to probe the hottest molecular gas  very close to the exciting source(s).     
   We confirm previous results that five 6.7~GHz maser clusters (labeled from "A" to "E") are distributed over a region 
  extended N--S across \ $\approx$1500~AU,  and are associated with  three components of the radio continuum emission. 
   We propose that these maser clusters identify three individual high-mass YSOs in \NGC1, named IRS1a (associated with clusters "B" and "C"), IRS1b (associated with cluster "A"), and IRS1c  (associated with cluster "E"). 
   We find that the 6.7~GHz masers distribute along a line with a regular variation of \Vlsr\ with position along the major axis of the distribution of maser cluster~"A" and the combined clusters~"B"+"C".   
   A similar \Vlsr\ gradient (although shallower) is also detected in the \nh3\ inversion lines.
   Interestingly, the variation of \Vlsr\ with projected     position is not linear but quadratic for both maser clusters.     
    We measure proper motions for 33 maser features, which 
    have an average amplitude ($4.8\pm0.6$~\kms) similar to the variation in \Vlsr\ across the maser cluster, 
    and are approximately parallel to the clusters' elongation axes. 
    By studying the time variation of the maser spectrum, we derive also l.o.s. 
    accelerations for 30 features, with typical amplitude of \ $\sim10^{-3}-10^{-2}$~\kmsy.   
    We model the masers in both clusters~"A" and "B"+"C" in terms of an edge-on disk in 
    centrifugal equilibrium. 
    Based on our modeling, masers of clusters~"B"+"C" may trace a quasi-Keplerian $\sim$1~M$_{\sun}$, thin disk, orbiting around a high-mass YSO, IRS1a, of up to $\approx$25~M$_{\sun}$. This YSO dominates the bolometric luminosity of the region. 
    The disk traced by the masers of cluster~"A" is both massive ($\lesssim$16~M$_{\sun}$ 
    inside a radius of $\approx$500~AU) and thick (opening angle $\approx$ 45\degree), 
    and the mass of the central YSO, IRS1b, is constrained to be at most a few M$_{\sun}$. 
    Towards cluster~"E", \nh3\ and 6.7~GHz masers trace more quiescent 
    dynamics than for the other clusters. The presence of a radio continuum peak  
    suggests that the YSO associated with the cluster~"E", IRS1c, may be an ionizing, massive YSO as well. 
    We present compelling evidence that \NGC1\ is not forming just one single high-mass YSO, but consists of a multiple system of high-mass YSOs, which are surrounded by accretion disks, and are probably driving individual outflows. This new model naturally explains all the  different orientations and disk/outflow structures proposed for the region in previous models.
}


%\titlerunning{Gas Dynamics on Scales 10--700 AU in \NGC1}
%\authorrunning{Moscadelli \& Goddi}

   \maketitle
%
%________________________________________________________________

\begin{table*}
\caption{6.7~GHz \meth\ maser EVN Observations}             % title of Table
\label{evn_par}      % is used to refer this table in the text
\centering                          % used for centering table
\begin{tabular}{c c c c c c}        % centered columns (4 columns)
\hline \hline                 % inserts double horizontal lines
%\noalign{\smallskip}
Epoch &  EVN Code & Observing Date & \multicolumn{1}{c}{Naturally-Weighted Beam} & Vel. Res. & Image rms \\    % table heading
      &     &      &  Maj. \ $\times$ \ Min. \ @ \ PA &   &  \\
      &     & (yy/mm/dd)  & (mas) \ $\times$ \ (mas) \ @ \  (deg)  &  (\kms) &   (mJy beam$^{-1}$) \\
%\noalign{\smallskip}
\hline                        % inserts single horizontal line
%\noalign{\smallskip}
   0 & EP039B & 2002/02/06  &  9.6 \ $\times$ \ 4.8 \ @ \ 37\degr  & 0.09  &  20 \\      % inserting body of the table
   1 & EC021 & 2004/11/05 & 7.7 \ $\times$ \ 6.5 \ @ \ $-$78\degr  &  0.09 &  4   \\
   2 & EP054 & 2006/02/24 & 7.8 \ $\times$ \ 6.6 \ @ \ $-$30\degr  & 0.09  & 9    \\
   3 & ES063B & 2009/11/03 & 7.1 \ $\times$ \ 5.0 \ @ \ $-$51\degr & 0.09   & 6     \\
%\noalign{\smallskip}
\hline   %inserts single line
%\noalign{\smallskip}
\end{tabular}
\tablefoot{Columns~1,~2~and~3 give
the epoch label, the EVN experiment code and the observing date, respectively; Col.~4 reports the FWHM Major 
and Minor axis, and the PA  of the naturally-weighted beam; Cols.~5~and~6 list the 
velocity resolution and the rms noise of the maser images.}
\end{table*}

\section{Introduction}

The formation of massive stars (O-B type) by the same accretion processes believed to form low-mass stars appears problematic, because the intense radiation pressure from the star luminosity and the thermal pressure from the \HII\ region around the young stellar objects (YSOs) may be sufficient to reverse the accretion flow and prevent matter from reaching the star \citep{Zin07}. The "standard" theory predicts that this occurs for stars having masses in excess of 8~M$_{\sun}$, leading to the paradoxical conclusion that stars above this limit should not exist \citep{Pal93}.
Recent theories have demonstrated that the radiation pressure problem can be solved if accretion occurs through a circumstellar disk \citep[e.g.,][]{Kui10, Kui11}, thus explaining the formation of stars up to 140~M$_{\sun}$. 
In a quite different scenario, \citet{Bon98} proposed that O-type stars may form through the merger of low-mass objects. 
Despite the theoretical evidence,  
while a handful of disk candidates in B-type ($M_{\star} < 20$~M$_{\sun}$) protostars has been reported in the literature in recent years \citep[][and references therein]{Ces06}, there has been no clear evidence for accretion disks around more massive O-type stars so far.  
The few rotating molecular structures detected around O-type stars (on scales $>$10000~AU) have been interpreted as
gravitationally-unstable, transient bodies, infalling and accreting  either to a central cluster of low-mass protostars 
\citep{Sol05,Bel11} or to a single massive protostar 
\citep{San03,Beu08}.   
Since O-type stars form  at large distances (a few kpc) and deeply embedded inside dense massive cores (likely containing protoclusters), both confusion/crowding and poor resolution in previous studies have precluded to discriminate between different star-formation scenarios and, ultimately, to establish whether the globally rotating and infalling material in the cloud actually accretes onto individual massive protostars.
This would require to resolve the structure and dynamics of accreting gas at small radii ($\lesssim$1000~AU) from massive YSOs but this has been challenging so far (present millimeter interferometers have typical resolutions of order of 1\arcsec, corresponding to 1000 AU at 1 kpc). 
Therefore, observational signatures of rotating disks around O-type forming stars are essential to progress in our understanding of the mass-accretion process and to constrain theoretical models of high-mass star formation (HMSF). 

In this context, one excellent diagnostic tool of gas kinematics within 10--1000~AU from YSOs is provided by multi-epoch very long baseline interferometric (VLBI) observations of interstellar masers \citep{God06a,God06b,God11a,Mat10,Mos07,Mos11a,Mos13,San10a,San10b,Tor03,Tor11}.
 Among different molecular masers, \meth\ is particularly interesting, because it is exclusively associated with HMSF and provides an excellent probe of accretion. 
Recently, \citet{God11a} reported  a convincing signature of infall of  a circumstellar molecular envelope with a radius of only 300~AU around a B-type forming star in AFGL~5142, by using multi-epoch VLBI observations of \meth\ masers spanning six years. 
Measurements of the 3D velocity field of the circumstellar gas has provided the most direct and least unbiased measurement (yet obtained) of infall of a  molecular envelope onto an intermediate- to high-mass protostar. 
Other interesting examples of 3D kinematics with methanol masers are reported in \citet{San10a,San10b} and \citet{Mos11a,Mos13}.

In order to characterize the accretion process in an O-type YSO,  in this paper we study one of the best high-mass accretion disk candidates in the northern hemisphere, \NGC1. 
This region is relatively nearby (2.7 kpc; \citealt{Mos09}), 
very luminous  ($\sim$10$^5$~L$_{\sun}$; e.g., \citealt{Aka05}), 
it contains an \HII\ region \citep[e.g.,][]{Wyn74}, 
and it has been suggested to be powered by an O6/7 star of about 30~M$_{\sun}$. 
Very Large Array (VLA) continuum observations revealed a double-peaked structure in the ionized gas 
within 0\farcs2 from the central core and a more extended ($\sim$1\arcsec) emission elongated N--S  
\citep[e.g.,][]{Gau95}. 
Radio recombination lines observed at cm- and mm-wavelengths show extremely broad line widths,
suggestive of expanding motions of the ionized gas \citep{Gau95,Ket08}. 
A multi-wavelength study of the radio continuum showed that the free-free emission from IRS1 is dominated by an ionized jet, rather than  a hyper-compact HII region \citep{Sand09}.
Recently, a number of interferometric studies conducted with increasing angular resolution,  
at 1.3~mm with the SMA (3\arcsec\ beamsize, \citealt{Qiu11}),
 at 1.3~and~3.4~mm with the SMA and CARMA (0\farcs7 beamsize, \citealt{Zhu13}),
 and at 0.8~mm with the PdBI (0\farcs2 beamsize, \citealt{Beu13}), 
detected  several typical hot-core species, showing inverse P-Cygni profiles, 
probing inward motion of the dense gas  on scales $\gtrsim$1000~AU with
a mass infall rate \ $\dot{M}\sim$10$^{-3}$~M$_{\sun}$~yr$^{-1}$. 
These radio centimeter and millimeter observations also identified several outflows emanating from \NGC1, 
along N--S \citep{Gau95,Sand09}, NW--SE \citep{Qiu11}, 
and NE--SW \citep{Beu13}.  
The simultaneous presence of jets/outflows and a strong accretion flow toward IRS\,1, 
led some authors to postulate the presence of an accretion disk surrounding IRS\,1.
\citet{Min98} observed a linear distribution of 6.7 and 12.2 GHz \meth\ masers with a position angle (PA) of 
about 112\degree \ and \citet{Pes04} proposed an edge-on Keplerian disk model 
to explain positions and l.o.s. velocities of maser spots. 
A mid-infrared (IR) study however questioned the edge-on disk model traced by the \meth\ masers, 
suggesting that the radio continuum emission (elongated N--S) traces 
an ionized wind emanating from the  surface of a disk with \ PA of $\sim$30\degree\  perpendicular to the NW--SE CO-bipolar outflow  \citep{DeBui05}. 
In an attempt to explain the different orientation of the elongated structures observed in the \meth\ masers 
and in the near-IR emission,  \citet{Krau06} proposed that the edge-on disk is driving a precessing jet.
\citet{Sur11} however pointed out that, 
in addition to the linear \meth\ maser cluster proposed to trace the edge-on Keplerian disk, 
there are additional maser clusters in the region within 1\arcsec,
and they proposed an alternative scenario 
where all the observed \meth\ maser clusters should mark the interface between the infalling envelope and a large-scale torus, having a rotation axis with the same PA (30\degree) \ of the elongated near-IR emission observed by \citet{DeBui05}.

While the individual competing models 
explain some properties of the system, either the ordered structure of one maser cluster \citep[e.g.,][]{Pes04},
or the global spatial distribution of all clusters \citep[e.g.,][]{Sur11}, or the complex pattern of molecular
outflows emerging from IRS\,1 \citep{Krau06}, 
they fail to provide a clear picture of accretion/outflow in terms of a simple disk/jet system, 
as expected in the context of a canonical picture of star formation. 
Likewise, despite the plethora of interferometric studies on the region at (sub)mm-wavelengths 
(with angular resolutions in the range \ 0\farcs2--2\arcsec),  
no clear evidence of a rotating disk has been found and only a confusing picture for the outflows 
has been drawn so far.

In this paper,  we overcome the shortcomings of previous works by analyzing a multi-epoch dataset of 6.7~GHz 
CH$_3$OH maser observations and  
complementing the maser data with new interferometric images of highly-excited inversion lines of \nh3, from (6,6) to (13,13). 
This approach has a two-fold advantage. 
First, the multi-epoch dataset (spanning almost eight years), enables us to measure proper motions and l.o.s. accelerations of \meth\ masers, besides positions and l.o.s. velocities (as done in previous works). 
Second,  highly-excited inversion lines of \nh3\ at cm-wavelengths, enable us to probe the hottest gas 
close the YSO(s) in an optically-thin regime 
and at the highest angular resolutions achievable with connected-element interferometers 
(previous works were conducted at mm-wavelengths and/or with poorer angular resolutions).  
A detailed analysis of the \nh3\ data will be presented in a forthcoming paper (Goddi et al. in prep.).
The main goal of this paper is to  investigate 
the 3D dynamics  of the  circumstellar molecular gas 
at the small scales (10--500~AU) probed by the masers and relate it to the large-scale motions (500--2000~AU) probed by the complementary interferometric thermal \nh3\ data. 

We describe observations and data reduction in Sect.~\ref{obs} and our observational results in Sect.~\ref{results}. 
In Sect.~\ref{vlsr_regu}, we examine the ordered velocity structures measured with the \meth\ masers and thermal \nh3\ lines.  
We present a dynamical model to explain our measurements in Sect.~\ref{mas_kin},
followed by a discussion on the nature of star formation in \NGC1\ in Sect.~\ref{phy_sen}.
Some implications of our model and a comparison with previous models are illustrated in Sect.~\ref{discu}. 
Finally, we summarize our main findings in Sect.~\ref{summary}.

\section{Observations}

\label{obs}

\subsection{EVN 6.7~GHz \meth\ Maser Archival Observations}
\label{obse_evn}
This work is based on archival data of 6.7~GHz \meth\ masers observed towards \NGC1\ with the European VLBI Network (EVN\footnote{The European VLBI Network is a joint facility of European, Chinese, South African and other radio astronomy institutes funded by their national research councils.}). 
In particular, we reduced and analyzed four individual datasets at four distinct observing epochs over the years \ 2002--2009. 
For each observation, Table~\ref{evn_par} reports the observing date, the
angular and velocity resolution, and the sensitivity of the reconstructed maser images.
The reduced sensitivity of the 2002 February run (EVN code: EP039B) 
is due to the lower number of observing antennae 
(six: Jodrell, Cambridge, Effelsberg, Medicina, Torun and Onsala) compared with the subsequent runs, which
employed either eight (EC021 and EP054: including also Noto and Westerbork) or nine (ES063B: including also Yebes)
telescopes. Details of the observational setup of ES063B can be found in \citet{Sur11}.

Data were reduced with the NRAO Astronomical Image Processing System (AIPS) package,
following the VLBI spectral line procedures. The time persistent emission of the strongest maser channel
(corresponding to feature \#1 in Table~\ref{tab_6.7}) has been self-calibrated and the complex gain 
corrections have been applied to all the emission channels before imaging. Maser images cover 
a field of view, \ $\Delta\alpha$ $\times$ $\Delta\delta$ $\approx$  8~\arcsec $\times$ 8~\arcsec, and a \Vlsr\
range, from $-$71~\kms\ to $-$52~\kms, adequate to recover all the 6.7~GHz maser emission previously detected towards \NGC1. 
For a description of the criteria used to identify individual masing clouds, derive
their parameters (position, intensity, flux and size), and measure their
relative proper motions, we refer to \citet{San10a}. 
In the following we use the term "spot" to indicate a compact emission centre
on a single-channel map, and the term "feature" to refer to a collection of spots
emitting in contiguous channels at approximately the same position in the sky
(within the beam FWHM).

Since none of the four EVN runs was observed in phase-reference mode, no accurate information on the absolute position
of the 6.7~GHz \meth\ masers can be derived directly from these data. However, \citet{Min98} and \citet{Pes04} have shown that
there is a good correspondence in position and \Vlsr\ between the 6.7~GHz and 12~GHz methanol masers in \NGC1.
In particular, the strongest features at both maser transitions have an elongated structure (cluster~"A",
see Sect~\ref{prop_mot}), and the agreement between 6.7 GHz and 12.2 GHz masers is within a few mas in relative position and 0.1~\kms\ in \Vlsr.  
For the 12.2~GHz methanol masers,  
\citet{Mos09} determined the absolute position (R.A.(J2000) = 23$^{\rm h}$ 13$^{\rm m}$ 45\fs3622 ,
DEC (J2000) = 61\degree 28$^{\prime}$ 10\farcs507) and the apparent proper motion ($\mu_{\rm RA}$ = $-$2.45~mas~yr$^{-1}$,
$\mu_{\rm DEC}$ = $-$2.45~mas~yr$^{-1}$) with high accuracy using multi-epoch, phase-referencing observations with the Very Long Baseline Array (VLBA). 
In the assumption that the most intense 6.7 and 12~GHz maser features are spatially coincident, we can then determine 
the absolute position of the 6.7~GHz masers with milliarcsecond accuracy as well. 
As a comparison, our derivation of the absolute position of the 6.7~GHz masers in \NGC1\ agrees with that determined by
\citet{Sur11} using Multi-Element Radio Linked Interferometer network (MERLIN) observations within their measurement error of 10~mas.  

The  EVN experiment EP039B (labeled \ "0" \ in Table~\ref{evn_par}) resulted in less sensitive images, 
and only a few, intense ($\ge$ 1~\Jyb), 6.7~GHz maser features could be detected. Maser proper motions have been derived using 
merely the subsequent three, more sensitive epochs (labeled \ 1 -- 3 in Table~\ref{evn_par}), 
showing similar (naturally-weighted) beams. In the following analysis, epoch \ "0" \ data are employed only to extend the time baseline 
and calculate the l.o.s. accelerations of the most intense maser features with higher accuracy (see Sect.~\ref{acc_los}).

\subsection{VLA Archival Observations}
\citet{Gau95} used the VLA in the most extended A-configuration to observe the H66$\alpha$ line (at 22364.174~MHz) towards \NGC1\ on 1992 December 12 and 20 (exp. code: AG360). 
We downloaded the data of this observation from the VLA archive and, following \citet{Gau95},
derived the continuum emission as the "channel zero" data, by averaging the inner 75\% of the 25~MHz 
observing band. Data have been calibrated following the standard procedure for VLA continuum data, as also
described in \citet{Gau95}. The continuum map, produced using uniform-weighting, has a FWHM beam of
\ 0\farcs077 $\times$ 0\farcs074 \ with PA = 90\degree.
We precessed the coordinate system of the continuum map from the B1950 to the J2000 equatorial system, and 
we also corrected for the apparent motion between the VLA observing epoch, December 1992, and the first epoch, September 2005, of the 12.2~GHz \meth\ maser VLBA observations. 
%, which have measured the maser absolute positions with milliarcsecond accuracy.  

\subsection{\nh3\ JVLA Observations}
Observations of NH$_3$ towards NGC7538~IRS1 were conducted using the Karl G. Jansky Very Large Array (JVLA) of the National
Radio Astronomy
Observatory (NRAO)
\footnote{NRAO is a facility of the National Science Foundation operated under cooperative agreement by Associated Universities, Inc.} in B configuration.
By using the broadband JVLA K- and Ka-band receivers, we
observed  a total of five metastable inversion transitions of NH$_3$: 
($J,K$)=(6,6), (7,7), (9,9), (10,10), (13,13)  at 1.3~cm, with frequencies going from \ $\approx$25.1~GHz for the (6,6) line to $\approx$33.2~GHz for the (13,13) line.   
Transitions were observed in pairs of (independently tunable) basebands 
during 6h tracks (two targets per track) on three different dates in 2012: 
the (6,6) and (7,7) lines on May 31 at K-band, the (9,9) and (13,13) lines on June 21, and the (10,10) transition on August 7,  at Ka-band. 
Each baseband had eight sub-bands with a 4~MHz bandwidth per sub-band ($\approx$40~\kms\ at 30~GHz), providing a total coverage of 32~MHz ($\approx$320~\kms\ at 30~GHz). Each sub-band consisted of 128 channels with a separation of 31.25~kHz ($\approx$0.4~\kms\ at 30~GHz). 
Typical on-source integration time was 53 min. Each transition was observed with 
"fast switching", where 60s scans on-target were alternated with 60s
scans of the nearby (3$^{\circ}$ on the sky) QSO J2339$+$6010  (measured flux density 0.2--0.3~Jy, depending on frequency). 
We derived  absolute flux calibration from observations of 3C~48 ($F_{\nu}$ = 0.5--0.7~Jy, depending on frequency), and bandpass calibration from observations of 3C~84 ($F_{\nu}$ = 24--28~Jy, depending on frequency).

The data were edited, calibrated, and imaged in a standard fashion using the Common Astronomy Software Applications (CASA) package. 
We fitted and subtracted continuum emission  from the spectral line data in the uv plane  
using CASA task UVCONTSUB, combining the continuum (line-free) signal from all eight sub-bands around the \nh3\ line. 
Before imaging the \nh3\ lines, we performed self-calibration of the continuum emission. 
The continuum images showed an integrated flux density  and a 1$\sigma$ rms varying in the range 400--600~mJy and 80--240~$\mu$Jy beam$^{-1}$, from 25 to 33~GHz, respectively.  
We then applied the self-calibration solutions from the continuum to the line datasets. 
Using the CASA task CLEAN, we imaged the NGC7538 region with a cell size of 0\farcs04, covering an 
8\arcsec~field around the phase-centre position: $\alpha(J2000) = 23^h 13^m 45\rlap{.}^s4$,   $\delta(J2000) = +61^{\circ} 28' 10"$. 
We adopted Briggs weighting with a ROBUST parameter set to 0.5 and smoothed the velocity resolution to 0.4~\kms\ for all transitions. 
The resulting synthesized clean beam FWHM were \ 0\farcs15--0\farcs27 \ and the rms noise level per  channel were \ $\approx$1.5--2.5~mJy~beam$^{-1}$, depending on frequency. 

Since the position of the phase calibrator, J2339$+$6010, was only accurate within 0\pas15, according to the The VLA Calibrator Manual\footnote{http://www.vla.nrao.edu/astro/calib/manual/csource.html}, 
we conducted a 30$^m$ test in December 2013 to check the astrometry of our \nh3\ maps, observing J2339$+$6010 and NGC7538 in fast-switching with QSO J2230+6946. We measured a positional offset of $\Delta \alpha \sim-$0\pas01, $\Delta \delta \sim$0\pas16 from the nominal position of J2339$+$6010 reported in the VLA catalog. 
All the \nh3\ maps of \NGC1\ reported in this paper were shifted accordingly by the measured offset. 
For the sake of comparison with the VLBI \meth\,maps, 
we also corrected for the apparent motion between December 2013, and the first epoch, September 2005, of the 12.2~GHz \meth\ maser VLBA observations (as done for the radio continuum image).

\begin{figure*}
\centering
\begin{minipage}{\textwidth}
\includegraphics[angle=0.0,width=1.0\textwidth]{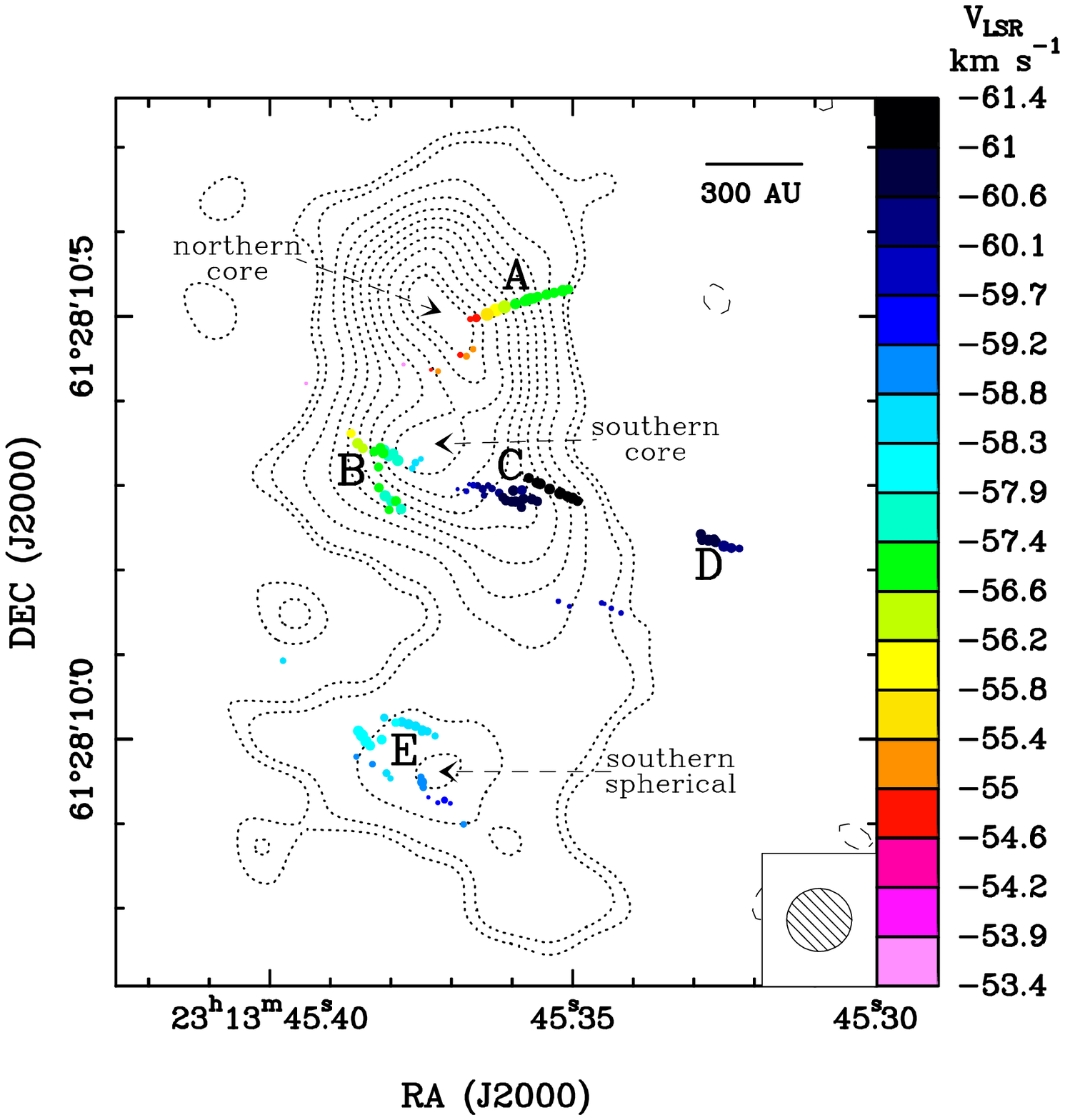}
\caption[]{6.7~GHz \meth\ masers detected over three epochs with the EVN, overlaid on the 1.3~cm continuum imaged with the VLA A-Array.  
{\it Colored dots} show the absolute position
of individual maser features, 
with color denoting the maser \Vlsr, according to the color-velocity conversion code
reported on the right side of the panel\footnotemark{}.
Maser absolute positions are relative to the epoch
2005 September 9, which is the first of the five VLBA epochs used by \citet{Mos09} 
to measure absolute positions and proper motions of the 12~GHz \meth\ masers in \NGC1.
Dot size is proportional to the logarithm of the maser
intensity. Masers are grouped in different clusters,
labeled using capital letters from "A" to "E".   
The 1.3~cm map ({\it Dotted contours})  was produced using archival data, originally reported by \citet{Gau95}.  
Plotted contours are \ 7\%, 10\% to 90\% (in steps of 10\%), and 95\% of the map
peak, 0.022~\Jyb. {\it Dotted arrows} point to the main 1.3~cm continuum peaks, 
which are named following the notation by \citet{Gau95}. 
The beam of the VLA-A 1.3~cm observations is reported in the insert
in the bottom right of the panel.
}
\footnotetext{$^{4}$The color used to plot a given maser feature can change across the figures of the article
depending on the plotted range of maser \Vlsr.}
\label{vlsr_pos}
\end{minipage}
\end{figure*}

\section{Results}
\label{results}

We discuss here gas dynamics in \NGC1\ 
based on the multi-epoch EVN dataset of 6.7~GHz \meth\ masers (\S~\ref{res_ch3oh}) 
and the \nh3\ inversion lines imaged with the JVLA B-Array (\S~\ref{res_nh3}).

\subsection{6.7~GHz \meth\ Maser Kinematics}
\label{res_ch3oh}

Figure~\ref{vlsr_pos} shows the spatial and \Vlsr\ distribution of the 6.7~GHz \meth\ masers in \NGC1,
overlaid on a  map of the 1.3~cm continuum emission from the VLA A-Array.
Spread over an area of \ $\Delta\alpha$ $\times$ $\Delta\delta$ $\approx$ 0\farcs4 $\times$ 0\farcs6,
most of the 6.7~GHz maser features are organized in five distinct clusters (with typical size $\le$100~mas), 
which, following the naming convention by \citet{Min00}, are identified with capital letters from "A" to "E".
For each cluster, Table~\ref{tab_6.7} reports the main properties of the detected maser features  
(epochs of detection, intensity, \Vlsr\ and position), labeled with 
integer numbers increasing reversely with the peak intensity.
As noted originally by \citet{Min98}, there is a good positional correspondence 
between the  maser clusters and the radio continuum peaks. Using the notation
by \citet{Gau95}, the maser cluster~"A" emerges to the NW of the "northern core" of the 1.3~cm continuum emission,
the "southern core" is found in between maser clusters~"B" and "C",
and the maser cluster~"E" is approximately coincident with the "southern, spherical" component of the radio continuum.

The kinematics of the 6.7~GHz masers towards \NGC1\ has been the subject 
of several papers \citep{Min98,Pes04,Sur11}. 
In the following, we focus on the two main achievements of the present
article, i.e. the measurement of the maser proper motions and the l.o.s. accelerations. 

\subsubsection{Proper Motions}
\label{prop_mot}

\begin{figure*}
\centering
\includegraphics[angle=0.0,width=0.85\textwidth]{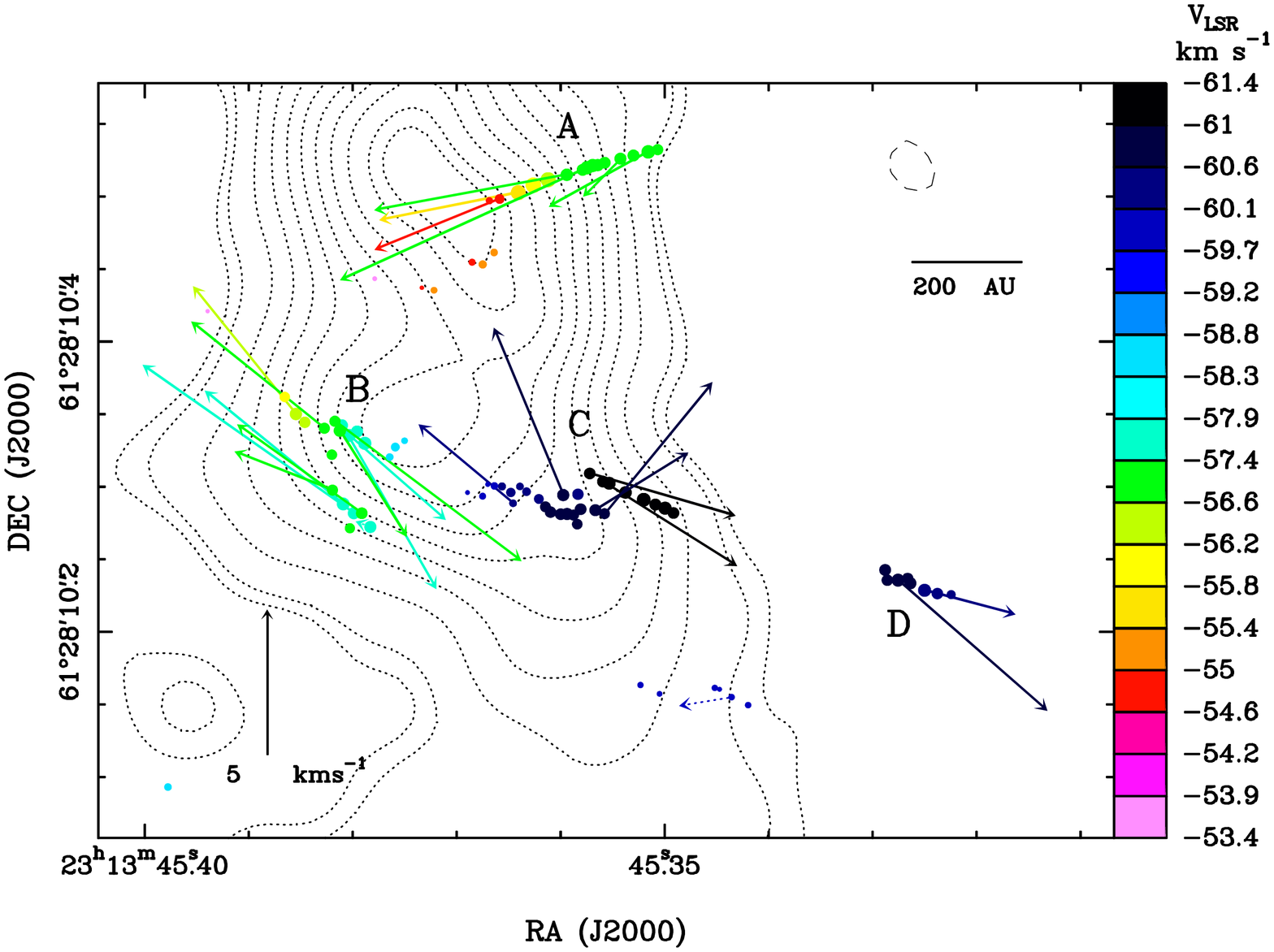}
\includegraphics[angle=0.0,width=0.85\textwidth]{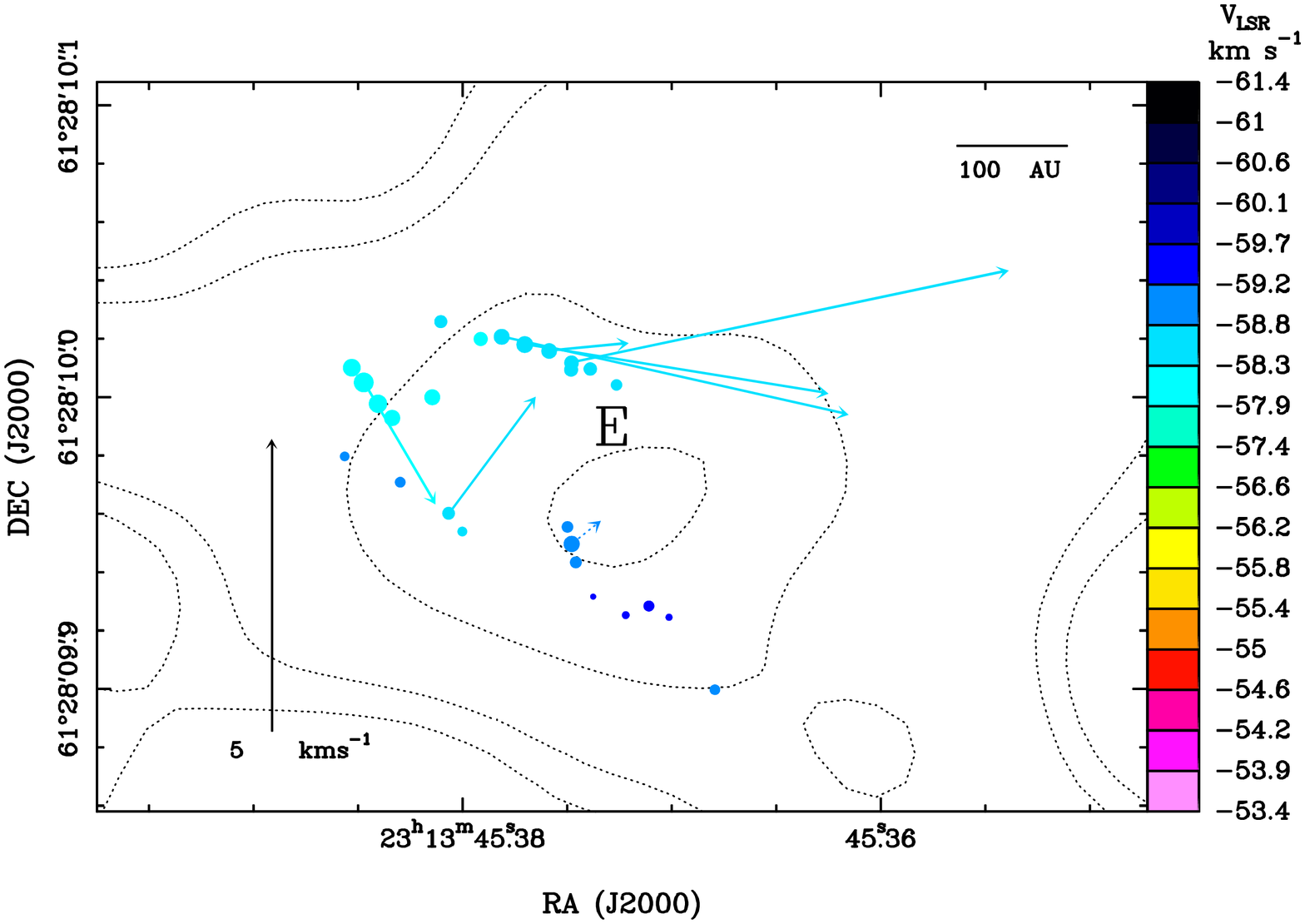}
\caption{Proper motions of 6.7~GHz \meth\ masers detected over three epochs with the EVN. 
Symbols, colors and contours have the same meaning as in Fig.~\ref{vlsr_pos}.
The plotted field of view includes the maser clusters "A", "B", "C" and "D" (top panel),
and cluster "E" (lower panel). 
{\it Colored arrows} show the measured maser proper motions, with {\it dotted arrows} 
denoting the most uncertain measurements. The scale for the proper motion
amplitude is given by the {\it black arrow} in the lower left corner of each panel.
}
\label{prmot}
\end{figure*}

Figure~\ref{prmot} presents the measured proper motions of the 6.7~GHz masers
for clusters "A" to "D" (top panel) and for the cluster~"E" (bottom panel), respectively. 
The last two columns of Table~\ref{tab_6.7}
list the proper motion components projected along the RA and DEC axes.
Proper motions are expressed relative to the "centre of Motion", 
whose position is derived as the average of the 33 maser features, 
persistent in time and with a stable spectral emission, for which
proper motions are derived. The amplitude of the proper motions varies in the range \ 1--9~\kms, with
average value and (1-$\sigma$) error of \ 4.8~\kms\ and 0.6~\kms, respectively.
It is interesting to note that proper motions of the features in cluster~"A" 
are parallel to the cluster elongation, and that the same holds for most
of the proper motions measured in the clusters "B", "C" and "D".
This good agreement between proper motion and maser cluster orientation hints at a physical origin.
We argue that the proper motions are not biased by the specific choice of the "centre of Motion" as the reference system for the proper motions. 
In fact, we estimate any systematic error related to the adoption of the reference system
 of the "centre of Motion" to be much smaller than the average proper motion amplitude (see Sect.~\ref{mas_kin_fit}).  
Therefore, we are confident that the derived proper motions can be used to describe the gas kinematics 
with respect to the star(s) exciting the maser emission.

\begin{longtab}
\begin{longtable}{cccccccc}
\caption{\label{tab_6.7} 6.7~GHz CH$_3$OH maser parameters}\\
\hline\hline
\multicolumn{1}{c}{Feature} & \multicolumn{1}{c}{Epochs of} & \multicolumn{1}{c}{I$_{\rm peak}$} & \multicolumn{1}{c}{$V_{\rm LSR}$} & \multicolumn{1}{c}{$\Delta~x$} & \multicolumn{1}{c}{$\Delta~y$} & \multicolumn{1}{c}{$V_{x}$} & \multicolumn{1}{c}{$V_{y}$}\\
\multicolumn{1}{c}{Number}  & \multicolumn{1}{c}{Detection} & \multicolumn{1}{c}{(Jy beam$^{-1}$)} & \multicolumn{1}{c}{(km s$^{-1}$)} & \multicolumn{1}{c}{(mas)} & \multicolumn{1}{c}{(mas)} & \multicolumn{1}{c}{(km s$^{-1}$)} & \multicolumn{1}{c}{(km s$^{-1}$)}\\
%Feature & Epochs of & I$_{\rm peak}$ & $V_{\rm LSR}$ & $\Delta~x$ & $\Delta~y$ & $V_{x}$ & $V_{y}$\\
%Number  & Detection & (Jy beam$^{-1}$) & (km s$^{-1}$) & (mas) & (mas) & (km s$^{-1}$) & (km s$^{-1}$)\\
\hline
\endfirsthead
\caption{continued.}\\
\hline\hline
\multicolumn{1}{c}{Feature} & \multicolumn{1}{c}{Epochs of} & \multicolumn{1}{c}{I$_{\rm peak}$} & \multicolumn{1}{c}{$V_{\rm LSR}$} & \multicolumn{1}{c}{$\Delta~x$} & \multicolumn{1}{c}{$\Delta~y$} & \multicolumn{1}{c}{$V_{x}$} & \multicolumn{1}{c}{$V_{y}$}\\
\multicolumn{1}{c}{Number}  & \multicolumn{1}{c}{Detection} & \multicolumn{1}{c}{(Jy beam$^{-1}$)} & \multicolumn{1}{c}{(km s$^{-1}$)} & \multicolumn{1}{c}{(mas)} & \multicolumn{1}{c}{(mas)} & \multicolumn{1}{c}{(km s$^{-1}$)} & \multicolumn{1}{c}{(km s$^{-1}$)}\\
%Feature & Epochs of & I$_{\rm peak}$ & $V_{\rm LSR}$ & $\Delta~x$ & $\Delta~y$ & $V_{x}$ & $V_{y}$\\
%Number  & Detection & (Jy beam$^{-1}$) & (km s$^{-1}$) & (mas) & (mas) & (km s$^{-1}$) & (km s$^{-1}$)\\
\hline
\endhead
\hline
\endfoot
 & & & & & & &\\
\multicolumn{8}{c}{Cluster A}\\
%& & & Cluster & A & & & \\
 & & & & & & &\\
    1 &          1,2,3 &    139.89 &   $-$56.0 &     0.00$\pm$0.00 &     0.00$\pm$0.00 &     ...  &     ...\\
    2 &          1,2,3 &     95.96 &   $-$56.4 &   $-$10.19$\pm$0.22 &     3.76$\pm$0.22 &     ...  &     ...   \\
    3 &          1,2,3 &     89.68 &   $-$55.7 &    10.36$\pm$0.23 &    $-$5.12$\pm$0.22 &    4.8$\pm$0.6 &   $-$0.9$\pm$0.6  \\
    4 &          1,2,3 &     16.11 &   $-$57.2 &   $-$79.19$\pm$0.22 &    22.96$\pm$0.22 &    3.4$\pm$0.5 &   $-$1.9$\pm$0.5  \\
    5 &              3 &     14.87 &   $-$56.8 &   $-$35.98$\pm$0.22 &    11.36$\pm$0.21 &     ...  &     ...   \\
    6 &            1,2 &     11.62 &   $-$56.9 &   $-$40.90$\pm$0.22 &    13.54$\pm$0.22 &     ...  &     ...   \\
    7 &          1,2,3 &      5.80 &   $-$56.7 &   $-$23.08$\pm$0.22 &     7.20$\pm$0.22 &    6.6$\pm$0.6 &   $-$1.2$\pm$0.5  \\
    8 &              3 &      4.77 &   $-$56.8 &   $-$43.33$\pm$0.22 &    13.16$\pm$0.22 &     ...  &     ...   \\
    9 &          1,2,3 &      4.07 &   $-$57.2 &   $-$60.01$\pm$0.23 &    18.28$\pm$0.22 &    1.3$\pm$0.6 &   $-$1.3$\pm$0.5  \\
   10 &          1,2,3 &      3.61 &   $-$57.3 &   $-$69.02$\pm$0.23 &    20.47$\pm$0.22 &     ...  &     ...   \\
   11 &          1,2,3 &      3.39 &   $-$56.9 &   $-$33.95$\pm$0.22 &    10.55$\pm$0.22 &    8.4$\pm$0.6 &   $-$3.8$\pm$0.6  \\
   12 &          1,2,3 &      3.38 &   $-$57.0 &   $-$49.22$\pm$0.22 &    15.47$\pm$0.22 &     ...  &     ...   \\
   13 &          1,2,3 &      1.79 &   $-$57.2 &   $-$85.69$\pm$0.24 &    24.49$\pm$0.23 &     ...  &     ...   \\
   14 &          1,2,3 &      0.62 &   $-$55.0 &    23.23$\pm$0.23 &    $-$9.49$\pm$0.22 &    4.3$\pm$0.7 &   $-$1.8$\pm$0.7  \\
   15 &            1,2 &      0.22 &   $-$55.3 &    34.88$\pm$0.26 &   $-$54.62$\pm$0.25 &     ...  &     ...   \\
   16 &              1 &      0.15 &   $-$55.4 &    27.03$\pm$0.56 &   $-$46.36$\pm$0.50 &     ...  &     ...   \\
   17 &              1 &      0.13 &   $-$54.8 &    30.26$\pm$0.33 &   $-$10.71$\pm$0.31 &     ...  &     ...   \\
   18 &              1 &      0.12 &   $-$54.9 &    42.18$\pm$0.31 &   $-$53.09$\pm$0.29 &     ...  &     ...   \\
   19 &              1 &      0.10 &   $-$55.2 &    68.44$\pm$0.35 &   $-$72.37$\pm$0.32 &     ...  &     ...   \\
   20 &              1 &      0.04 &   $-$53.5 &   109.29$\pm$0.47 &   $-$64.42$\pm$0.42 &     ...  &     ...   \\
   21 &              1 &      0.04 &   $-$54.8 &    76.82$\pm$0.51 &   $-$70.66$\pm$0.45 &     ...  &     ...   \\
\hline
& & & & & & & \\
\multicolumn{8}{c}{Cluster B} \\
%& & & Cluster & B & & & \\
 & & & & & & & \\
    1 &          1,2,3 &     10.88 &   $-$57.8 &   116.11$\pm$0.22 &  $-$177.87$\pm$0.22 &     ...  &     ...   \\
    2 &          1,2,3 &      8.57 &   $-$57.5 &   126.83$\pm$0.22 &  $-$172.41$\pm$0.22 &   $-$3.3$\pm$0.5 &   $-$2.9$\pm$0.5  \\
    3 &          1,2,3 &      8.45 &   $-$57.6 &   130.97$\pm$0.22 &  $-$219.63$\pm$0.22 &    4.8$\pm$0.5 &    3.9$\pm$0.5  \\
    4 &          1,2,3 &      5.39 &   $-$57.2 &   133.33$\pm$0.22 &  $-$169.14$\pm$0.22 &   $-$2.3$\pm$0.5 &   $-$3.7$\pm$0.5  \\
    5 &          1,2,3 &      5.34 &   $-$56.2 &   163.58$\pm$0.28 &  $-$157.64$\pm$0.26 &    3.5$\pm$0.7 &    4.4$\pm$0.6  \\
    6 &          1,2,3 &      4.55 &   $-$57.7 &   112.29$\pm$0.22 &  $-$235.45$\pm$0.22 &    0.5$\pm$0.6 &    0.2$\pm$0.6  \\
    7 &              1 &      4.19 &   $-$57.9 &   121.44$\pm$0.22 &  $-$169.72$\pm$0.22 &     ...  &     ...   \\
    8 &          1,2,3 &      4.15 &   $-$57.6 &   123.73$\pm$0.22 &  $-$226.02$\pm$0.22 &    7.3$\pm$0.6 &    5.1$\pm$0.5  \\
    9 &          1,2,3 &      3.66 &   $-$57.5 &   131.83$\pm$0.22 &  $-$165.43$\pm$0.22 &   $-$3.3$\pm$0.6 &   $-$5.6$\pm$0.6  \\
   10 &          1,2,3 &      2.98 &   $-$57.4 &   118.21$\pm$0.24 &  $-$226.02$\pm$0.23 &    4.3$\pm$0.6 &    3.1$\pm$0.6  \\
   11 &          1,2,3 &      2.26 &   $-$56.5 &   157.45$\pm$0.33 &  $-$163.39$\pm$0.30 &     ...  &     ...   \\
   12 &          1,2,3 &      1.90 &   $-$57.4 &   136.69$\pm$0.25 &  $-$162.66$\pm$0.24 &   $-$6.4$\pm$0.6 &   $-$4.8$\pm$0.6  \\
   13 &          1,2,3 &      1.82 &   $-$56.8 &   143.98$\pm$0.23 &  $-$167.49$\pm$0.23 &    4.6$\pm$0.6 &    3.7$\pm$0.6  \\
   14 &          1,2,3 &      1.72 &   $-$57.3 &   138.38$\pm$0.23 &  $-$210.10$\pm$0.22 &    3.4$\pm$0.6 &    1.3$\pm$0.6  \\
   15 &            1,2 &      1.12 &   $-$57.4 &   138.81$\pm$0.24 &  $-$185.71$\pm$0.24 &     ...  &     ...   \\
   16 &              3 &      1.01 &   $-$55.8 &   172.35$\pm$0.34 &  $-$146.55$\pm$0.33 &     ...  &     ...   \\
   17 &            1,2 &      0.85 &   $-$57.0 &   126.38$\pm$0.25 &  $-$236.36$\pm$0.24 &     ...  &     ...   \\
   18 &            1,2 &      0.34 &   $-$58.7 &    95.10$\pm$0.23 &  $-$180.44$\pm$0.23 &     ...  &     ...   \\
   19 &              1 &      0.16 &   $-$58.7 &    98.95$\pm$0.29 &  $-$187.32$\pm$0.27 &     ...  &     ...   \\
   20 &              1 &      0.09 &   $-$58.7 &    88.72$\pm$0.36 &  $-$176.09$\pm$0.32 &     ...  &     ...   \\
\hline 
& & & & & & & \\
\multicolumn{8}{c}{Cluster C}  \\
%& & & Cluster & C & & & \\
 & & & & & & & \\
    1 &          1,2,3 &     22.45 &   $-$61.3 &   $-$76.06$\pm$0.22 &  $-$216.73$\pm$0.22 &     ...  &     ...   \\
    2 &          1,2,3 &     11.73 &   $-$61.2 &   $-$90.79$\pm$0.22 &  $-$222.56$\pm$0.22 &     ...  &     ...   \\
    3 &          1,2,3 &      8.90 &   $-$61.4 &   $-$52.32$\pm$0.22 &  $-$205.38$\pm$0.22 &     ...  &     ...   \\
    4 &          1,2,3 &      6.70 &   $-$61.4 &   $-$63.52$\pm$0.22 &  $-$211.73$\pm$0.22 &     ...  &     ...   \\
    5 &          1,2,3 &      4.67 &   $-$60.6 &   $-$20.59$\pm$0.22 &  $-$213.51$\pm$0.22 &    2.4$\pm$0.6 &    5.8$\pm$0.6  \\
    6 &              1 &      4.47 &   $-$60.9 &   $-$23.21$\pm$0.22 &  $-$226.53$\pm$0.22 &     ...  &     ...   \\
    7 &          1,2,3 &      3.97 &   $-$60.8 &   $-$42.72$\pm$0.23 &  $-$223.93$\pm$0.22 &   $-$3.2$\pm$0.6 &    2.0$\pm$0.6  \\
    8 &            2,3 &      3.65 &   $-$61.2 &   $-$84.90$\pm$0.26 &  $-$220.33$\pm$0.27 &     ...  &     ...   \\
    9 &          1,2,3 &      3.16 &   $-$60.8 &   $-$32.59$\pm$0.22 &  $-$223.26$\pm$0.22 &     ...  &     ...   \\
   10 &          1,2,3 &      2.88 &   $-$61.4 &   $-$47.95$\pm$0.23 &  $-$204.22$\pm$0.22 &   $-$4.6$\pm$0.6 &   $-$2.9$\pm$0.6  \\
   11 &          1,2,3 &      2.76 &   $-$60.8 &   $-$18.82$\pm$0.22 &  $-$226.70$\pm$0.22 &     ...  &     ...   \\
   12 &          1,2,3 &      2.74 &   $-$61.2 &   $-$96.57$\pm$0.22 &  $-$225.85$\pm$0.22 &     ...  &     ...   \\
   13 &          1,2,3 &      2.73 &   $-$61.0 &   $-$38.89$\pm$0.22 &  $-$198.64$\pm$0.22 &   $-$5.0$\pm$0.6 &   $-$1.5$\pm$0.6  \\
   14 &              1 &      2.44 &   $-$60.8 &   $-$11.87$\pm$0.23 &  $-$225.24$\pm$0.23 &     ...  &     ...   \\
   15 &            1,2 &      2.18 &   $-$60.5 &   $-$30.75$\pm$0.23 &  $-$212.96$\pm$0.23 &     ...  &     ...   \\
   16 &              2 &      1.66 &   $-$60.9 &   $-$27.59$\pm$0.25 &  $-$227.45$\pm$0.25 &     ...  &     ...   \\
   17 &          1,2,3 &      1.52 &   $-$60.8 &   $-$49.05$\pm$0.23 &  $-$226.32$\pm$0.23 &   $-$3.7$\pm$0.6 &    4.5$\pm$0.6  \\
   18 &            2,3 &      1.50 &   $-$60.6 &    $-$7.96$\pm$0.29 &  $-$221.73$\pm$0.30 &     ...  &     ...   \\
   19 &              1 &      1.21 &   $-$60.9 &   $-$30.26$\pm$0.23 &  $-$233.55$\pm$0.22 &     ...  &     ...   \\
   20 &              1 &      0.67 &   $-$60.4 &    $-$3.82$\pm$0.26 &  $-$216.19$\pm$0.25 &     ...  &     ...   \\
   21 &            1,2 &      0.46 &   $-$60.2 &    15.53$\pm$0.23 &  $-$211.69$\pm$0.23 &     ...  &     ...   \\
   22 &              1 &      0.29 &   $-$60.2 &     4.66$\pm$0.24 &  $-$211.16$\pm$0.23 &     ...  &     ...   \\
   23 &              1 &      0.21 &   $-$60.2 &    21.63$\pm$0.24 &  $-$207.60$\pm$0.23 &     ...  &     ...   \\
   24 &              1 &      0.16 &   $-$60.2 &     9.22$\pm$0.29 &  $-$207.58$\pm$0.27 &     ...  &     ...   \\
   25 &          1,2,3 &      0.16 &   $-$60.1 &    13.98$\pm$0.25 &  $-$219.16$\pm$0.24 &    3.3$\pm$0.8 &    2.7$\pm$0.8  \\
   26 &              1 &      0.15 &   $-$60.0 &    26.78$\pm$0.29 &  $-$207.22$\pm$0.27 &     ...  &     ...   \\
   27 &              1 &      0.11 &   $-$60.0 &    34.94$\pm$0.34 &  $-$214.29$\pm$0.31 &     ...  &     ...   \\
   28 &              1 &      0.05 &   $-$59.9 &    31.41$\pm$0.52 &  $-$206.13$\pm$0.45 &     ...  &     ...   \\
   29 &              1 &      0.04 &   $-$59.8 &    45.34$\pm$0.50 &  $-$211.78$\pm$0.44 &     ...  &     ...   \\
\hline
& & & & & & & \\
\multicolumn{8}{c}{Cluster D}   \\
%& & & Cluster & D & & & \\
 & & & & & & & \\
    1 &          1,2,3 &     10.50 &   $-$60.5 &  $-$269.61$\pm$0.23 &  $-$279.16$\pm$0.22 &   $-$3.1$\pm$0.5 &   $-$0.8$\pm$0.5  \\
    2 &          1,2,3 &      9.30 &   $-$60.6 &  $-$251.30$\pm$0.22 &  $-$272.12$\pm$0.22 &   $-$5.2$\pm$0.5 &   $-$4.5$\pm$0.5  \\
    3 &          1,2,3 &      4.92 &   $-$60.6 &  $-$259.74$\pm$0.22 &  $-$274.23$\pm$0.22 &     ...  &     ...   \\
    4 &              1 &      3.48 &   $-$60.9 &  $-$242.43$\pm$0.22 &  $-$265.12$\pm$0.22 &     ...  &     ...   \\
    5 &          1,2,3 &      2.49 &   $-$60.4 &  $-$278.43$\pm$0.22 &  $-$281.39$\pm$0.22 &     ...  &     ...   \\
    6 &          1,2,3 &      2.45 &   $-$60.7 &  $-$243.93$\pm$0.23 &  $-$272.15$\pm$0.22 &     ...  &     ...   \\
    7 &              3 &      2.33 &   $-$60.8 &  $-$256.89$\pm$0.23 &  $-$271.67$\pm$0.23 &     ...  &     ...   \\
    8 &            1,2 &      0.39 &   $-$60.3 &  $-$287.93$\pm$0.26 &  $-$282.13$\pm$0.25 &     ...  &     ...   \\
\hline
& & & & & & & \\
\multicolumn{8}{c}{Cluster E}   \\ 
%& & & Cluster & E & & & \\
 & & & & & & & \\
    1 &          1,2,3 &     27.20 &   $-$58.0 &   158.51$\pm$0.22 &  $-$502.88$\pm$0.22 &   $-$1.2$\pm$0.6 &   $-$2.1$\pm$0.6  \\
    2 &          1,2,3 &      9.99 &   $-$58.1 &   153.63$\pm$0.22 &  $-$510.19$\pm$0.22 &     ...  &     ...   \\
    3 &          1,2,3 &      6.59 &   $-$58.0 &   162.60$\pm$0.22 &  $-$497.84$\pm$0.22 &     ...  &     ...   \\
    4 &          1,2,3 &      3.59 &   $-$58.3 &   103.34$\pm$0.22 &  $-$489.90$\pm$0.22 &   $-$5.2$\pm$0.6 &   $-$0.8$\pm$0.6  \\
    5 &          1,2,3 &      2.58 &   $-$58.2 &   148.77$\pm$0.23 &  $-$515.05$\pm$0.22 &     ...  &     ...   \\
    6 &          1,2,3 &      2.57 &   $-$59.0 &    87.26$\pm$0.22 &  $-$558.22$\pm$0.22 &   $-$0.5$\pm$0.6 &    0.4$\pm$0.6  \\
    7 &              1 &      2.23 &   $-$58.1 &   135.02$\pm$0.23 &  $-$507.95$\pm$0.22 &     ...  &     ...   \\
    8 &          1,2,3 &      2.10 &   $-$58.3 &   111.24$\pm$0.22 &  $-$487.24$\pm$0.22 &   $-$5.9$\pm$0.6 &   $-$1.3$\pm$0.6  \\
    9 &          1,2,3 &      1.87 &   $-$58.4 &    94.98$\pm$0.23 &  $-$492.10$\pm$0.22 &   $-$1.4$\pm$0.6 &    0.1$\pm$0.6  \\
   10 &          1,2,3 &      1.08 &   $-$58.5 &    87.32$\pm$0.23 &  $-$496.15$\pm$0.23 &   $-$7.5$\pm$0.7 &    1.6$\pm$0.7  \\
   11 &            1,2 &      0.81 &   $-$58.3 &   118.43$\pm$0.23 &  $-$487.99$\pm$0.23 &     ...  &     ...   \\
   12 &              3 &      0.76 &   $-$58.4 &    88.50$\pm$0.24 &  $-$499.17$\pm$0.24 &     ...  &     ...   \\
   13 &            1,2 &      0.57 &   $-$58.5 &    80.88$\pm$0.23 &  $-$498.27$\pm$0.22 &     ...  &     ...   \\
   14 &              1 &      0.48 &   $-$58.4 &   132.10$\pm$0.25 &  $-$482.04$\pm$0.24 &     ...  &     ...   \\
   15 &          1,2,3 &      0.42 &   $-$58.7 &   129.42$\pm$0.25 &  $-$547.72$\pm$0.25 &   $-$1.5$\pm$0.7 &    2.0$\pm$0.7  \\
   16 &          1,2,3 &      0.28 &   $-$59.0 &    85.84$\pm$0.26 &  $-$564.54$\pm$0.25 &     ...  &     ...   \\
   17 &          1,2,3 &      0.25 &   $-$59.0 &    88.70$\pm$0.26 &  $-$552.39$\pm$0.25 &     ...  &     ...   \\
   18 &            1,2 &      0.23 &   $-$58.6 &    71.88$\pm$0.25 &  $-$503.73$\pm$0.24 &     ...  &     ...   \\
   19 &              1 &      0.19 &   $-$59.3 &    60.79$\pm$0.26 &  $-$579.49$\pm$0.25 &     ...  &     ...   \\
   20 &            1,2 &      0.18 &   $-$58.8 &   146.00$\pm$0.27 &  $-$537.06$\pm$0.26 &     ...  &     ...   \\
   21 &              1 &      0.17 &   $-$59.0 &    38.18$\pm$0.30 &  $-$608.17$\pm$0.28 &     ...  &     ...   \\
   22 &            1,2 &      0.13 &   $-$58.7 &   124.73$\pm$0.27 &  $-$553.96$\pm$0.26 &     ...  &     ...   \\
   23 &              1 &      0.11 &   $-$58.9 &   165.04$\pm$0.33 &  $-$528.22$\pm$0.31 &     ...  &     ...   \\
   24 &              1 &      0.07 &   $-$59.4 &    68.73$\pm$0.39 &  $-$582.62$\pm$0.35 &     ...  &     ...   \\
   25 &              1 &      0.05 &   $-$59.3 &    53.89$\pm$0.44 &  $-$583.35$\pm$0.39 &     ...  &     ...   \\
   26 &              1 &      0.04 &   $-$59.2 &    79.90$\pm$0.54 &  $-$576.28$\pm$0.48 &     ...  &     ...   \\
\hline
& & & & & & & \\
\multicolumn{8}{c}{Not in clusters}  \\
%& & & Not in & clusters & & & \\
 & & & & & & & \\
    1 &              1 &      0.15 &   $-$58.8 &   251.76$\pm$0.28 &  $-$414.67$\pm$0.27 &     ...  &     ...   \\
    2 &          1,2,3 &      0.09 &   $-$59.9 &  $-$136.61$\pm$0.32 &  $-$352.74$\pm$0.30 &    1.8$\pm$1.1 &   $-$0.3$\pm$1.0  \\
    3 &              1 &      0.09 &   $-$60.0 &  $-$147.99$\pm$0.36 &  $-$358.23$\pm$0.33 &     ...  &     ...   \\
    4 &              3 &      0.08 &   $-$59.7 &  $-$123.97$\pm$0.40 &  $-$347.00$\pm$0.38 &     ...  &     ...   \\
    5 &              1 &      0.08 &   $-$59.7 &   $-$73.83$\pm$0.36 &  $-$344.40$\pm$0.32 &     ...  &     ...   \\
    6 &              1 &      0.06 &   $-$59.7 &   $-$86.97$\pm$0.40 &  $-$350.47$\pm$0.36 &     ...  &     ...   \\
    7 &              1 &      0.04 &   $-$59.9 &  $-$128.31$\pm$0.53 &  $-$347.34$\pm$0.47 &     ...  &     ...   \\
    8 &              1 &      0.04 &   $-$53.6 &   224.43$\pm$0.59 &   $-$86.81$\pm$0.51 &     ...  &     ...   \\
\hline
& & & & & & & \\
\multicolumn{8}{c}{Center of Motion} \\
%& & & Center of & Motion & & & \\
 & & & & & & & \\
    0 &          1,2,3 &       ... &   $$-$$58.3 &  37.16$\pm$0.16  &   $$-$$237.33$\pm$0.16    &   0.0$\pm$0.0  &    0.0$\pm$0.0    \\
\end{longtable}
\tablefoot{
Column~1 gives the feature label number; column~2
lists the observing epochs at which the feature was detected;
columns~3~and~4 provide the intensity of the strongest spot
and the intensity-weighted LSR velocity, respectively, averaged over the
observing epochs; columns~5~and~6 give the position offsets (with
the associated errors) along the R.A. and Dec. axes, relative to the
feature~\#1 of cluster~A, measured at the first epoch of detection; columns~7~and~8 give the components of the relative
proper motion (with the associated errors) along the R.A. and Dec.
axes, measured with respect to the reference feature~\#0 (the
``center of motion'').\\
%\tablefoottext{a}{Feature with not reliable proper motion, even if observed at three or four epochs.}
}
\end{longtab}

\subsubsection{L.O.S. Acceleration}

\label{acc_los}

\begin{figure*}
\centering
\includegraphics[angle=0.0,width=14cm]{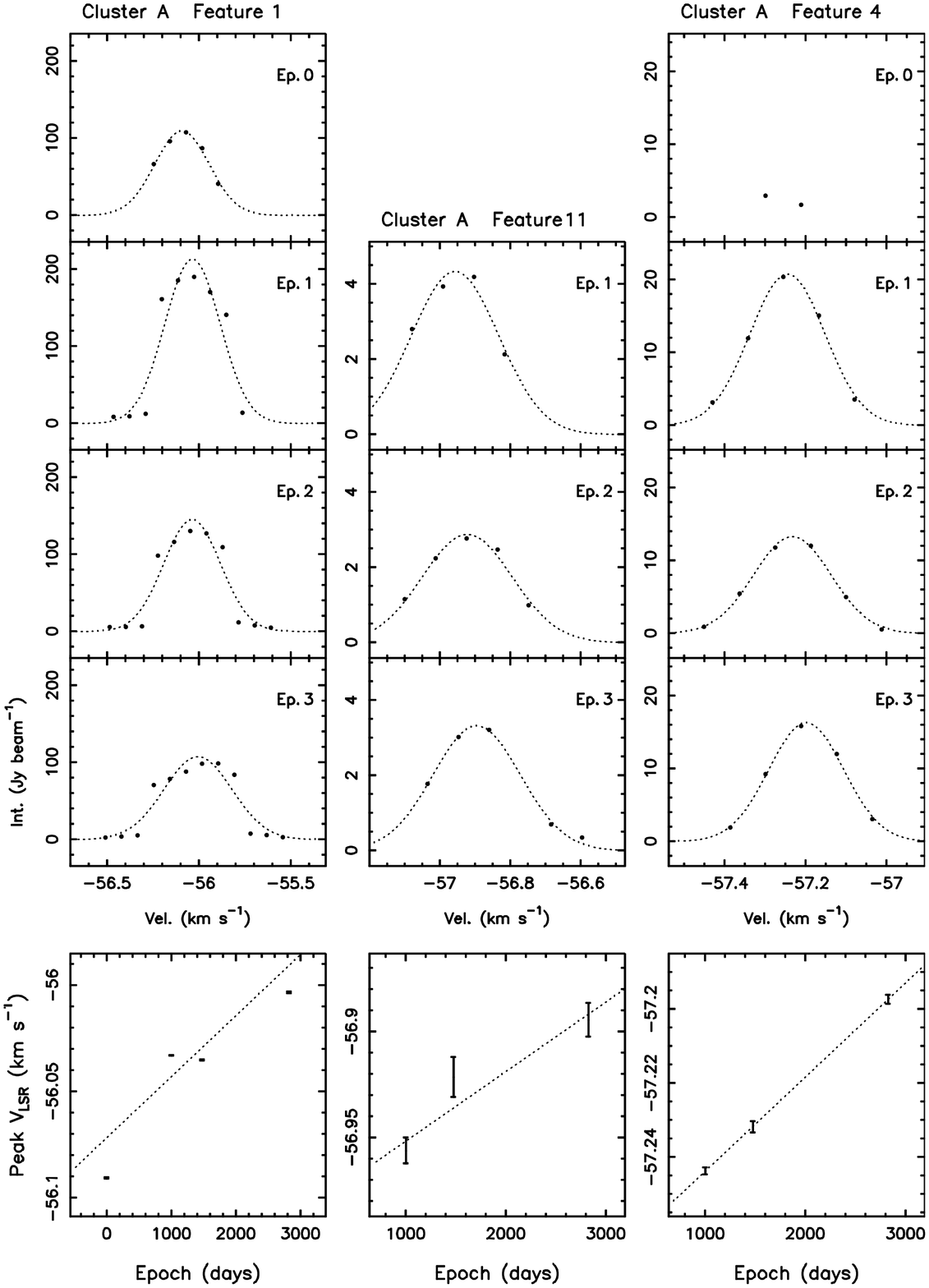}
\caption{Spectral profiles and time variation of peak velocities of selected maser features of cluster~"A".  
The plots in each column show a distinct maser feature in the cluster, labeled  according to Table~\ref{tab_6.7}.
({\it Top panels}) Spectral profiles of maser features at different epochs,
from Epoch~0 or Epoch~1 down to Epoch~3. 
The spectra ({\it black dots}) are produced by
plotting the spot intensity  vs. the spot channel velocity. For sufficiently well sampled maser spectra,
the {\it dotted curve} shows the Gaussian profile fitted to the spectral emission.
({\it Bottom panels}) Plot of the maser peak velocity
(with the associated errorbar) 
vs. the observing epoch (expressed in days elapsed from Epoch~0). 
The {\it dotted line} gives the least-square linear fit of velocities vs. time.
}
\label{acc_clA}
\end{figure*}

\begin{figure*}
\centering
\includegraphics[angle=0.0,width=14cm]{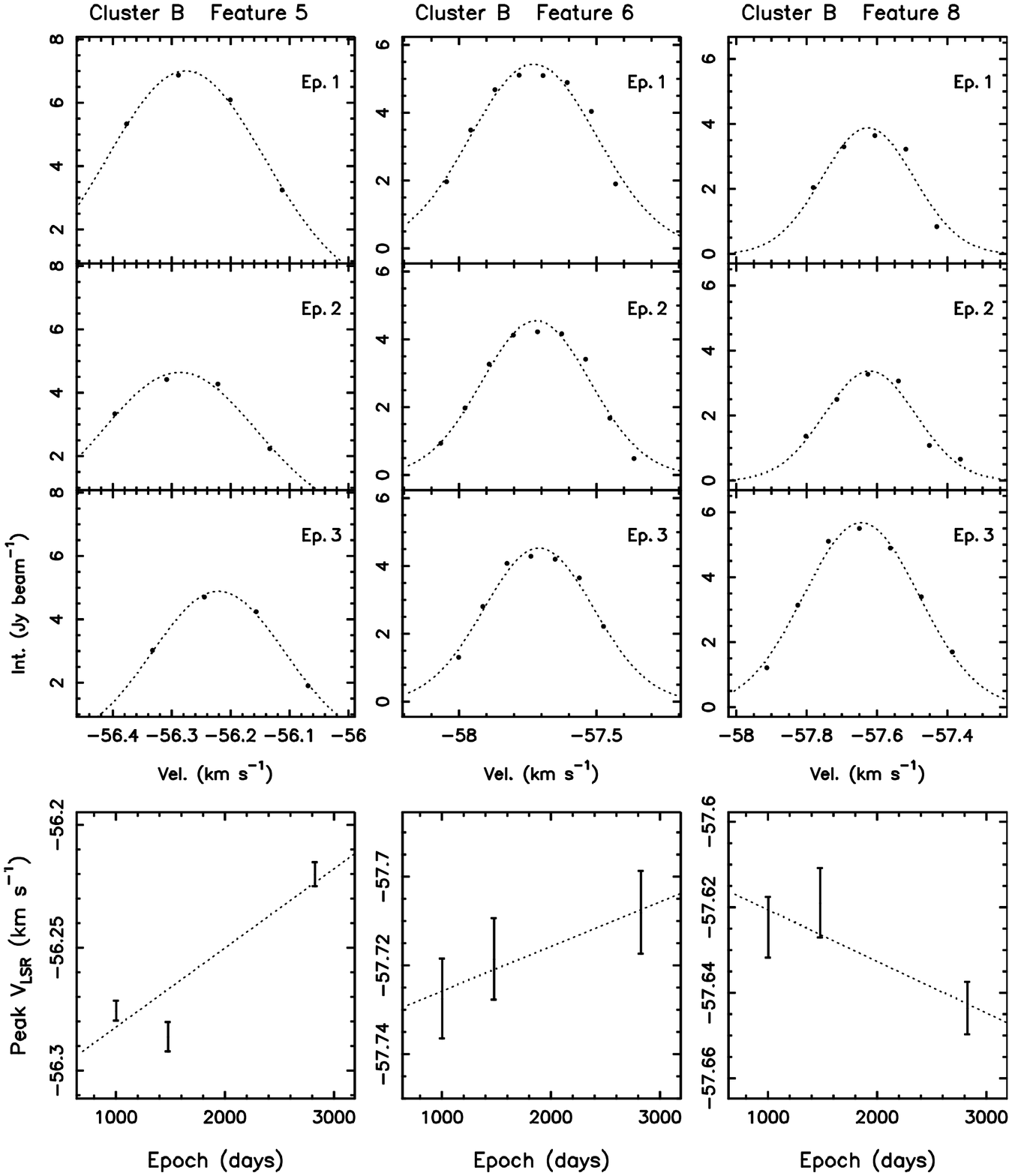}
\caption{Same as for Fig.~\ref{acc_clA} for the cluster~"B".}
\label{acc_clB}
\end{figure*}

\begin{figure*}
\centering
%\resizebox{12cm}{!}{\includegraphics[angle=0.0]{6.7_pos_vel.eps}}
\includegraphics[angle=0.0,width=14cm]{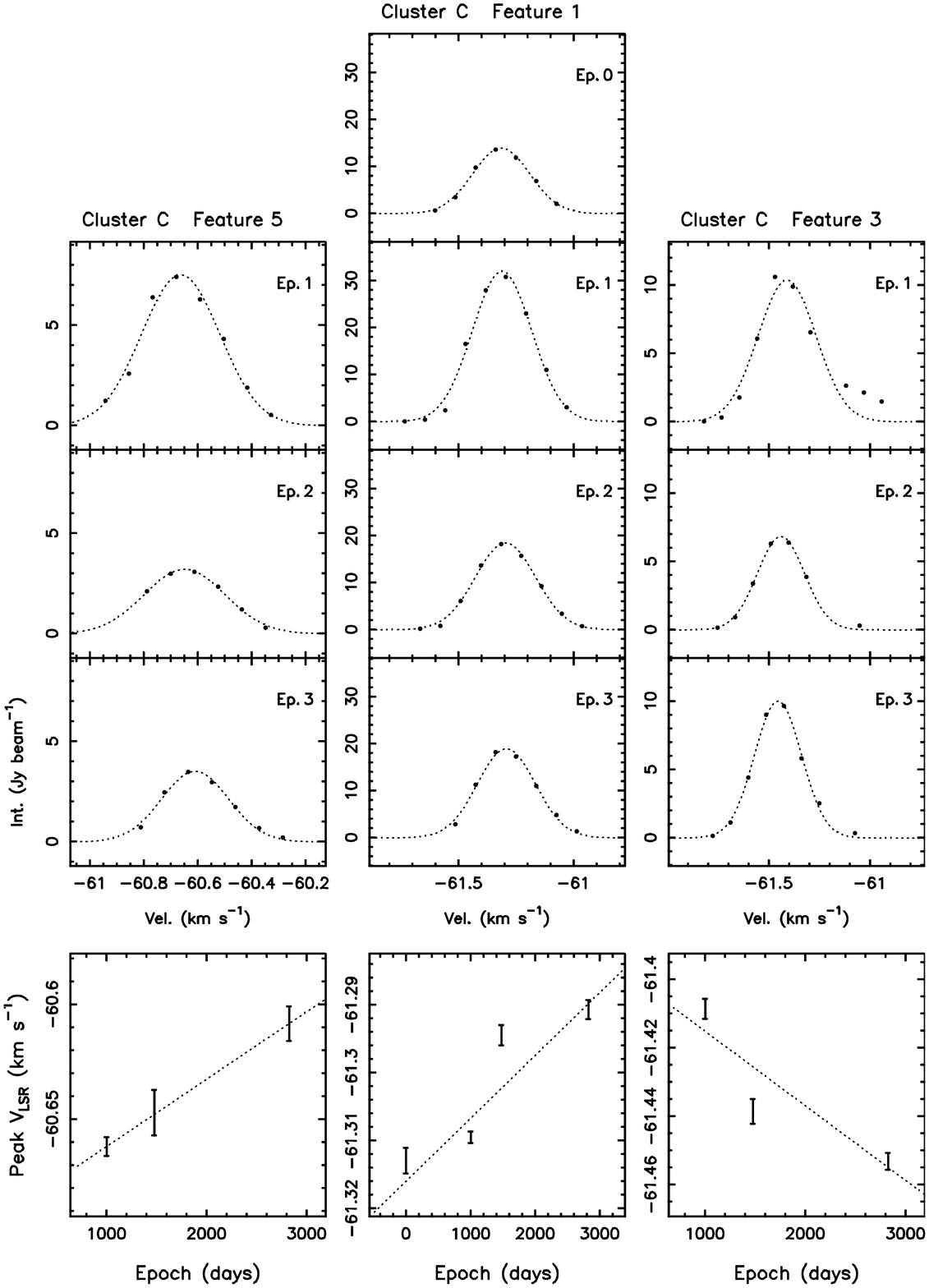}
\caption{Same as for Fig.~\ref{acc_clA} for the cluster~"C".}
\label{acc_clC}
\end{figure*}

\begin{figure*}
\centering
%\resizebox{12cm}{!}{\includegraphics[angle=0.0]{6.7_pos_vel.eps}}
\includegraphics[angle=0.0,width=14cm]{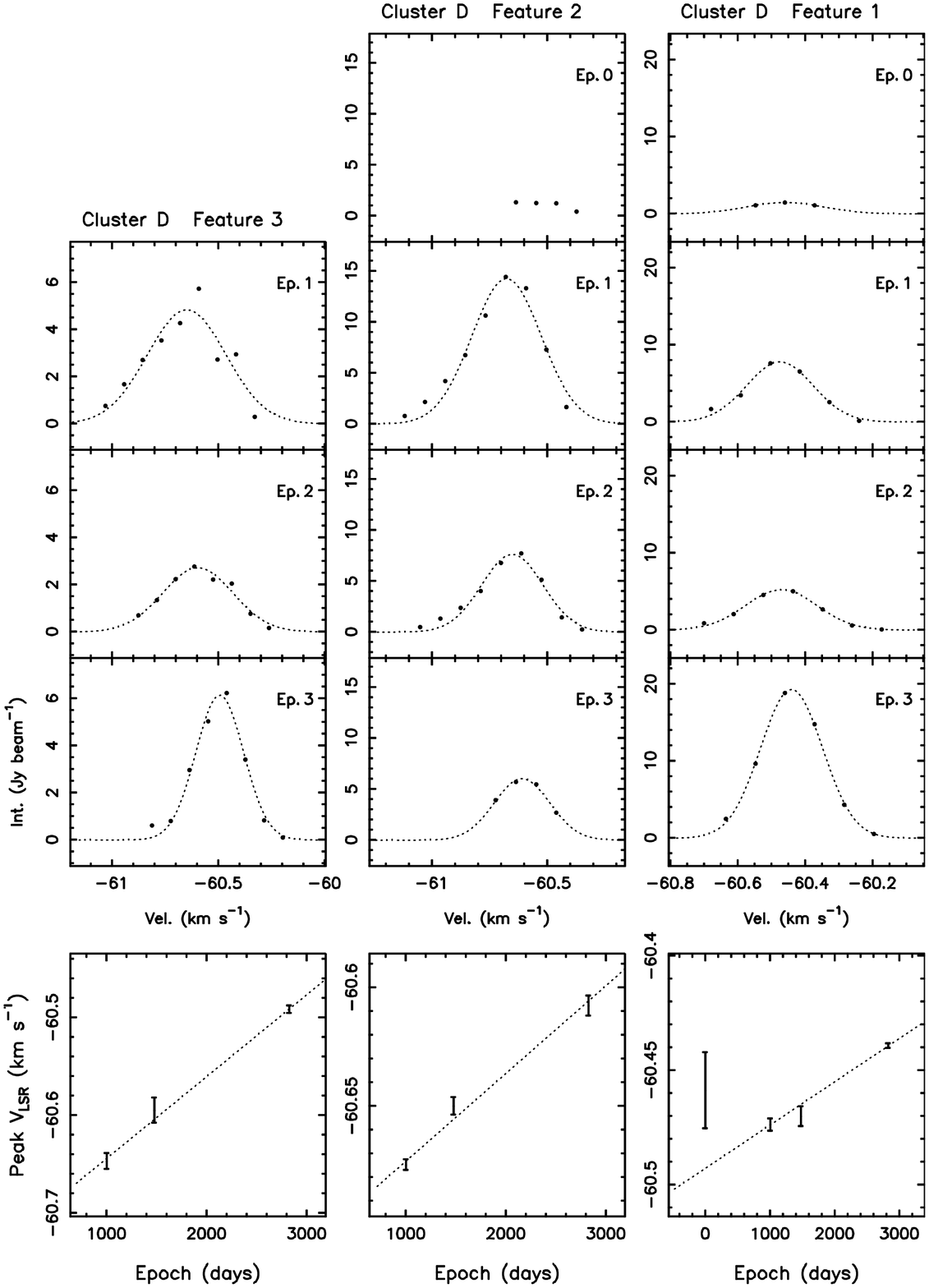}
\caption{Same as for Fig.~\ref{acc_clA} for the cluster~"D".}
\label{acc_clD}
\end{figure*}

For a selection of features in each maser cluster, Figs.~\ref{acc_clA}~to~\ref{acc_clD} show the Gaussian fit
of the maser spectral profiles and the linear fit of the variation of the peak velocity with the observing epoch. 
These figures show that the spectrum
of relatively strong maser features is reasonably well fitted with a Gaussian profile, allowing us to derive
the maser peak velocity at a given epoch with a typical accuracy of $\sim$0.01~\kms, about ten times smaller than
the velocity resolution of \ 0.09~\kms. 
Remarkably, these plots show that the peak velocity of many 6.7~GHz features vary linearly with time, 
providing a measurement of maser l.o.s. accelerations of typical amplitude
of \ $\sim$10$^{-3}$ -- 10$^{-2}$~\kmsy. For each maser cluster, Table~\ref{tab_acc} reports the values of 
the measured l.o.s. accelerations for features which are persistent in time, with peak flux densities $>$3~\Jyb, and 
with a sufficiently well sampled maser spectrum. Figure~\ref{acc_pos} shows the spatial distribution 
of maser l.o.s. accelerations inside each cluster. 
We note that features in cluster~"A" have similar values of l.o.s. acceleration around \ 0.01~\kmsy, while in clusters "B" and "C"
there is a larger scatter, with both positive and negative values from \ $-$0.019 to 0.016~\kmsy.
Finally, in cluster~"E" most of the measurements are compatible with a null value of maser l.o.s. acceleration
(see Table~\ref{tab_acc}). 
We discuss the implications of these accelerations in terms of gas dynamics in Sect.~\ref{mas_kin}.

\begin{figure*}
\centering
%\resizebox{12cm}{!}{\includegraphics[angle=0.0]{6.7_pos_vel.eps}}
\includegraphics[angle=0.0,width=14cm]{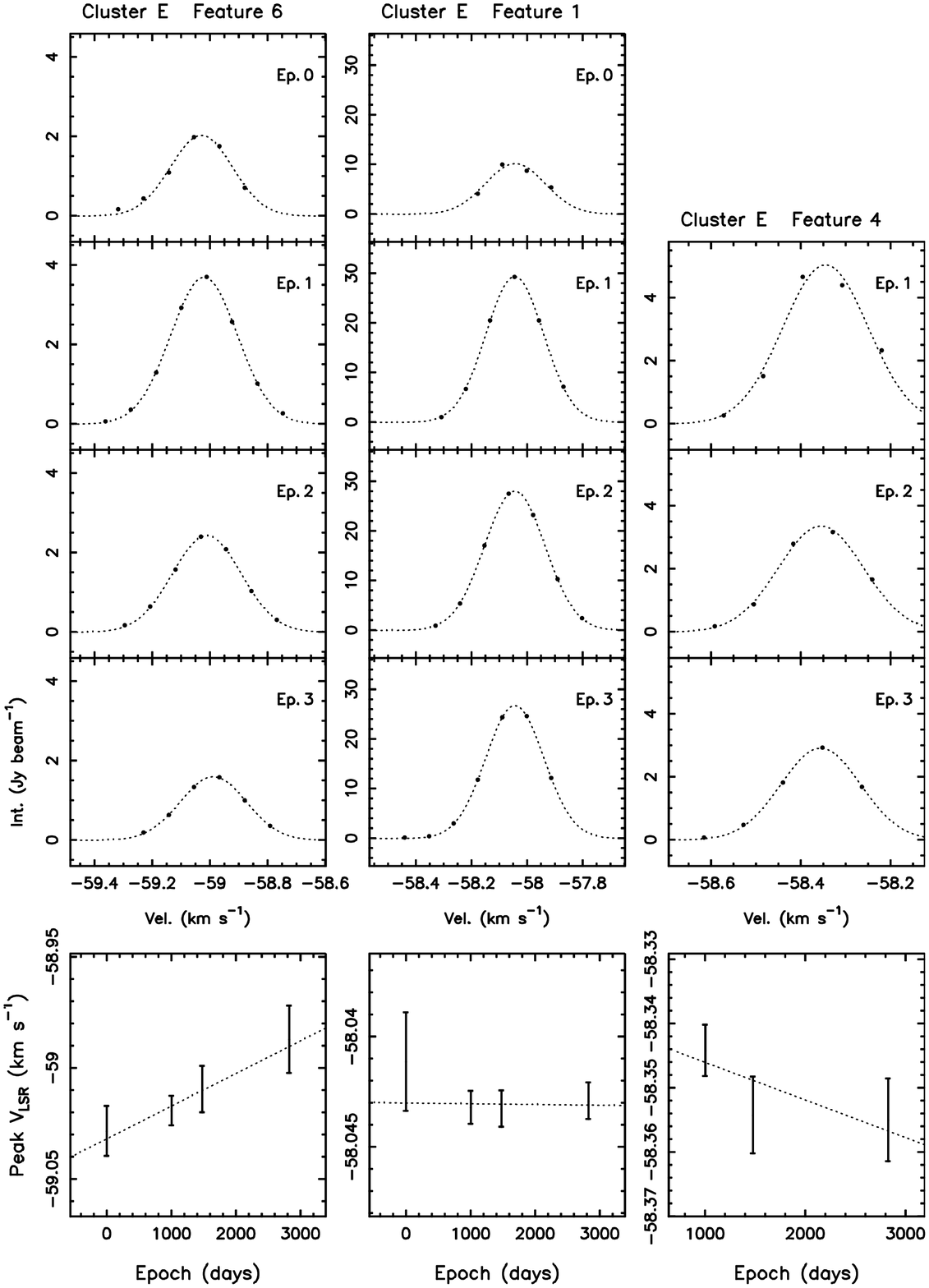}
\caption{Same as for Fig.~\ref{acc_clA} for the cluster~"E".}
\label{acc_clE}
\end{figure*}

%\begin{table*}
\begin{sidewaystable}
\caption{Maser Feature L.O.S. Acceleration}             
\label{tab_acc}      
\centering          
\begin{tabular}{cr|cr|cr|cr|cr}     % 7 columns 
\hline\hline   
\multicolumn{2}{c}{Cluster~A} &  \multicolumn{2}{c}{Cluster~B} &    \multicolumn{2}{c}{Cluster~C} &
\multicolumn{2}{c}{Cluster~D} & \multicolumn{2}{c}{Cluster~E} \\
\hline
\multicolumn{1}{c}{Feature} & \multicolumn{1}{c}{Acc.} & \multicolumn{1}{c}{Feature} & \multicolumn{1}{c}{Acc.} &
\multicolumn{1}{c}{Feature} & \multicolumn{1}{c}{Acc.} & \multicolumn{1}{c}{Feature} & \multicolumn{1}{c}{Acc.} &
\multicolumn{1}{c}{Feature} & \multicolumn{1}{c}{Acc.} \\
\multicolumn{1}{c}{Number}  & \multicolumn{1}{c}{(km s$^{-1}$ yr$^{-1}$)} &
\multicolumn{1}{c}{Number}  & \multicolumn{1}{c}{(km s$^{-1}$ yr$^{-1}$)} &
\multicolumn{1}{c}{Number}  & \multicolumn{1}{c}{(km s$^{-1}$ yr$^{-1}$)} &
\multicolumn{1}{c}{Number}  & \multicolumn{1}{c}{(km s$^{-1}$ yr$^{-1}$)} &
\multicolumn{1}{c}{Number}  & \multicolumn{1}{c}{(km s$^{-1}$ yr$^{-1}$)} \\
\hline
  1  &   0.01048$\pm$6.E-05 & 1   & 0.0036$\pm$5.E-04    & 1 & 0.0034$\pm$3.E-04        
  & 1 & 0.0069$\pm$5.E-04 &  1  &  $-$0.00002$\pm$20.E-05  \\
  3  &     0.010$\pm$1.E-04 & 2   & $-$0.008$\pm$1.E-03  & 2 & $-$0.00006$\pm$48.E-05
  & 2 & 0.0124$\pm$9.E-04 &  4  &  $-$0.002$\pm$2.E-03  \\
  4  &   0.0093$\pm$3.E-04  & 3  & 0.011$\pm$1.E-03      & 3 & $-$0.0080$\pm$7.E-04
  & 3 & 0.031$\pm$2.E-03  &  6  &  0.005$\pm$2.E-03   \\ 
  9  &    0.009$\pm$1.E-03  & 4  & $-$0.019$\pm$2.E-03   & 5 & 0.011$\pm$2.0E-03 
  &  &                    &  8  & 0.001$\pm$3.E-03    \\
 11  &   0.012$\pm$2.E-03   & 5  & 0.012$\pm$1.E-03      & 7 & 0.001$\pm$1.E-03
  &  &                    &  9  & 0.0004$\pm$45.E-04  \\
     &                      & 6  & 0.004$\pm$3.E-03      &   &      
  &  &                    &  10  &  $-$0.013$\pm$8.E-03   \\
     &                      & 8  & $-$0.004$\pm$2.E-03   &   &
  &  &                    &   15  &    0.006$\pm$22.E-03     \\
     &                      & 10  & $-$0.003$\pm$3.E-03  &   &
  &  &                    &      &                       \\
     &                      & 13  &   0.013$\pm$8.E-03   &   &
  &  &                    &       &                      \\
     &                      & 14  & 0.016$\pm$5.E-03     &   &
  &  &                      &      &                     \\
\hline                  
\end{tabular}
\tablefoot{
The table reports the l.o.s. acceleration for the features of each maser cluster, which are persistent over three or four
epochs and with a sufficiently well sampled spectral profile. \\
 For each maser cluster, features are indicated using the
same label numbers as in Table~\ref{tab_6.7}.
}
\end{sidewaystable}
%\end{table*}

\begin{figure*}
\begin{center}
\begin{tabular}{cc}
\epsfig{file=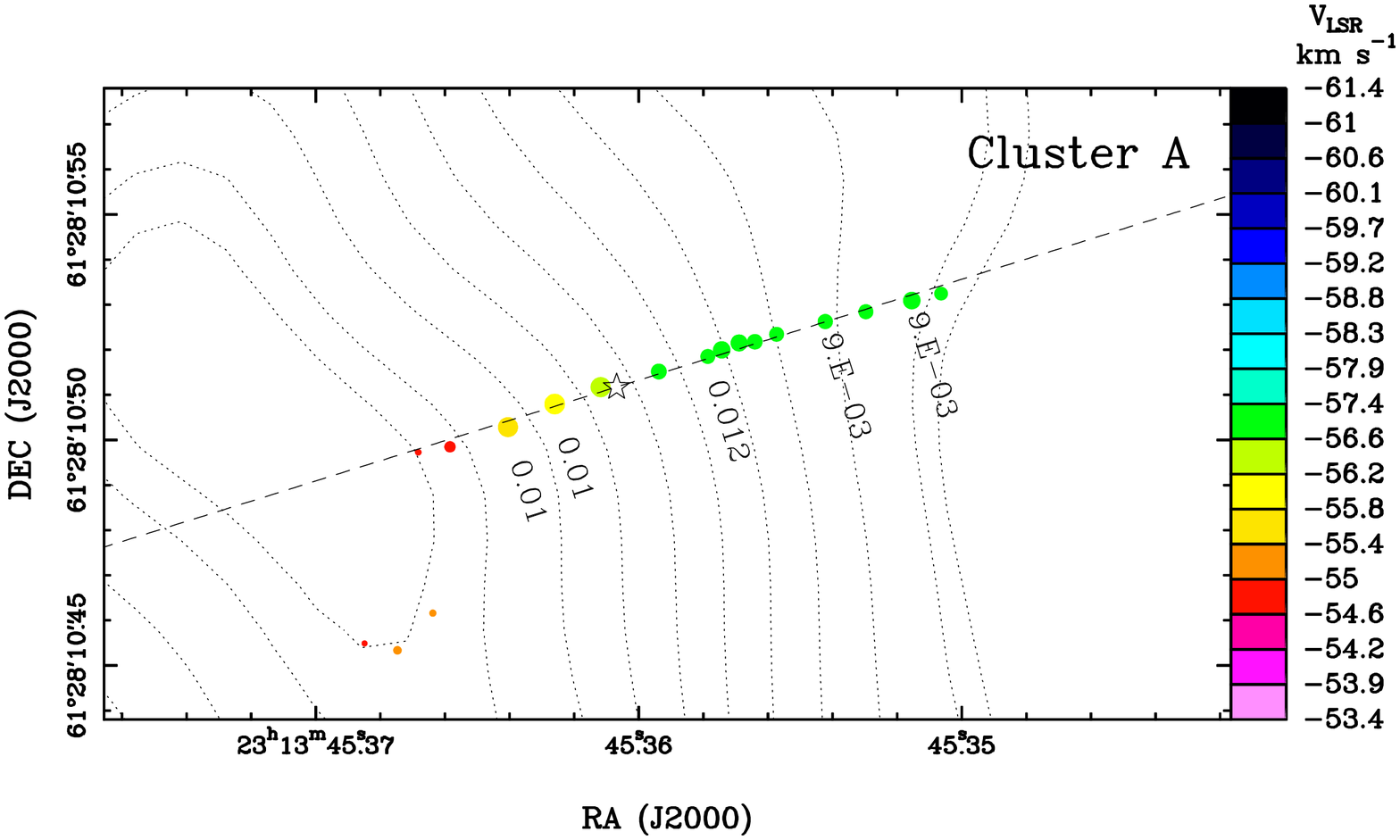,width=0.5\linewidth,clip=} &
\epsfig{file=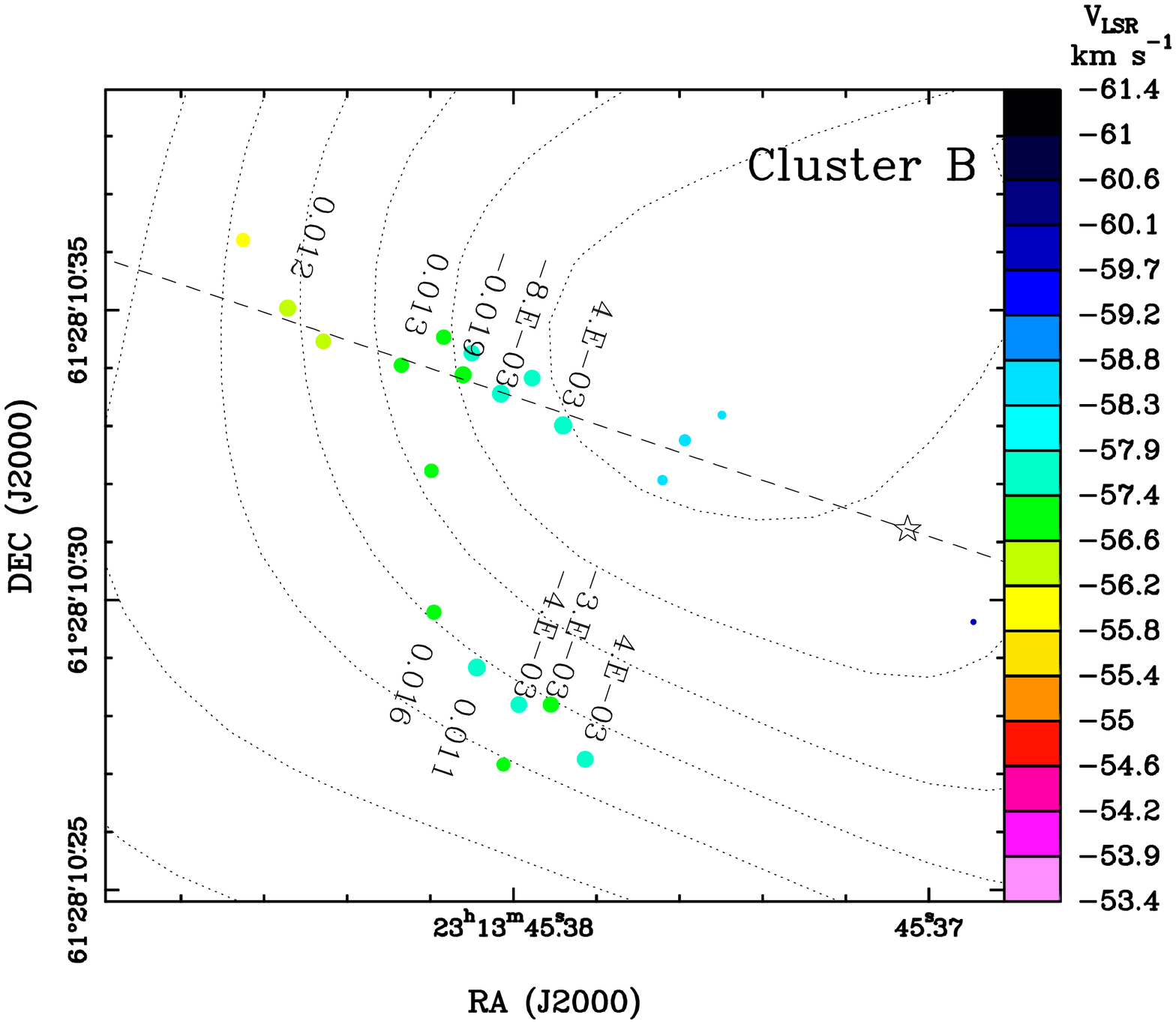,width=0.5\linewidth,clip=} \\
\epsfig{file=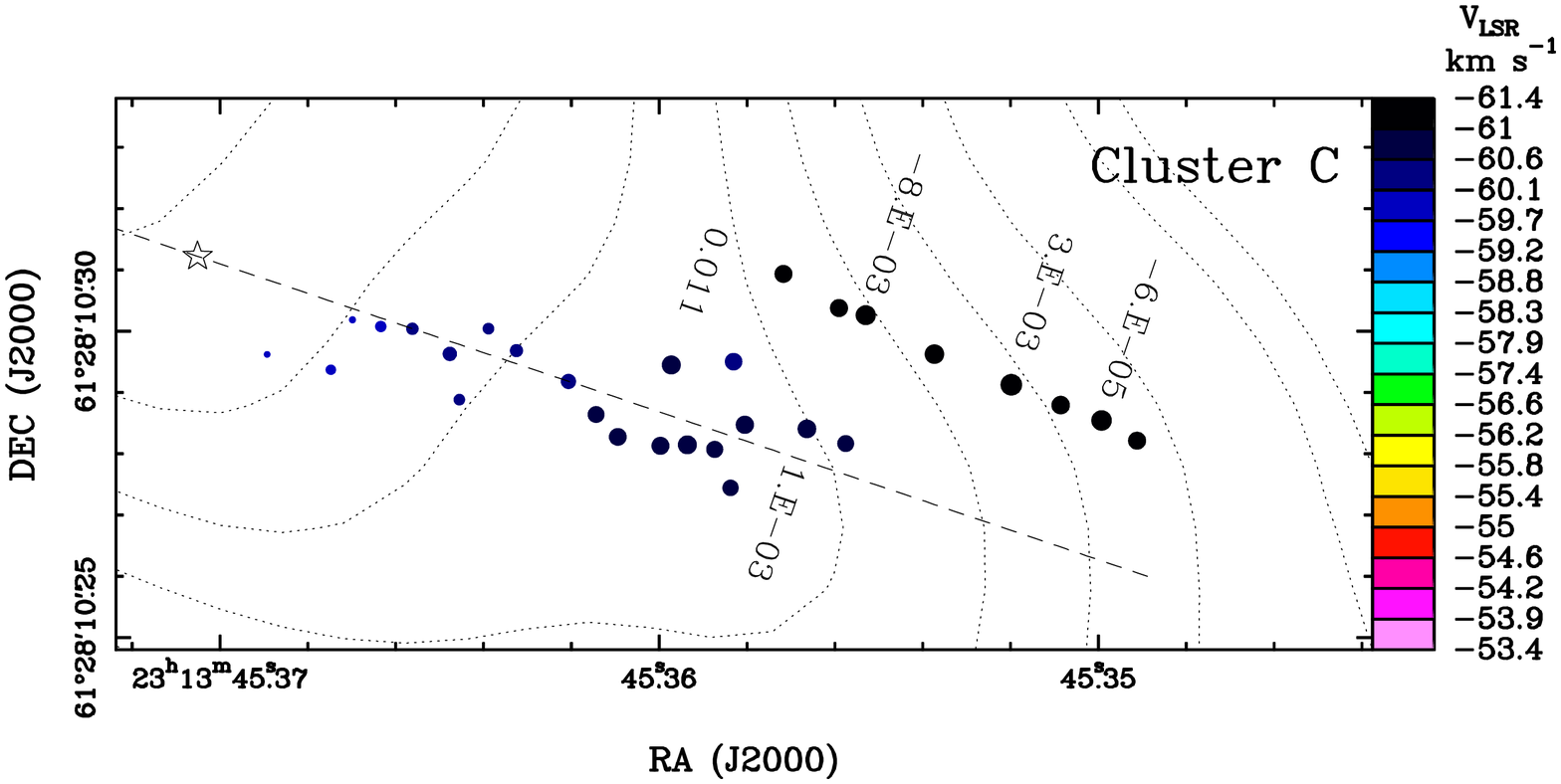,width=0.5\linewidth,clip=} &
\epsfig{file=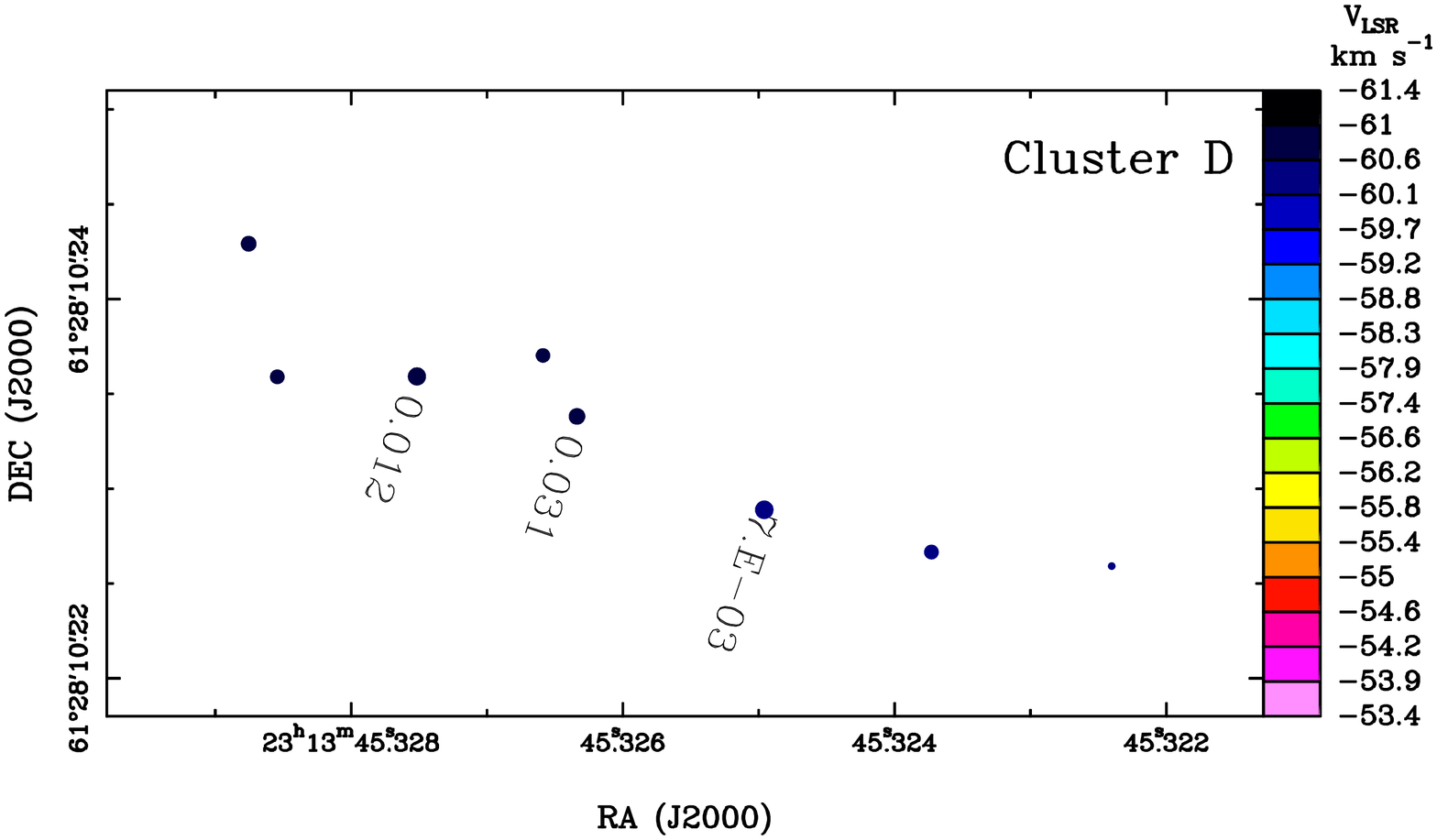,width=0.5\linewidth,clip=} \\
\multicolumn{2}{c}{
\epsfig{file=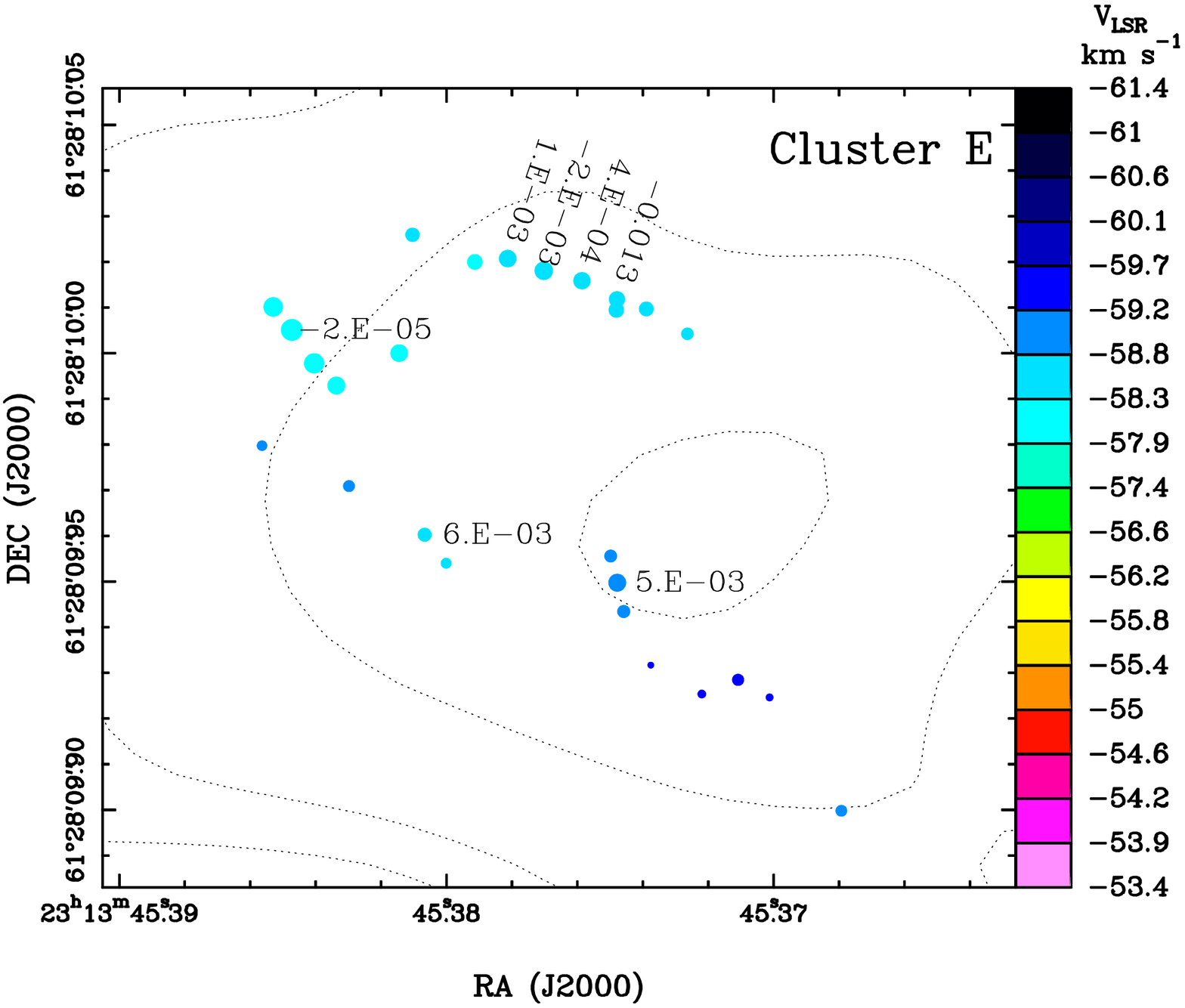,width=0.5\linewidth,clip=} 
}
\end{tabular}
\caption{l.o.s. accelerations measured for individual 6.7 GHz maser features. 
Each panel shows a different cluster, from "A" to "E".
For each maser feature, the value of the l.o.s. acceleration is quoted in \ km~s$^{-1}$~yr$^{-1}$. 
Symbols, colors and contours have the same meaning as in Fig.~\ref{vlsr_pos}.
In panels of the maser clusters "A", "B' and "C", the {\it dashed line} 
indicates the major axis of the maser feature' spatial distribution,
and the {\it star} marks the reference point to measure axis-projected offsets
(see Sect.~\ref{6.7_vlsr_regu}).
}
\label{acc_pos}
\end{center}
\end{figure*}

%-------------------------------------------------------------------------
\begin{figure}
\centering
\includegraphics[width=0.36\textwidth]{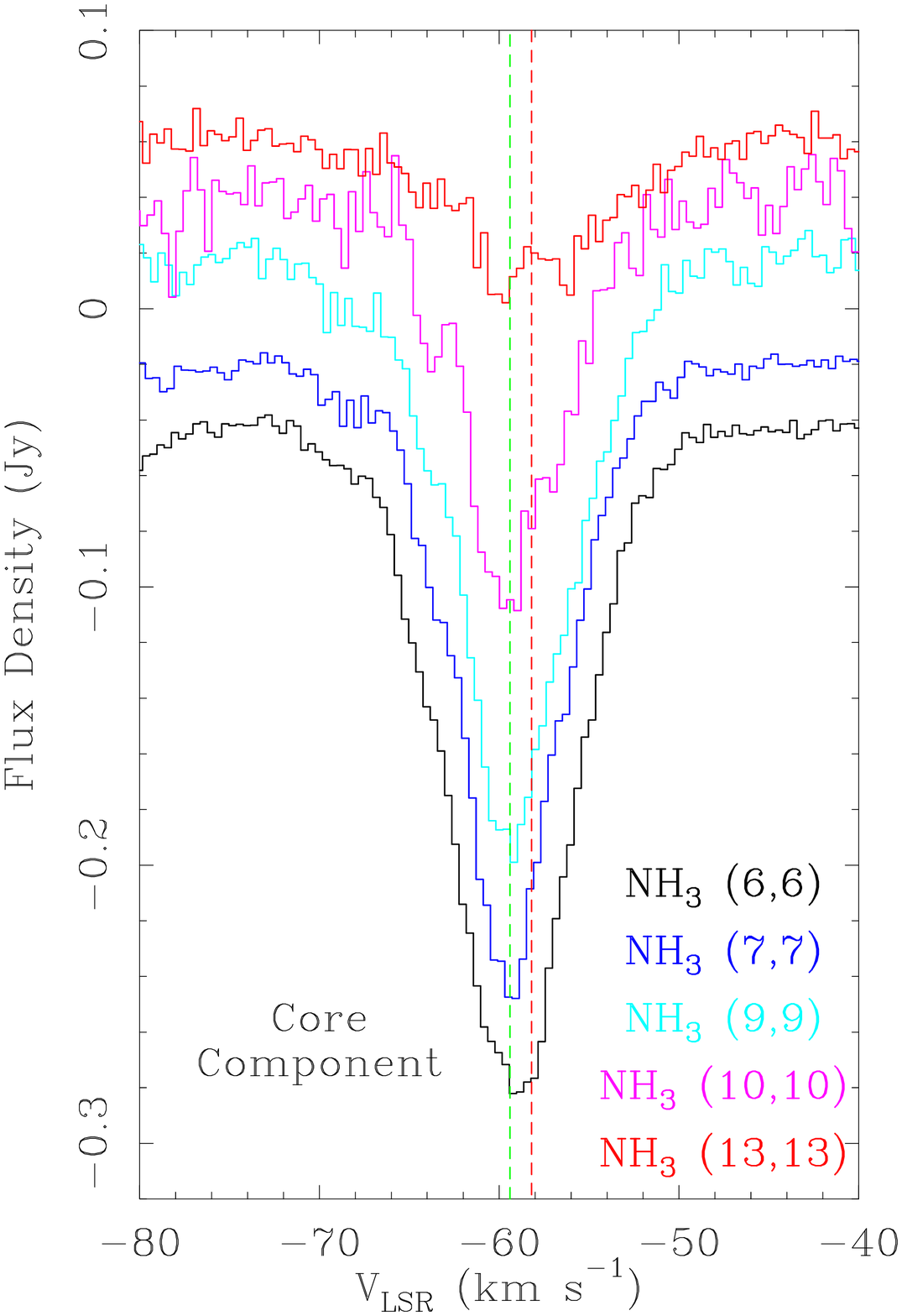}
\includegraphics[width=0.36\textwidth]{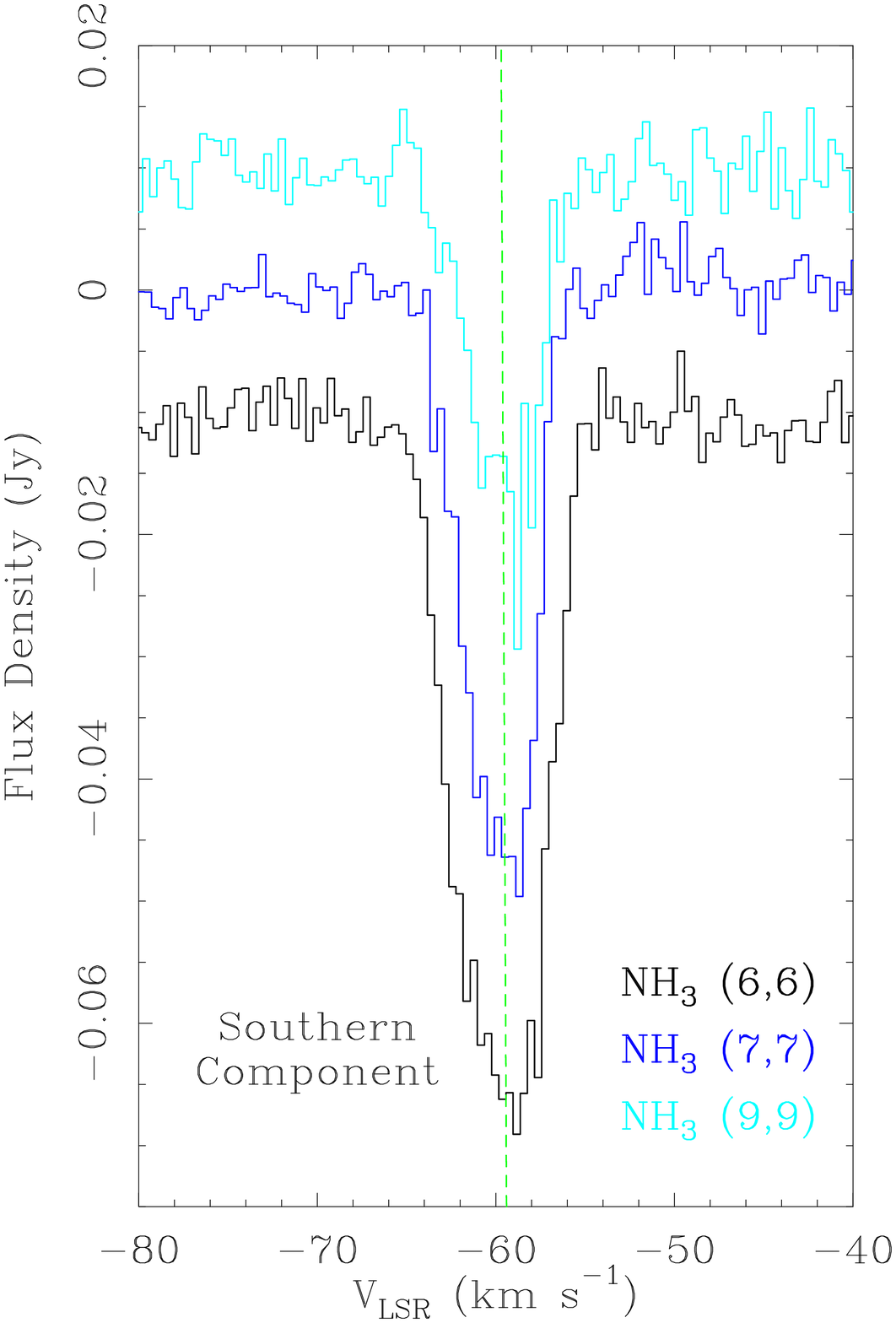}
\caption{Spectral profiles for \nh3\ inversion transitions (6,6), (7,7), (9,9), (10,10), and (13,13) observed toward \NGC1\ with the JVLA B-Array. The upper panel is showing the spectra integrated over the core component of the 1.3~cm continuum emission (see Fig.~\ref{vlsr_pos}), and the lower panel the spectra integrated towards the southern spherical component. 
An offset in flux density was applied to better evidence their spectral profiles.  
The velocity resolution is 0.4~\kms\ and the l.o.s. velocities are with respect to the local standard of rest (LSR). 
The dashed green (at $-$59.4~\kms) and red (at $-$58.2~\kms) lines indicate the central velocity of the least, (6,6), and most, (13,13), excited transitions, respectively, determined by fitting a single Gaussian to the transition spectral profile.  
In the southern component only the (6,6), (7,7), and (9,9) lines are clearly detected. 
The upper state energy levels of transitions shown here are \ $\approx$408--1693~K.
}
\label{nh3_spectra}
\end{figure}
%%--------------------------------------------------------------------------------------

%\subsection{\nh3 \Vlsr\ Distribution}
\subsection{Kinematics of \nh3\ Inversion Lines}
\label{res_nh3}

\begin{figure*}
\centering
%\resizebox{12cm}{!}{\includegraphics[angle=0.0]{6.7_pos_vel.eps}}
\includegraphics[angle=0.0,width=19cm]{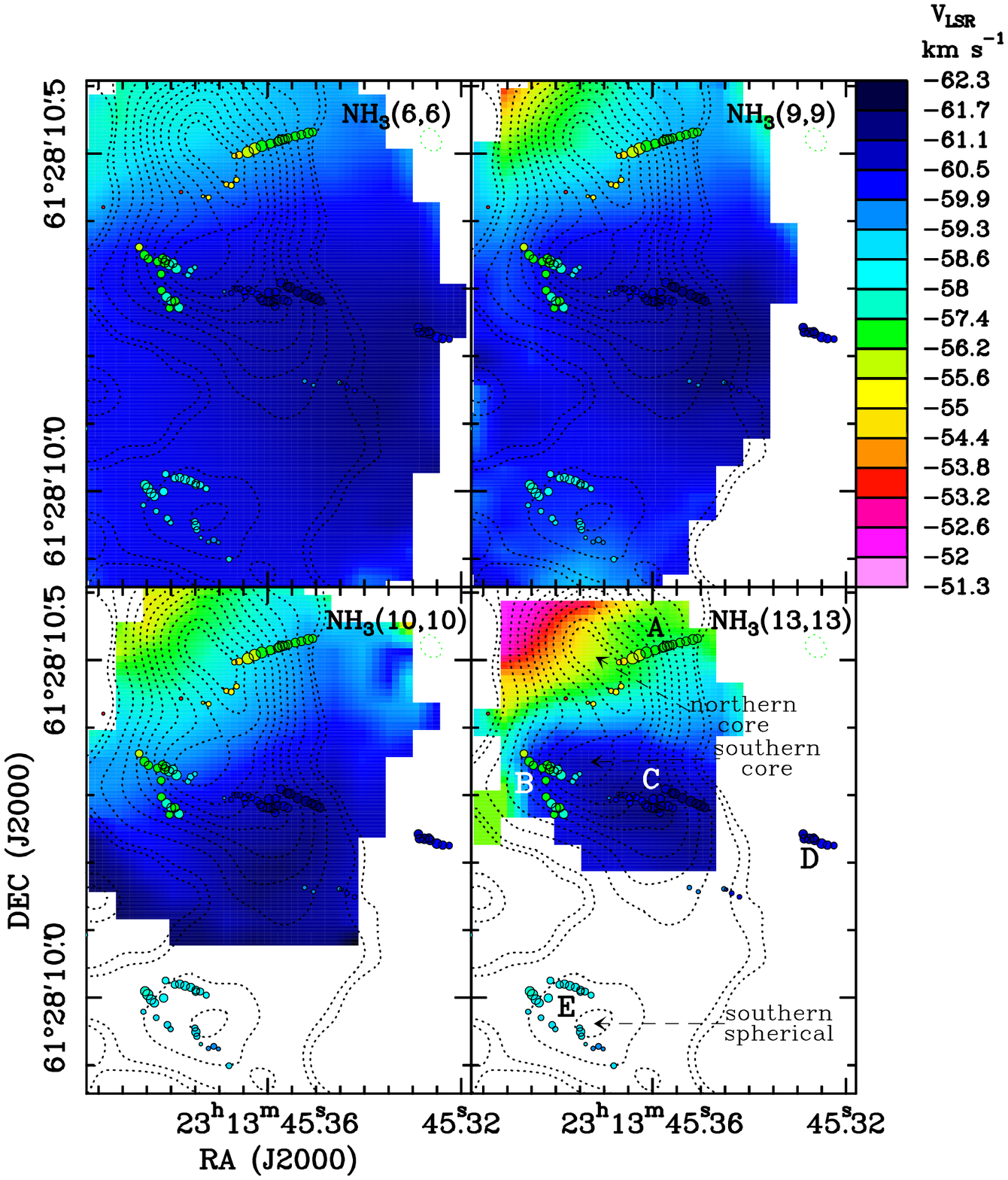}
\caption{Velocity fields of four inversion transitions of  NH$_3$, (6,6), (9,9), (10,10), and (13,13), 
as measured towards \NGC1\ with the JVLA B-Array.
In each panel, the positions of the 6.7~GHz \meth\ masers ({\it Black-circled, colored dots}) are overlaid on the \nh3\ images. 
The conversion code between colors and \Vlsr\ is indicated in the wedge to the right side
of the upper right panel. Dot size is proportional to the logarithm of the maser
intensity. {\it Dotted contours} have the same meaning as in Fig.~\ref{vlsr_pos}.
In the bottom right panel, the labels of the maser clusters and the names of the main
1.3~cm continuum peaks are given.}
\label{nh3_mom1}
\end{figure*}

We have mapped the hot \nh3\ gas from metastable transitions (6,6) up to (13,13) with 0\farcs2 resolution towards NGC7538~IRS1. These transitions are observed in absorption against the strong ultra-compact \HII\ region  
and cover upper-state energies, $E_{\mathrm {up}}$, from 400~K to 1700~K,  probing the hottest molecular gas in the region.
For each transition, we derived spectral profiles (Fig.~\ref{nh3_spectra}) and maps of  the velocity field (Fig.~\ref{nh3_mom1}).
The full analysis of this dataset, including kinematics and physical condition estimates, will be presented in a separate paper (Goddi et al., in prep.). 
In this paper, we will focus on the kinematics of the thermal molecular gas as probed by \nh3 on scales of 500--1500~AU,
to complement the small scale dynamics probed by \meth\ masers.

For each transition, we produced spectra by mapping each spectral channel and summing the flux density in each channel map, separately for the core and the southern component (top and bottom panels of Figures~\ref{nh3_spectra}, respectively). Towards the core, multiple transitions show similar line profiles, central velocities ($V_c$ from $-$58.2 to $-$59.4~\kms), and velocity widths ($\Delta W$ = 6.8--9.9~\kms) of the main hyperfine component,  as determined from single-Gaussian fits. 
Despite showing similar values, there are non-negligible changes in $\Delta W$ and $V_c$ with $(J,K)$, as compared with the velocity resolution (0.4~\kms).  
The central velocity of the optically thick lines, (6,6) and (7,7), gives a good estimate of the systemic velocity, 
$-$59.4~\kms, the same value quoted by \citet{Qiu11} from several molecular lines with \ $E_{\mathrm {up}}$=16--133~K \ imaged at 1.3~mm with the SMA. 
Interestingly, based on CARMA and SMA observations, \citet{Zhu13} report a similar value for the systemic velocity employing less excited lines at 1.3~mm, but also find an average peak velocity of \ $-$58.6~\kms, using more highly excited (and more optically thin) lines at 0.86~mm. Similarly, we find that the central velocity of the (13,13) line, $-$58.2~\kms, is higher than that of the (6,6) line. 
We also find the largest linewidth (9.9~\kms) for the (13,13) line. 

 In Figure~\ref{nh3_mom1}, we show the intensity-weighted l.o.s. velocity fields (or first moment maps) for four \nh3\ transitions, overlaid on the 1.3 cm continuum emission.  The positions and \Vlsr\ of \meth\ maser features are also overlaid.  
The \nh3\  absorption follows closely the continuum emission, as expected. 
In particular, for lower excitation transitions, (6,6) and (9,9), the absorption is extended N--S across \ $\approx$1\arcsec (2700~AU), and reveals two main condensations of hot molecular gas 
associated  with the core and the southern spherical component identified by \citet{Gau95} at 1.3 cm (Fig.~\ref{vlsr_pos}). 
The highest excitation \nh3\ lines, (10,10) and  (13,13), originate from the core of the radio continuum, 
and probe the hottest gas associated with clusters "A", "B", and "C" of \meth\  masers. 
The southern spherical component, associated with cluster "E" of \meth\ masers, has the weakest integrated absorption and it is not detected in the highest-$JK$ transitions.  

Remarkably, the \nh3\ absorption towards the core shows a distinct velocity gradient in each line,  with redshifted absorption towards NE and slightly blueshifted absorption towards SW with respect to the hot core centre, where the largest values of velocity dispersion are also observed. 
The velocity gradient is at a \ PA $\approx$ 30--40\degree, and has a magnitude \ $\Delta V$ $\approx$ 5 $\rightarrow$ 9~\kms, going from the lower-excitation to the higher-excitation lines. 
A similar PA ($\approx$43\degree) is measured by Zhu et al. 2013 from SMA images of the OCS(19-18), 
CH$_3$CN(12-11), and $^{13}$CO(2-1) lines with 0\farcs7 angular resolution. 

Despite a qualitative agreement, there are some significant differences in the velocity field probed by \nh3\,and \meth\,masers. 
In Sect.~\ref{nh3_vlsr_regu}, we will show that this is a direct consequence of the limited angular resolution of the \nh3\,maps which do not resolve the northern and southern components of the radio continuum.

\section{Ordered Kinematical Structures}
\label{vlsr_regu}

\subsection{6.7~GHz \meth\ Masers}
\label{6.7_vlsr_regu}

\begin{figure*}
\centering
%\resizebox{12cm}{!}{\includegraphics[angle=0.0]{6.7_pos_vel.eps}}
\includegraphics[angle=0.0,width=\textwidth]{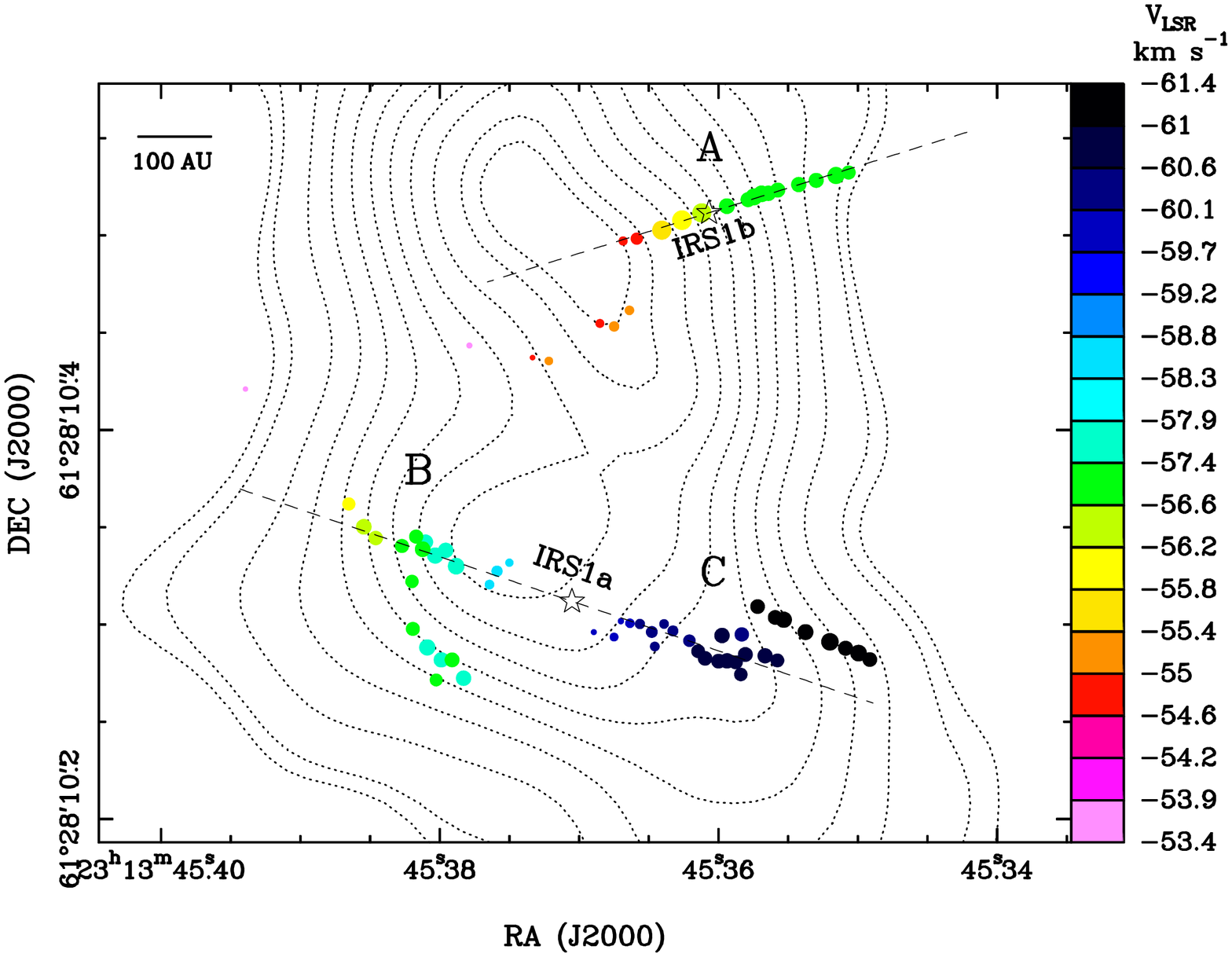}
\caption{Linear fits to the spatial distributions of maser features in cluster~"A", and
in the combined clusters "B"+"C" ({\it dashed lines}). 
Symbols, colors and contours have the same meaning as in Fig.~\ref{vlsr_pos}.
To fit the maser spatial distribution in these clusters, we have selected a subset of features 
closely aligned along the cluster major axis,
excluding the subgroup of weak features scattered to the SE of cluster~"A",
the subgroup of features of cluster "B" lying further South, and the subgroup
in cluster "C" more detached to NW. 
The {\it stars} labeled IRS1a and IRS1b mark the YSO positions, as discussed in Sect.~\ref{phy_sen}. 
}
\label{vlsr_pos_z}
\end{figure*}

\begin{figure*}
\includegraphics[angle=0.0,width=9.60cm]{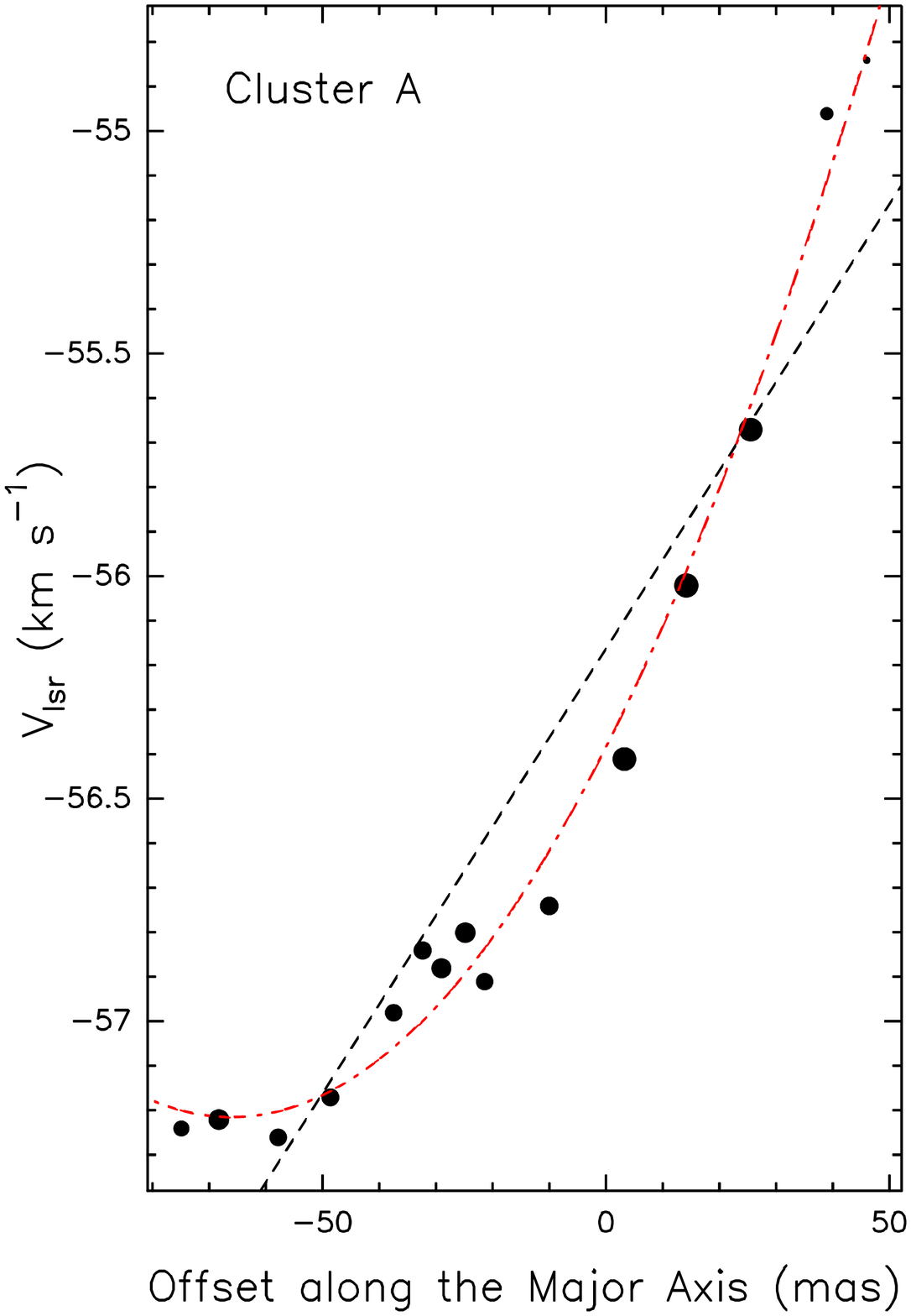}
\hspace{0.5cm}
\includegraphics[angle=0.0,width=8.9cm]{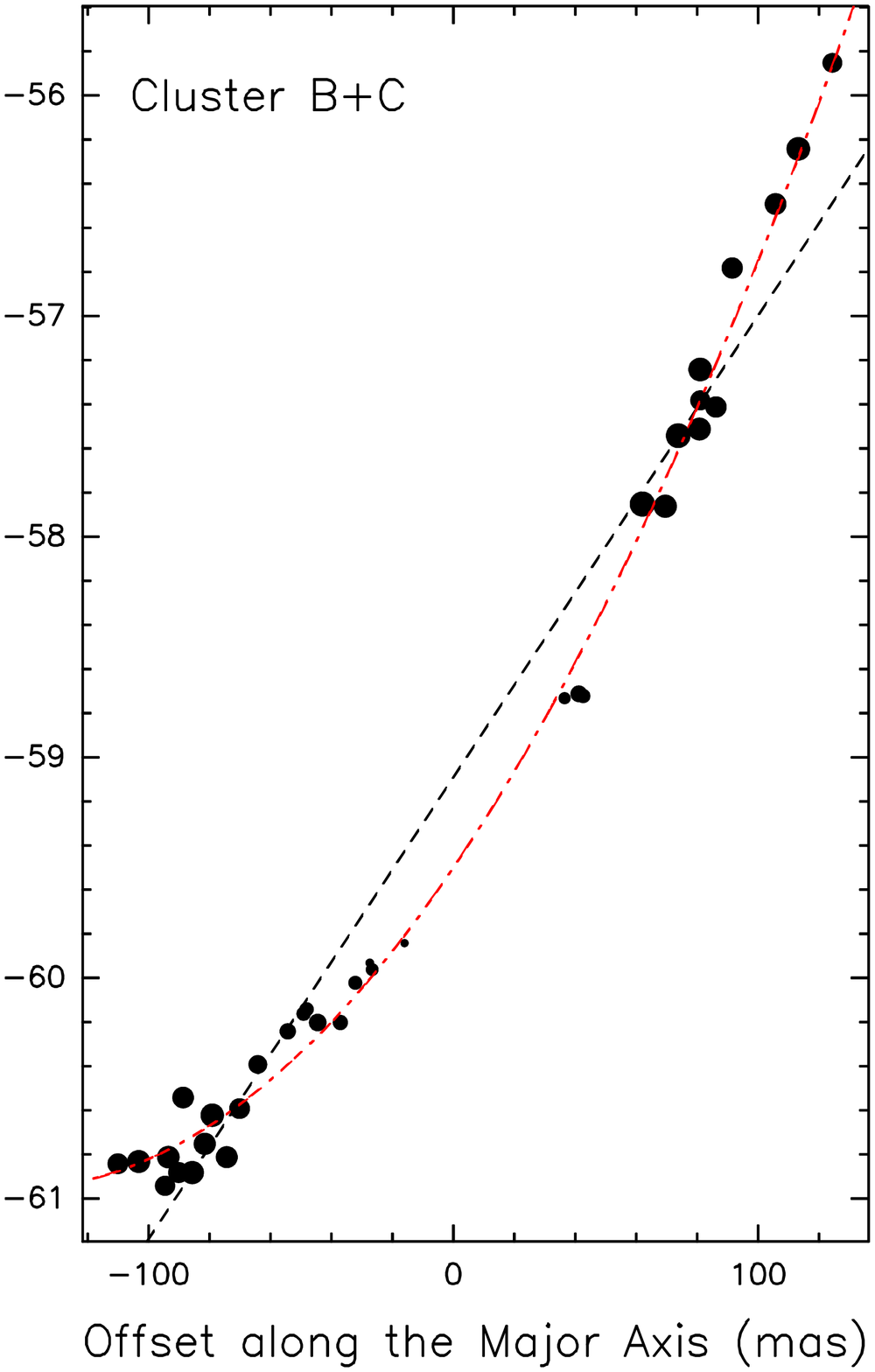}
\caption{Plot of the maser \Vlsr\ versus position projected 
along the major axis of the spatial distribution for the maser cluster(s)~"A" {\it (left~panel)}
and "B"+"C" {\it (right panel)}. 
The positional offsets are measured with respect to the YSO positions shown in Fig.~\ref{vlsr_pos_z}. 
Maser (relative) positions and \Vlsr\ are known with an accuracy better than 1~mas (see Table~\ref{tab_6.7})
and 0.1~\kms\ (see Table~\ref{evn_par}), respectively. 
These plots are produced considering only the features closely aligned
along the cluster major axis.
Dot size is proportional to the logarithm of the maser
intensity. The linear and quadratic fit to the plotted distribution
is indicated with a {\it black dashed} and {\it red dot-dashed line},
respectively.}
\label{vlsr_off}
\end{figure*}

Figure~\ref{vlsr_pos_z} evidences an ordered  spatial and \Vlsr\ distribution
of maser features in each of clusters "A", "B" and "C". 
The maser features of cluster~"A" show a remarkable positional alignment 
as well as a regular change in \Vlsr\ with position. 
Figure~\ref{vlsr_pos_z} shows that the majority of the maser features of clusters "B" and "C" is also
distributed close to a line with a monotonic variation of the maser \Vlsr\ 
along the major axis of the distribution. 
A least-square fit to the positions of maser features in clusters~"A" and 
the combined clusters "B"+"C", gives PA of the major axis of
\ 107\degree \ and  \ 71\degree, respectively.
For each group of features, Fig.~\ref{vlsr_off}
plots the maser \Vlsr\ versus positions projected along the axis of the distribution.
We fitted  the position-velocity distribution of masers with both a linear and quadratic curve, 
demonstrating that the change of \Vlsr\  with position is better represented by  a quadratic curve than a line.

In particular, indicating with \ $s$ \ the axis-projected position of a maser feature, 
we can write \Vlsr\ as a quadratic polynomial of the variable \ $s$:
\begin{equation}
V_{\rm LSR}(s) = \alpha s^2 + \beta s + \gamma 
\label{Q_cf} 
\end{equation}

\noindent The coefficients we derived from the least-square fit are the following:
\begin{eqnarray}
\text{Cluster~"A":} \begin{cases} \; \alpha = 1.9 \, 10^{-4} \; \text{km~s$^{-1}$~mas$^{-2}$}  \\
     \; \beta = 0.025 \; \text{km~s$^{-1}$~mas$^{-1}$} \\ 
     \; \gamma = -56.39 \; \text{km~s$^{-1}$} \label{Q_cf_A}  \end{cases} 
     \\
\text{Clusters~"B"+"C":} \begin{cases} \; \alpha = 7.1 \, 10^{-5} \; \text{km~s$^{-1}$~mas$^{-2}$}  \\
    \; \beta = 0.020 \; \text{km~s$^{-1}$~mas$^{-1}$} \\
    \; \gamma = -59.50 \; \text{km~s$^{-1}$}  \label{Q_cf_BC} \end{cases}  
\end{eqnarray}

\noindent It is worth noting that while the fitted values of \ $\alpha$ \ do not depend on the choice of the \ $s = 0$ \ 
position along the major axes of the distributions, \ $\beta$ \ and \ $\gamma$ \ vary, respectively, linearly and quadratically with the offset in the origin of the coordinate \ $s$.

For a physical meaning of these parameters, see Section~\ref{mas_kin_dis}.

\subsection{\nh3}
\label{nh3_vlsr_regu}

\begin{figure}
\includegraphics[angle=0.0,width=9cm]{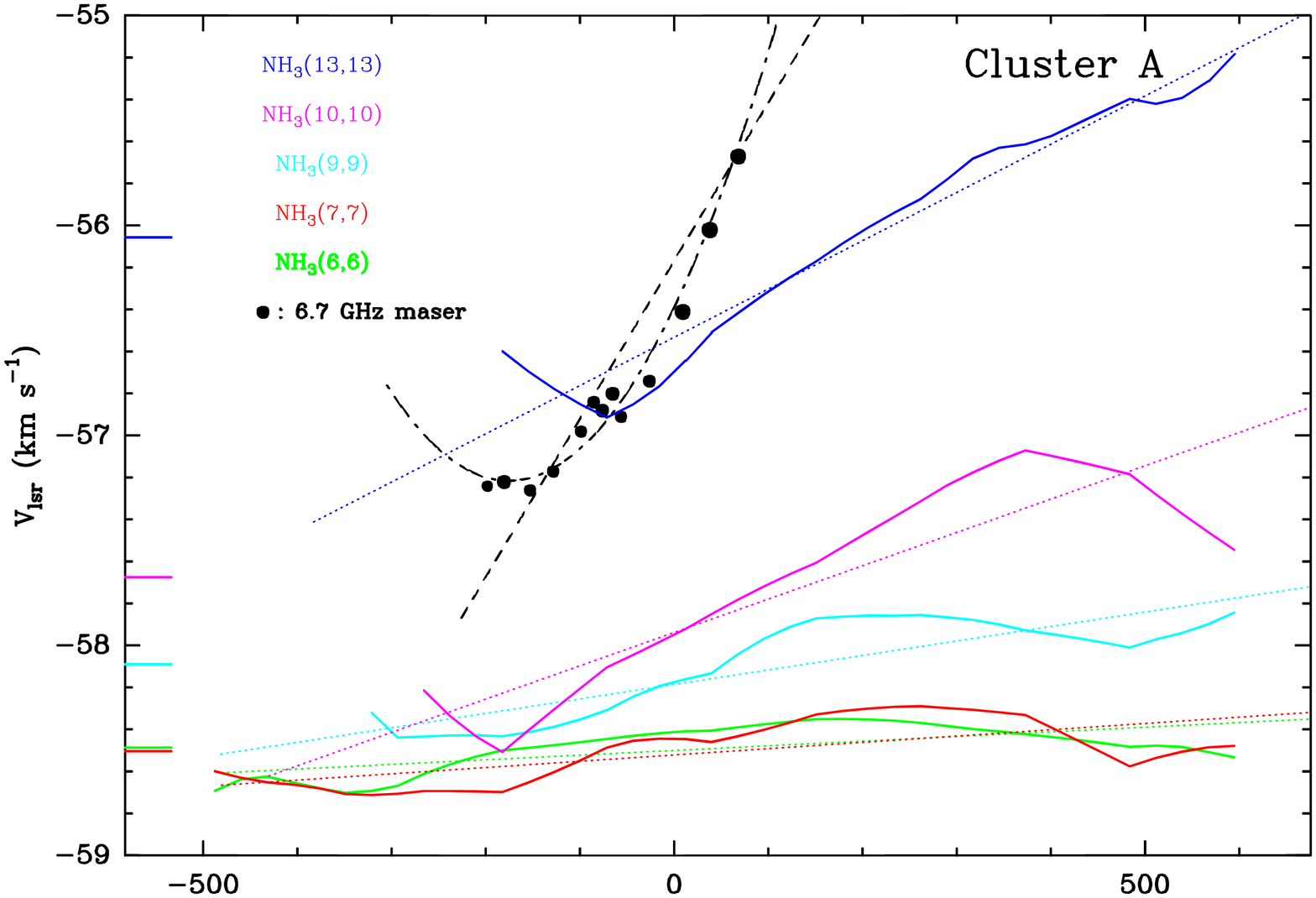} \\

\includegraphics[angle=0.0,width=9cm]{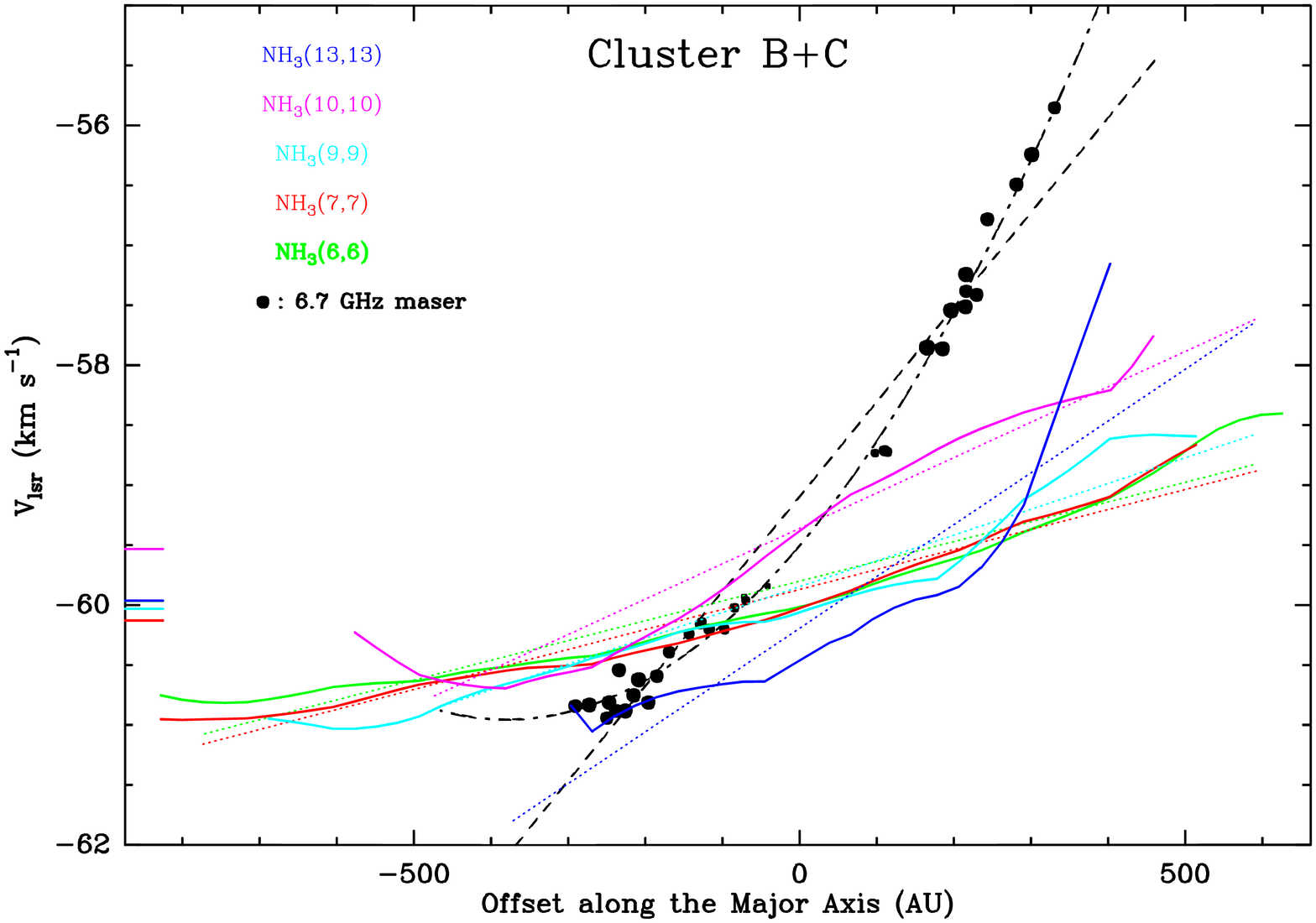}
\caption{{\it Colored curves} show strips of the first-moment map of 
different \nh3\ transitions, taken along the major axis of the maser clusters~"A" ({\it upper panel})
and "B"+"C"({\it lower panel}) .
{\it Colored bars} along the "Y" axis mark the mean \Vlsr\ of the \nh3\ lines. 
The association color--\nh3\ transition is indicated in the upper left corner of the panel.
{\it Black dots} report the 6.7~GHz masers of the clusters, selecting only the features
closely aligned along the cluster major axis.
Dot size is proportional to the logarithm of the maser
intensity. The linear and quadratic fits to the maser distribution
are indicated with a {\it black dashed} and {\it black dot-dashed line},
respectively.
{\it Colored dotted lines} show the linear fits to the first-moment strips of the \nh3\ transitions.   
}
\label{grad_B+C}
\end{figure}

The velocity field of the hot molecular gas traced by the \nh3\ inversion lines shows  
a \Vlsr\ gradient at PA (30\degree -- 40\degree), intermediate between the PA  of the major axes of the maser clusters~"B"+"C" (71\degree) and "A" (-17\degree).
While towards the clusters~"B"+"C", the absorption occurs at approximately the same \Vlsr\ in all \nh3\ lines ($-$59 to $-$61~\kms), a large change in the absorption velocity ($-$58 to $-$52~\kms),  is observed to the E--NE of the cluster~"A", going from the (6,6) to  the (13,13) line.

To better compare the  \nh3\ and 6.7~GHz maser \Vlsr\ gradients, we extracted a strip from the \nh3\ first-moment maps
along the major axis of the clusters~"A" and "B"+"C", and 
we plotted in Fig.~\ref{grad_B+C} the \nh3\ average \Vlsr\ vs 
the axis-projected positional offset. Since the major axis of both maser clusters is oriented close to \ E--W, 
the strip value is calculated taking the average velocity of three pixels (0\farcs04 in size), 
the central one along the major axis and the other two to the North and the South of the first pixel, respectively.

Figure~\ref{grad_B+C} reveals two elements. First, there is a clear trend for a steepening of the
slopes of the \Vlsr\ profiles going from the (6,6) to the (13,13) inversion lines of \nh3, for both maser clusters. 
In fact, the \Vlsr\ gradient steadily increases with the excitation of the \nh3\ inversion transition,
from \ $6 \times10^{-4}$~\kmo\ (for the (6,6) line) to \ $ 6\times10^{-3}$~\kmo\ (for the (13,13) line) in the cluster~"A",
and from \  $4 \times10^{-3}$~\kmo\ (for the (6,6) line) to \ 10$^{-2}$~\kmo\ (for the (13,13) line) in the clusters~"B"+"C".
This shows that towards both maser clusters, a \Vlsr\ gradient is detected also in the \nh3\ inversion lines. 
The second element is that the gradients traced in \nh3\ are significantly shallower than those measured with
the 6.7~GHz masers ($\approx2\times10^{-2}$~\kmo), particularly for cluster~"A".  
In the rest of this section, we show that this is an effect of the lower angular resolution of the \nh3\ maps with respect to the VLBI measurements of \meth.  
In particular, the angular resolution of the \nh3\ JVLA observations  ($\approx$0\farcs2) is comparable or 
larger than both the separation and the size of the maser clusters "A", "B", and "C". 
Therefore, the \nh3\ first-moment maps, over regions of weaker signal, could be heavily contaminated by nearby (within the synthesized beam) strong absorption at different \Vlsr. 

\begin{figure*}
\includegraphics[angle=0.0,width=\textwidth]{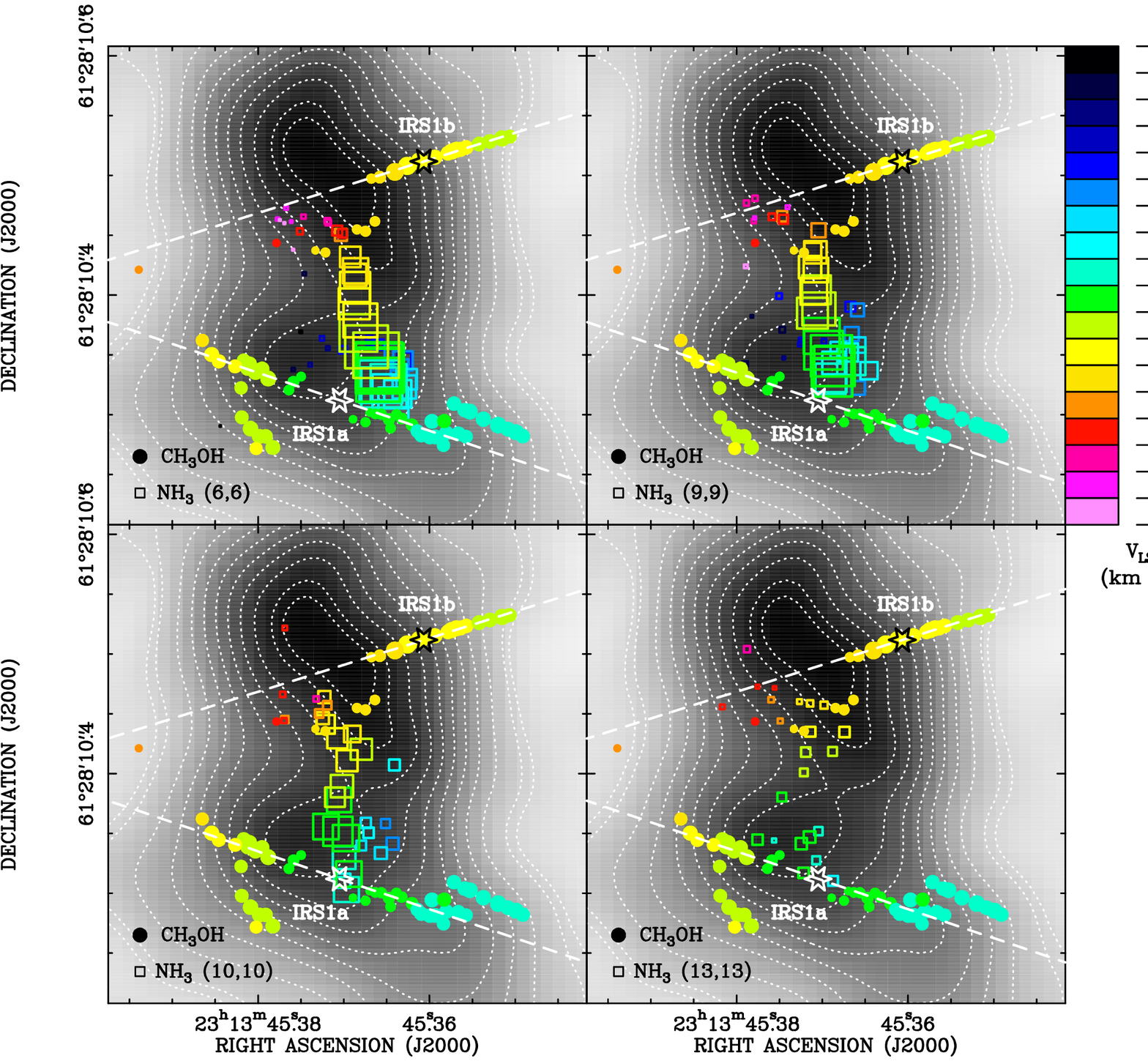}
\caption{ Emission centroids of \nh3\ fitted as a function of velocity ({\it open squares}) and \meth\ masers ({\it filled circles})  overlaid on the the 1.3~cm continuum map  ({\it black image and white contours}). 
 {\em Color} denotes  \Vlsr\ (color scale on the right-hand side). The sizes of squares  and circles  scale linearly and logarithmically with the flux density of \nh3\ and \meth\ maser emission, respectively. 
 The relative alignment between \nh3\ and \meth\ is s accurate to $\sim$30~mas. 
 Note that \nh3\ emission distributes between IRS1a and IRS1b, shows a velocity gradients roughly N--S, and is strongest towards IRS1a.}
\label{nh3_pvI_mas}
\end{figure*}

To investigate this problem, we determined the \Vlsr\ distribution of the \nh3\ inversion lines in an
alternative way, by Gaussian-fitting the position of the compact absorption feature in individual spectral channels 
with good SNR ($\ge$ 5$\sigma$), and producing plots of channel peak positions, collected in Figure~\ref{nh3_pvI_mas}, where also the channel \Vlsr\ and the intensity are reported.  
Figure~\ref{nh3_pvI_mas} illustrates three points: 
1)~the absorption of all the \nh3\ lines is much stronger towards the 
clusters~"B"+"C" than towards the cluster~"A"; \ 
2)~the absorption occurs at well separated velocities,
 at \ \Vlsr\ $\la$ $-$58~\kms\ towards the clusters~"B"+"C" and  at \ \Vlsr\ $\ga$ $-$55~\kms\
towards the cluster~"A"; \ 
3)~for any of the \nh3\ lines, no channel is found with absorption peaking nearby the
linear maser distribution of cluster~"A". 
Looking at Fig.~\ref{nh3_mom1}, we stress now two effects: 
1)~the reddest pixels of the \nh3\ first-moment maps are found to the NE of the cluster~"A", i.e.,
at the largest distance from the clusters~"B"+"C"; \ 
2)~in correspondence of the linear maser distribution of cluster~"A", the intensity-weighted velocities of \nh3\ are biased to more negative values than the maser \Vlsr\ (see also Fig.~\ref{grad_B+C}, upper panel). 
We ascribe the differences between the velocity field shown by the \nh3\ first-moment maps in  Fig.~\ref{nh3_mom1} and the  channel peak maps in Fig.~\ref{nh3_pvI_mas} to the contamination of the stronger absorption at more negative \Vlsr\ of the clusters~"B"+"C" over the region of the cluster~"A" (the two clusters are unresolved in our \nh3\ maps). 

The most remarkable difference between the plots of Fig.~\ref{nh3_pvI_mas} and the first-moment maps,  is that the (weak) reddest \nh3\ absorption appears to trace the extension of the linear maser distribution of the cluster~"A" to SE, rather than emerging from a region to the NE of the maser line, as shown in the first moment maps.    

We will discuss more in detail the relation between the \meth\ maser and \nh3\ line \Vlsr\ distributions in Sect.~\ref{phy_sen}.

\section{Edge-on Disk Model}
\label{mas_kin}

\subsection{Qualitative Assessment}
\label{mas_kin_qua}

In this Section, we present a kinematical model to interpret our observational findings, 
in particular the regular \Vlsr\ patterns described in Sect.~\ref{vlsr_regu}.
Here, we focus on the clusters "A" and "B"+"C", which share several geometrical and kinematical properties: \\
1) linear or elongated spatial distribution;  \\
2) regular variation of \Vlsr\ with position along the major axis of the distribution;\\
3) proper motions approximately parallel to the elongation axis; \\
4) average amplitude of proper motions ($\approx$5~\kms) similar to the variation in \Vlsr\ (4--6~\kms) across the maser cluster. \\
We argue that these four (independent) pieces of evidence are strongly suggestive of edge-on rotation traced by both clusters.

For cluster "A", \citet{Pes04} previously proposed a model of edge-on (Keplerian) disk to explain the regular velocity structure identified with the 6.7 and 12.2~GHz masers by \citet{Min98}. 
Recently, \citet{Pes12} determined the internal proper motions for four intense, 12~GHz CH$_3$OH masers in cluster "A", 
finding that they are aligned with the cluster orientation and have amplitudes in the
range \ 1 -- 9~\kms, in good agreement with our results for six 6.7~GHz masers (see Table~\ref{tab_6.7}). 
The two findings that the maser \Vlsr\ increases from NW to SE and all the features move
concomitantly to SE (see Fig.~\ref{prmot}), constrain all the features of cluster~"A" 
to reside on the near-side of the disk.
 
We report here for the first time evidence of rotation for clusters~"B"+"C". 
The distribution of maser positions and proper motions in clusters~"B"+"C" is less
regular than for the cluster~"A". Several features of the clusters~"B"+"C" have 
positions not closely aligned with the major axis of the cluster (see Fig.~\ref{vlsr_pos_z})
and a few proper motions are oriented at a large angle from the cluster axis. 
Both elements suggest a small deviation from edge-on rotation. 
Indicating with \ $i_d$ \ the angle between the disk plane and the l.o.s.,  
the components of velocities and accelerations along the l.o.s. would decrease by a factor \ $\cos(i_d)$ \ 
while the components on the plane of the sky transversal to the disk major axis 
would be proportional  to the factor \ $\sin(i_d)$. 
Looking at Tables~\ref{tab_6.7}~and~\ref{tab_acc}, typical relative errors
for the proper motion components and l.o.s. accelerations are of 10--20\%, 
therefore a deviation from the edge-on geometry by less than \ $i_d \approx 20$\degr \ cannot be revealed with our data. 
In the following, when we use the term "edge-on" referred to the maser clusters~"B"+"C", 
we mean a deviation by less than \ $\approx$20\degr\, from an exactly edge-on geometry.
Another difference with respect to cluster~"A" is that in the clusters~"B"+"C"
nearby maser features appear to be moving in opposite directions (NE vs SW; see Fig.~\ref{prmot}).
We argue however that, considering that the maser \Vlsr\ increases from SW to NE, 
the observed pattern of proper motions in clusters~"B"+"C" 
could still be consistent with rotation (seen about edge-on)
if the masers moving to NE and SW were, on the near- and far-side of the rotating structure, respectively. 

Figure~\ref{acc_pos} and Table~\ref{tab_acc} show that the l.o.s. acceleration 
for individual masers in cluster~"A" is always positive. If we interpret these
measurements in terms of centripetal acceleration, a positive value of acceleration is indeed
expected if the maser emerges from the near-side of the disk, as indicated by the orientation
of the measured proper motions. 
For the maser clusters~"B"+"C", the situation is more complex, 
because their l.o.s. acceleration varies from \ $-$0.019~\kmsy\ to 0.016~\kmsy. 
For centripetal acceleration, we expect the masers rotating 
on the near- and far-side of the disk to have positive and negative values of l.o.s. acceleration, respectively.
The comparison between Table~\ref{tab_6.7}~and~\ref{tab_acc} shows that this is indeed the case.  
Accurate (SNR > 3$\sigma$) measurements of l.o.s. acceleration
are derived for the (intense) features with label numbers from \ \#1 to \#5 \ in cluster~"B" 
and \ \#1, \#3 and \#5 \ in cluster~"C" (see Table~\ref{tab_acc}). 
Out of these eight features,  
proper motions are measured for five (see Table~\ref{tab_6.7}). 
Consistently with our hypothesis, features moving either to NE or SW, 
i.e., either on the near- or far-side of the disk, have either positive or negative values of l.o.s. acceleration.
Therefore, for the small subset of features for which both proper motions and l.o.s. accelerations are well determined,
our measurements appear to be qualitatively consistent with a model of rotation.

In an alternative scenario, linear distributions of 6.7~GHz masers with regular l.o.s. velocity gradients could trace collimated outflows, 
as  proposed by \citet{DeB03}~and~\citet{DeB09} based on H$_2$ near-IR observations and  interferometric mapping of the SiO 2-1 line emission in \meth\ maser sources, respectively. 
 In that case, maser proper motions would still be preferentially parallel to the maser distribution axis, and the observed maser l.o.s. accelerations could characterize accelerating protostellar outflows. 
And in fact, this has been shown to be the case in some high-mass YSOs using VLBI measurements of \meth\,masers \citep[e.g.,][]{Mos11a,Mos13}. 
We argue however that the observed patterns of maser \Vlsr\, proper motions and l.o.s. accelerations in NGC7538 cannot be explained in terms of a collimated jet neither for the cluster~"A" nor for clusters "B"+"C". 
In cluster~A", 
the maser \Vlsr\ increases concertedly with the proper motions, and, if
masers are observed in foreground of the continuum emission, that indicates that 
the masers move and are accelerated  {\it towards} and not {\it away from} the continuum emission, 
that is the putative location of the exciting protostar. 
This evidence obviously contrasts with an interpretation in terms of a jet ejected from the protostar.
For clusters "B"+"C", a collimated flow cannot explain neither the opposite orientation of velocity vectors nor the accelerations with both positive and negative sign  in nearby maser features. 
In fact, a collimated jet requires that the gas particles in the same lobe would have velocities with the same orientation and, if the flow is accelerated, similar accelerations. 
Besides these compelling elements, we measure maser velocities of only 5~\kms\ (on average,), which are clearly too low with respect to typical velocities of protostellar jets close to their axis, that are measured to vary from tens to hundreds of kilometer per second, for example using water masers \citep[e.g., ][]{God05}.

In summary, our measurements of velocity and acceleration towards both clusters~"A" and "B"+"C" are qualitatively consistent with edge-on rotation, and inconsistent with expansion in collimated jets.  
In the rest of this Section, we make a more quantitative analysis presenting a best-fit model to the data. 
 
\subsection{Maser Distribution on the Edge-on Disk} 
 \label{mas_kin_dis}
 
If all the 6.7~GHz masers in each cluster rotated at the same radial distance from the star, one would expect: 
\begin{equation}
\label{rot_k}
\frac{dV_{\rm LSR}}{ds} = \Omega 
\end{equation}
where \ $\Omega$ \ is the (constant) angular velocity, and $s$ is the sky-projected distance from the star 
along the maser elongation axis.

The simplest explanation for the quadratical (rather than linear) dependence of \Vlsr\ with $s$ (see Fig.~\ref{vlsr_off}) is that
$\Omega$ is {\em not} constant but varies (as first approximation)  linearly with $s$, i.e.:
\begin{equation}
\label{Ome}
\Omega = \alpha s + \beta
\end{equation}  
where $\alpha$ and $\beta$ are the second and first order coefficients of the quadratic curve fitted to the
change of \Vlsr\ with $s$ (see Sect.~\ref{vlsr_regu}).

Knowing the maser \Vlsr\  and \ $dV_{\rm LSR}/ds $, assuming edge-on rotation and 
using Equation~\ref{Ome}, we show in the Appendix
that it is possible to estimate both the sky-projected position of the centre of rotation (i.e., the star) and the
LSR systemic velocity. 
The  stellar positions derived this way for individual maser clusters are indicated with a {\it star} symbol 
in Figs.~\ref{acc_pos}~and~\ref{vlsr_pos_z}.  
We also used the stellar position to calculate the offset along the major-axis of the maser distributions in Fig.~\ref{vlsr_off}.
Therefore, the fitted quadratic coefficients \ $\alpha$, $\beta$ \ and \ $\gamma$, introduced in Sect.~\ref{vlsr_regu},
have the physical meaning of the derivative of \ $\Omega$ \ with \ $s$, the value of \ $\Omega$ \ at the star position,
and the LSR systemic velocity, respectively. 

We notice that the estimates for the stellar positions and LSR velocities from our analysis agree with other independent measurements. 
In particular, Fig.~\ref{vlsr_pos_z} illustrates that  for both clusters~"A" and "B"+"C" the derived stellar position falls close to the associated 1.3~cm continuum peak (the northern 
and southern core, respectively; see Fig.~\ref{vlsr_pos}), despite the 1.3~cm continuum beam is about ten times bigger than the 6.7~GHz maser beam. 
The continuum peak could effectively mark the star position if the continuum comes from a hypercompact \HII\ region. In addition, the model-inferred LSR systemic velocity for the maser clusters~"B"+"C", -59.5~\kms\ (see Sect.~\ref{6.7_vlsr_regu}), 
is in very good agreement with the LSR systemic velocity of the dense clump probed by the lower excitation \nh3\ lines, -59.4~\kms\,  (see Sect.~\ref{res_nh3}). 
If  the star exciting the maser clusters~"B"+"C" is the most massive 
in the region (see Sect.~\ref{mas_kin_fit}) and the lower excitation \nh3\ lines probe the molecular core from which the young star is forming, then we expect to observe similar velocities for both the star and the natal core.

For an edge-on disk in centrifugal equilibrium, one can write:
\begin{align}
\label{ed_cen_1}
V_{rot} &= \frac{\sqrt{GM(R)}}{R^{0.5}} \\
\label{ed_cen_2}
\Omega &= \frac{\sqrt{GM(R)}}{R^{1.5}} \\
\label{ed_cen_3}
A_c &= \frac{GM(R)}{R^2}
\end{align}
where \ $G$ \ is the gravitational constant, $M(R)$ \ is the mass enclosed within the radius \ $R$,
and \ $V_{rot}$ \ and \ $A_c$ \ are the rotational velocity and the centripetal acceleration, respectively.
  
We can express the dependence of the disk mass with radius in terms of a power-law:
\begin{equation}
 M(R) \propto R^q.
\label{MR}
\end{equation}

We do not expect the gas density to increase with radius (from the star), 
so we can exclude values of  \ $q$>3. The value \ $q$=0 \ describes the Keplerian case,
while the value \ $q$=2 \ corresponds to a flat disk with constant density.

Employing Equation~\ref{MR}, we can rewrite Equations~\ref{ed_cen_1}~to~\ref{ed_cen_3} as:
\begin{align}
\label{MR_1}
V_{rot} &\propto R^{q/2-0.5} \\
\label{MR_2}
\Omega &\propto R^{q/2-1.5} \\
\label{MR_3}
A_c &\propto  R^{q-2}
\end{align}

Finally, indicating with \ $R_0$ \ the maser radius along the l.o.s. to the star (i.e., at \ $s = 0$),
and combining Equations~\ref{Ome}~and~\ref{MR_2}, one can express \ $R$ \ in terms of \ $s$:
\begin{align}
\label{Ome_R}
\Omega &= \beta \left(\frac{R}{R_0}\right)^{q/2-1.5}  \\
\label{R_s}
 R &= \frac{R_0}{(1+\frac{\alpha}{\beta}s)^{2/(3-q)}}
\end{align}

\begin{figure}
% \sidecaption
\includegraphics[angle=0.0,width=10cm]{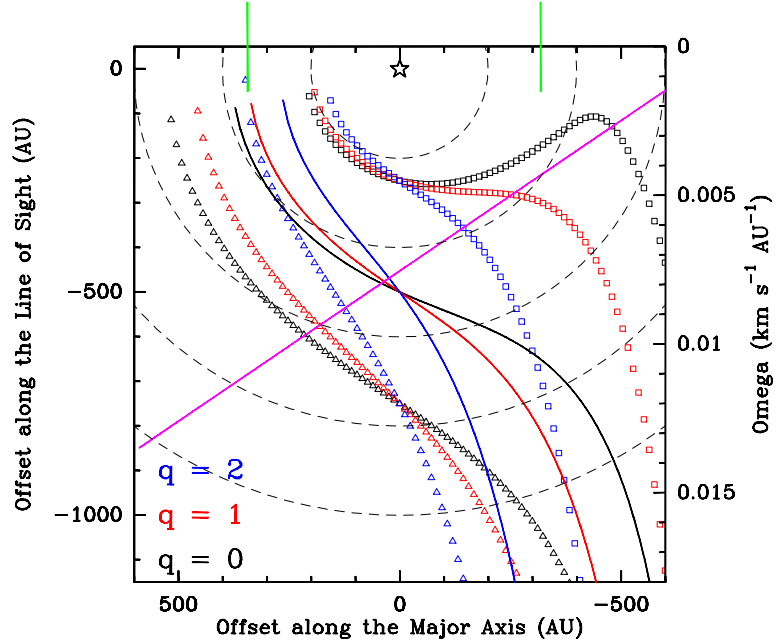}
\caption{The colored curves are the loci of the 6.7~GHz features, as derived from the edge-on disk model described in Sect.~\ref{mas_kin_dis}. We assume that the masers emerge from the near-side of the disk. 
Maser positions are described by the Equation~\ref{R_s} (see Sect.~\ref{mas_kin}),
using the coefficients \ $\alpha = 7.1 \, 10^{-5}$~km~s$^{-1}$~mas$^{-2}$ \ and 
\ $\beta = 0.020$~km~s$^{-1}$~mas$^{-1}$, derived in Sect.~\ref{6.7_vlsr_regu} fitting the change with position of
maser \Vlsr\ in the clusters~"B"+"C".
The colors \ {\it black}, {\it red} and {\it blue} identify the maser patterns 
corresponding to the values of the exponent \ $q$ \ equal to \ 0, 1, and 2, respectively.
{\it Light squares}, {\it heavy solid line} and {\it light triangles} are used to plot the curves corresponding
to the values of the maser radius at the star position, $R_0$ = 250, 500 and 750~AU, respectively. 
The {\it star} marks the location of the star at the disk centre.
The {\it magenta solid line} indicates the linear change of the maser angular velocity \ $\Omega$ \ 
with the position projected along the cluster major axis.
The two {\it green bars} mark the range of major-axis projected offsets over which
the 6.7~GHz masers of the clusters~"B"+"C" distribute.
{\it Dashed arcs} indicate circular orbits at steps of 200~AU in radial distance.  
}
\label{mas_pat}
\end{figure}

\noindent Equation~\ref{R_s} sets the maser positions onto the near or far-side of the rotating, edge-on disk.
A comparison between the orientation of the \Vlsr\ gradient and the proper motions suggests that all (most of) the maser features of the cluster~"A" ("B"+"C") reside on the near-side of the disk (see Sect.~\ref{mas_kin_qua}).
Figure~\ref{mas_pat} illustrates the maser distribution patterns on the disk for a range of plausible values of \ $R_0$ \ and \ $q$. 

\subsection{Fit of the Model Parameters}
\label{mas_kin_fit}

In Sect.~\ref{mas_kin_dis} we have proposed an interpretation for 
the observed regular variation of the maser \Vlsr\ with sky-projected position
based on the assumption that the maser emission emerges from an edge-on disk in centrifugal equilibrium. 
We have shown that the regular pattern in \Vlsr\  can be reproduced if masers distribute on the edge-on disk
along specific curves described by the two parameters \ $R_0$ \ and \ $q$.
Now, we can use our measurements of maser l.o.s. accelerations and proper motions
to constrain the model of edge-on rotation and derive the best values of \ $R_0$ \ and \ $q$ \ 
for each of the two maser clusters~"A" and "B"+"C".

Denoting with 
\begin{align}
\label{z}
z &= \sqrt{R^2-s^2} \\
\label{M0}
M_0 &= \beta^2 R_0^3 / G \\
\label{A0}
A_0 &= G M_0 / R_0^2
\end{align}
the position along the l.o.s., the mass within the radius \ $R_0$, and
the acceleration at \ $R_0$ (with $G$ indicating the Gravitational Constant), 
respectively, 
the sky-projected velocity, $V_{s}$, and the l.o.s. acceleration, $A_z$, at the  
location identified by the spatial coordinates \ $R$ \ and \ $z$ \, is given by:
\begin{align}
\label{Vs}
V_s &= (\alpha s + \beta) R \; \frac{|z|}{R}  \\
\label{Az}
A_z &= A_0 \left(\frac{R}{R_0}\right)^{q-2.0} \; \frac{|z|}{R}
\end{align}

Before describing the procedure adopted to fit maser velocities and accelerations, we verify the consistency 
of our model of centrifugally supported edge-on rotation for the case of the maser cluster~"A". 
Equations~\ref{ed_cen_2}~and~\ref{ed_cen_3} imply  \ $ R = A_c/\Omega^{2}$, which at \ $R = R_0$, following
our definitions (see Equations~\ref{Ome_R},~\ref{A0}~and~\ref{Az}), can be rewritten as \  $ R_0 = A_0/\beta^{2}$.
$\beta = 0.025$~\kmo \ for the cluster~"A" (see Sect.~\ref{6.7_vlsr_regu}). 
Since, over the cluster~"A", the measured l.o.s. accelerations are all in the range \ 0.01$\pm$0.001~\kmsy\
(see Fig.~\ref{acc_clA}~and~Table~\ref{tab_acc}), we can confidently take \ $ A_0 = 0.01$~\kmsy.
Thus, we find \ $R_0\approx$570~AU (215~mas at the distance of \NGC1) and from the product of \ $R_0$ \ and \ $\beta$ \ 
 a rotational velocity of \ 5.4~\kms. That compares well with the average maser velocity 
projected along the major axis of the cluster"A", which is \ 5.1~\kms. Such a remarkable agreement
between the value of rotational velocity derived from the maser \Vlsr\ and l.o.s. accelerations,
in the framework of the model of edge-on rotation, and the direct measurement of maser proper motions, 
makes us  confident that the adopted model is able to correctly reproduce the maser kinematics.

The best values for  \ $R_0$ \ and \ $q$ \ are derived by minimizing the \ $\chi^2$ \ expression:
\begin{equation}
\label{Chiq}
\chi^2 = \frac{1}{N_{free}} \; \left( \sum_{i} \left(\frac{A^i_z-\Upsilon^i}{\Delta\Upsilon^i}\right)^2 + \sum_{j} \left(\frac{V^j_s-\Lambda^j}{\Delta\Lambda^j}\right)^2 \right)
\end{equation}
where \ $\Upsilon$ ($\Delta\Upsilon$) \ and \ $\Lambda$ ($\Delta\Lambda$) \ denote the amplitudes (and corresponding errors), of the {\em measured} l.o.s. accelerations
and velocity components along the major axis of the maser cluster, respectively, and the index \ $i$ \ and \ $j$ \ run over the
features of the cluster with measured acceleration and proper motion, respectively (see Tables~\ref{tab_6.7}~and~\ref{tab_acc}). $N_{free}$ is the degree of freedom of the model, defined
by the difference between the number of measurements and the number (2) of free parameters. 
For the cluster~"A", the \ $\chi^2$ \ is calculated using all the measured accelerations and proper motions
(i.e., $i = 5$ \ and \ $j = 6$). For the clusters~"B"+"C"', beside excluding the subset of features more detached 
from the major axis of the maser distribution (see Figs.~\ref{vlsr_off}~and~\ref{acc_pos}), we did not consider the features \ \#7 and \#17 of cluster~"C", whose
proper motions are directed at large angle from the cluster axis (see Fig.~\ref{prmot}), and feature \ \#1 of cluster~"B", whose l.o.s. acceleration deviates very much (by 15$\sigma$) from the average value of 
nearby features (with this selection, we have  \ $i = 5$ \ and \ $j = 8$).
The errors of the l.o.s. accelerations, $\Delta\Upsilon$,  are those reported in Table~\ref{tab_acc}.
For the proper motions, besides the formal errors derived from the least-square fit of position offsets with time (reported in Table~\ref{tab_6.7}), we also need to consider a systematic error on the choice of the reference system, the "centre of Motion"
(see Sect.~\ref{prop_mot}), which could not represent adequately the star/disk system. 
We have shown above that, at least for the masers of the cluster~"A", the sky-projected velocity inferred from the model (using the maser \Vlsr\ and l.o.s. accelerations) agrees with the average amplitude 
of measured proper motions within a few tenths of kilometer per seconds.
Therefore we can conservatively estimate  the systematic error on the proper motions to be \ $\la$1~\kms.
To perform the model-fit, the uncertainties on the velocity, $\Delta\Lambda$, are calculated by adding in quadrature to the measurement errors (read from Table~\ref{tab_6.7}) a systematic error plateau of \ 1~\kms.

\begin{figure}
\includegraphics[angle=0.0,width=9cm]{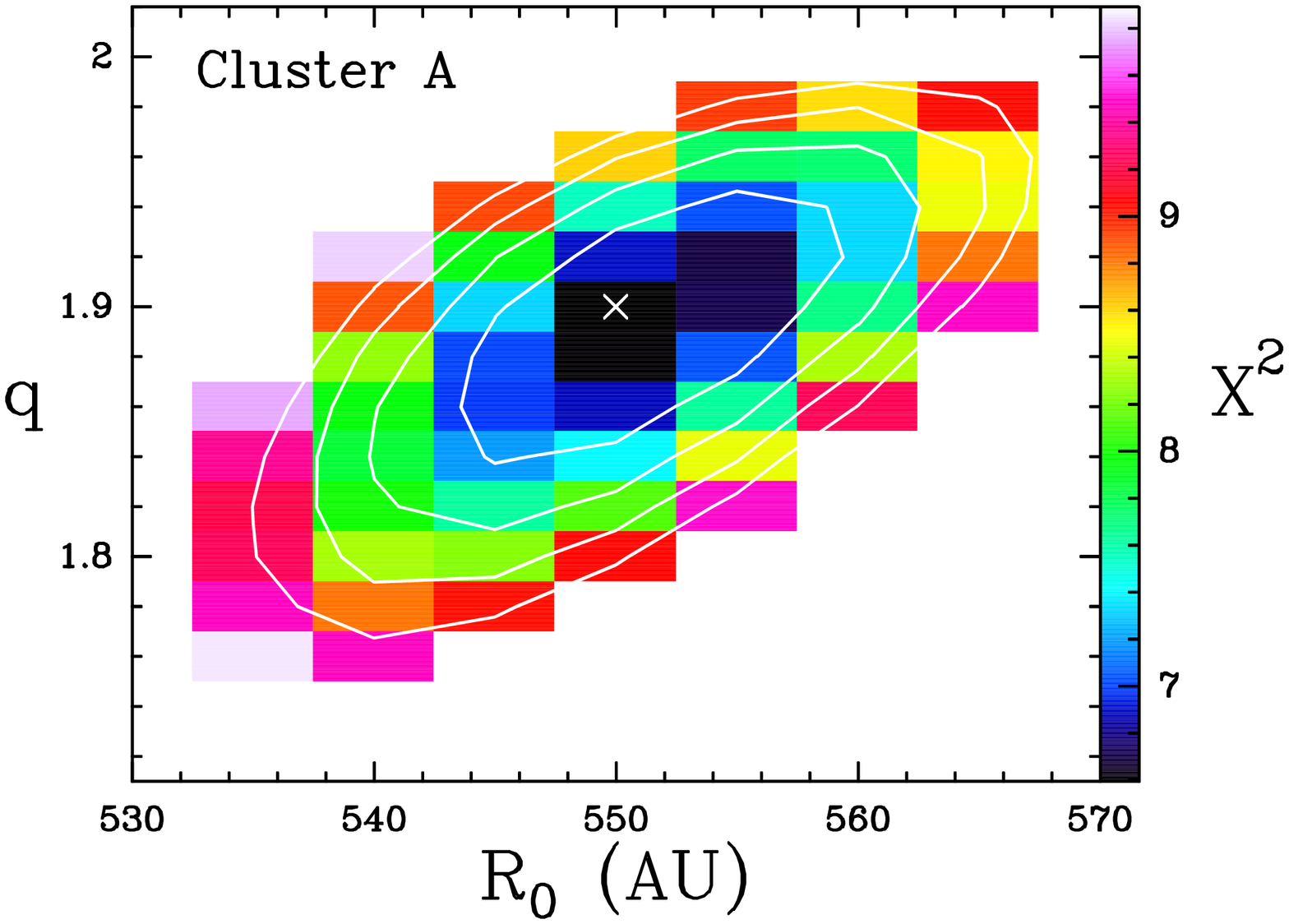} \\

\includegraphics[angle=0.0,width=9cm]{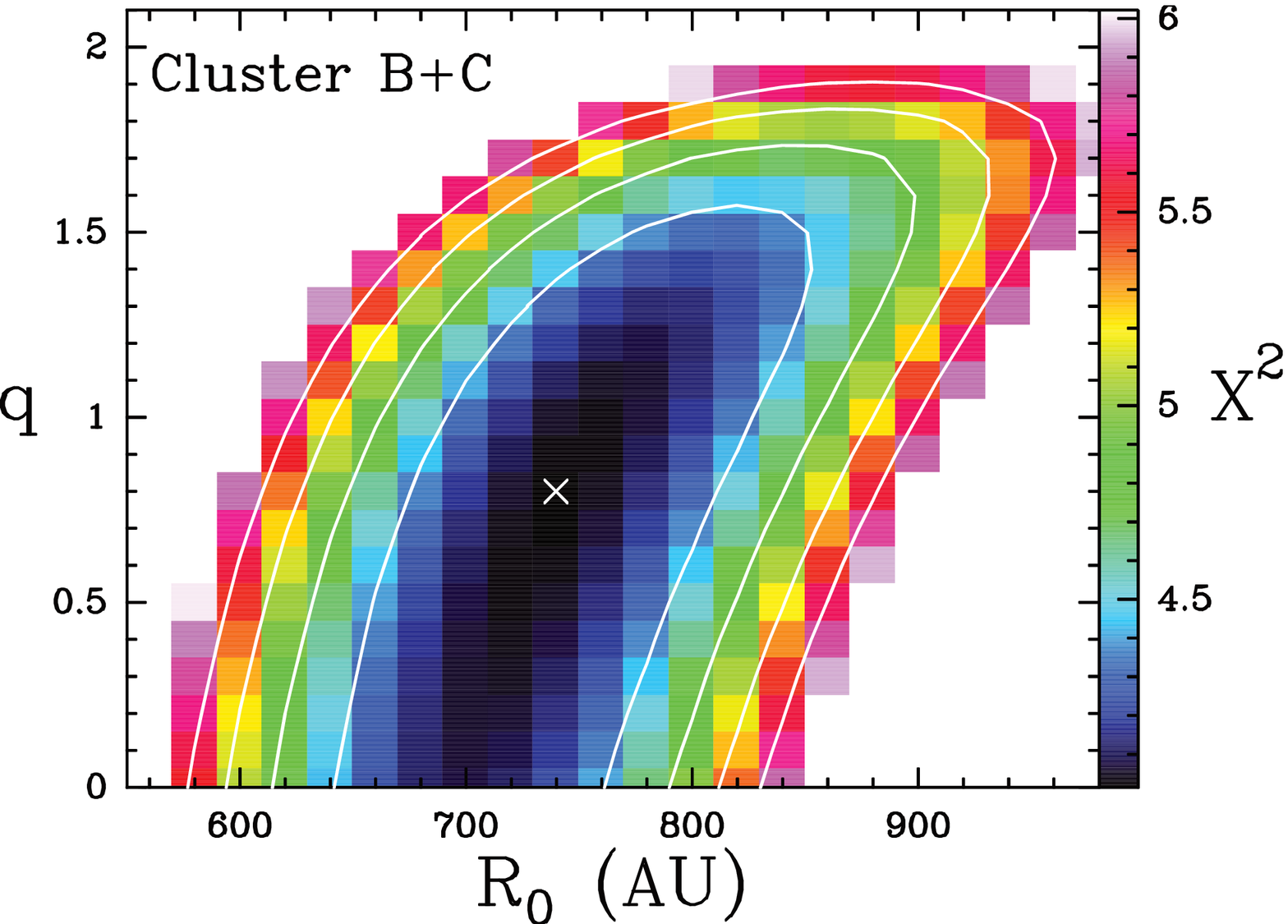}
\caption{Result of the model-fit for the cluster~"A" ({\it upper panel}) and clusters~"B"+"C" ({\it lower panel}).
 The {\it colored map} shows the distribution of \ $\chi^2$ (calculated using Equation~\ref{Chiq}) around the best-fit position,
 plotting values from the minimum of \ $\chi^2$ \ up to 50\% above the minimum.  
 The color-value conversion code is shown by the wedge on the right of the panel. 
 The {\it white cross} shows the position of the minimum of \ $\chi^2$. The {\it full white contours} show
 levels of 10\%, 20\%, 30\% and 40\% above the minimum value of \ $\chi^2$. 
 }
\label{Chiq_fit}
\end{figure}

We have searched for the minimum of \ $\chi^2$ \ over the parameter space 
\ $100 \le R_0 \le 1100 $ AU and \ $0 \le q \le 3$, in steps of 20~AU for $R_0$ and \ 0.1 \ for \ $q$.
For the cluster~"A" only, we have optimized the search using a narrower parameter window with steps
of 5~AU for $R_0$ and \ 0.02 \ for \ $q$.
The derived best-fit values are:
\begin{eqnarray}
\text{Cluster~"A":} \begin{cases} \; R_0 = 550 \pm 10 \; \text{AU}  \\
     \; q = 1.9 \pm 0.06  \\ 
     \; M_0 = 16 \pm 1 \; \text{M$_{\sun}$} \label{fit_clA}  \end{cases} 
     \\
\text{Clusters~"B"+"C":} \begin{cases} \; R_0 = 740 \pm 100 \; \text{AU}  \\
    \; q = 0.8 \pm 0.8 \\
    \; M_0 =  25 \pm 10 \; \text{M$_{\sun}$}  \label{fit_clBC} \end{cases}  
\end{eqnarray}
The quoted uncertainties are formal fit errors evaluated taking the displacement from the parameter best-value in correspondence of which 
the \ $\chi^2$ \ increases by $\approx$10\% above the minimum. Fig.~\ref{Chiq_fit} shows the distribution of \ $\chi^2$ \ around the position of minimum for the model-fit of both maser clusters.

\begin{figure*}
\includegraphics[angle=0.0,width=8cm]{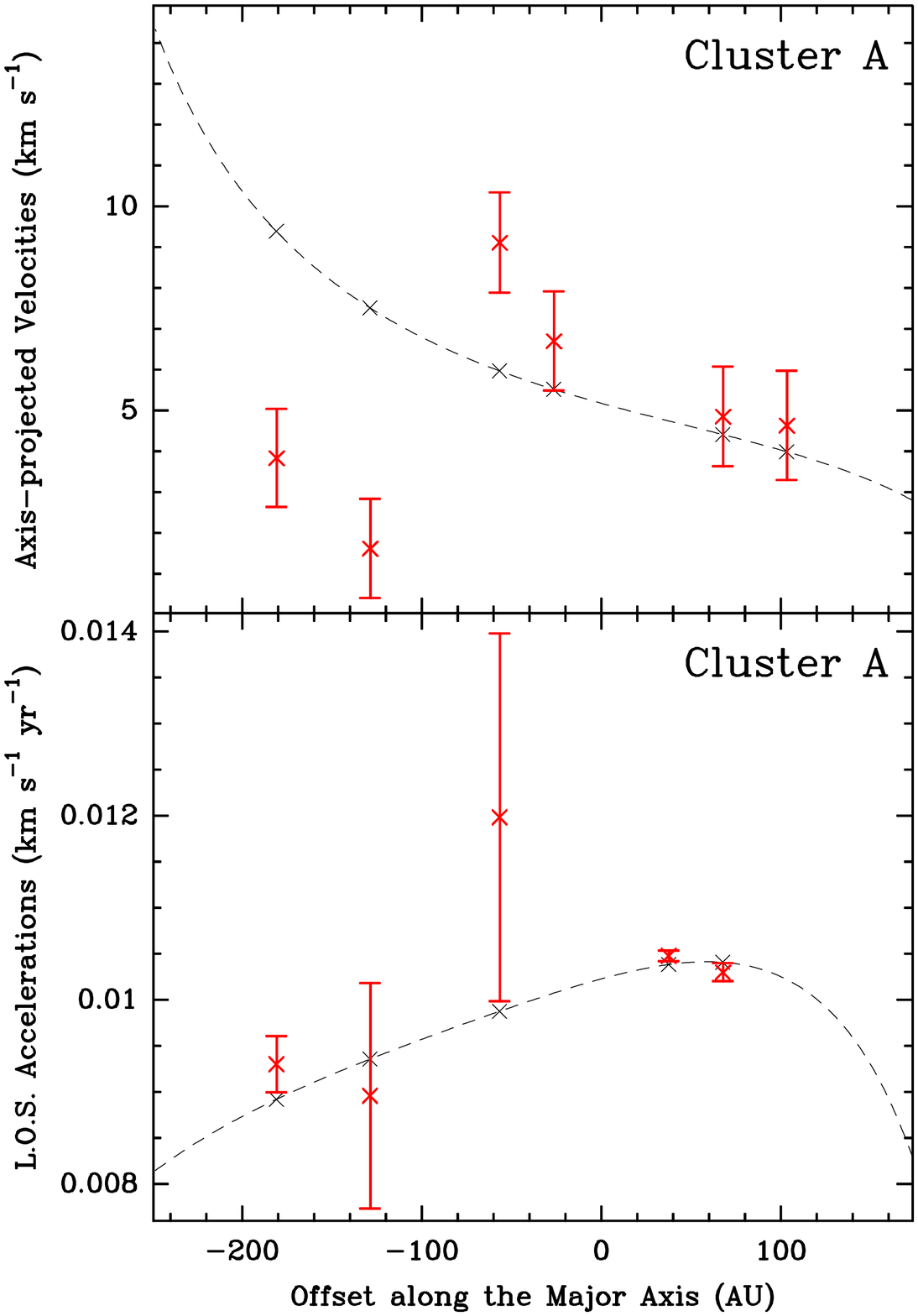} 
\hspace{1cm}
\includegraphics[angle=0.0,width=8cm]{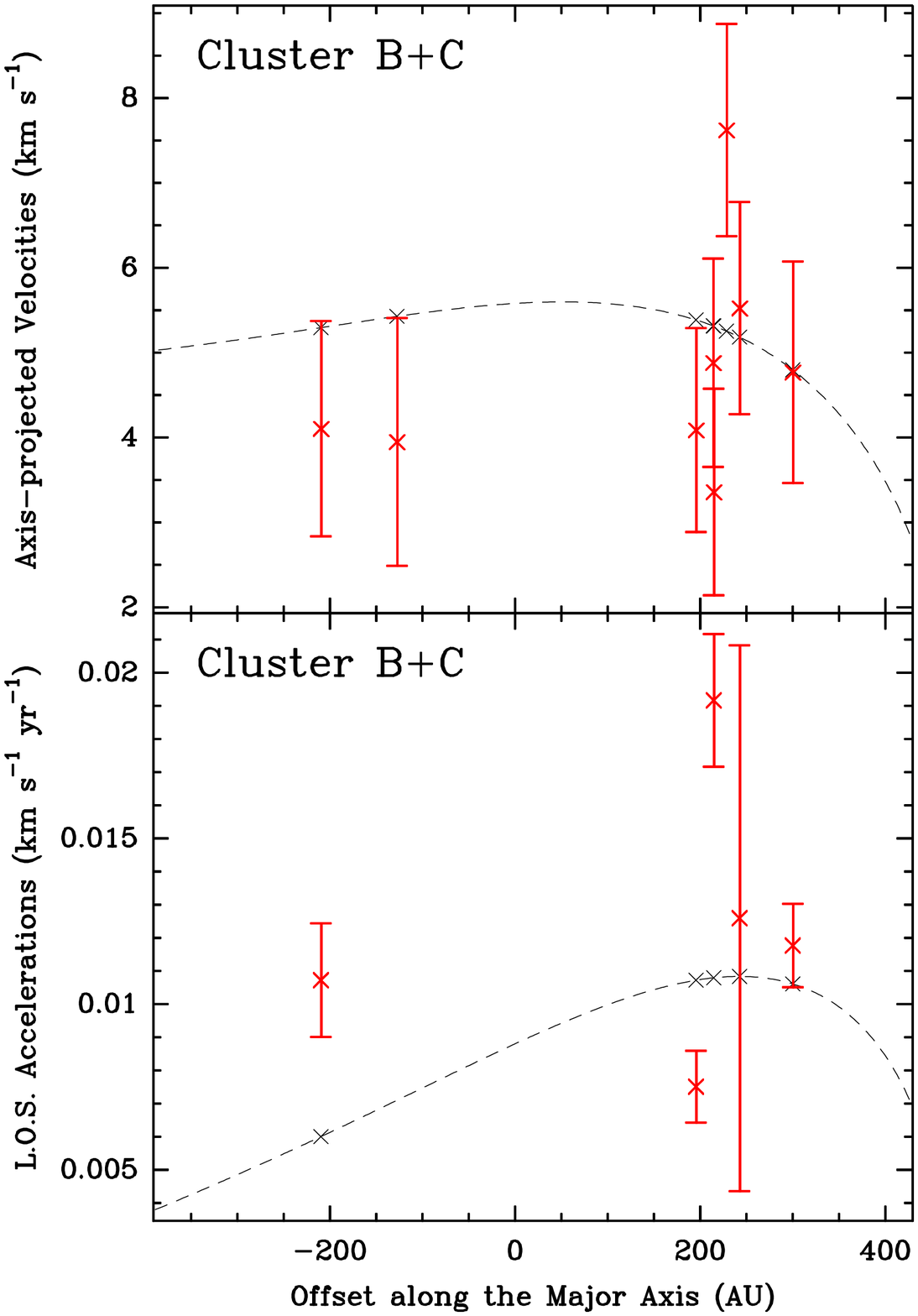}
\caption{Comparison of the measurements and best-fit values for the cluster~"A" ({\it left panels}) and clusters~"B"+"C" ({\it right panels}).
{\it Upper} and {\it lower} panels refer to the velocities (projected along the cluster major-axis) and the l.o.s. 
accelerations, respectively.
 {\it Red crosses} and {\it errorbars} give the measurements and corresponding errors, plotted vs the maser 
 position.
 The {\it black dashed line} shows the change of the model-predicted quantities along the maser pattern,
 with {\it black crosses} denoting the values in correspondence of the maser positions.  
 }
\label{ac_pm_fit}
\end{figure*}

\begin{figure*}
% \sidecaption
\includegraphics[angle=0.0,width=7.1cm]{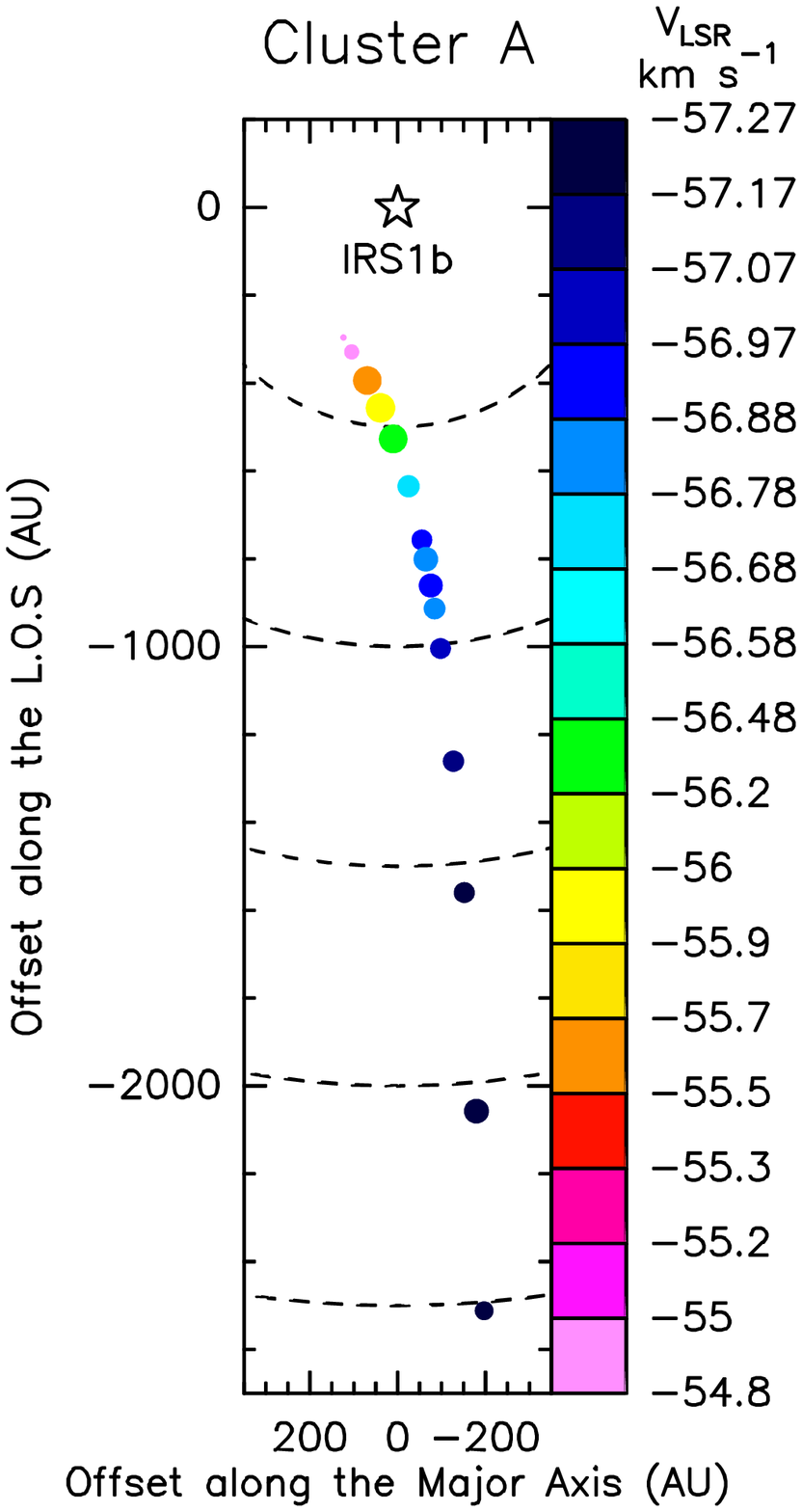}
\hspace{0.5cm}
\includegraphics[angle=0.0,width=11.4cm]{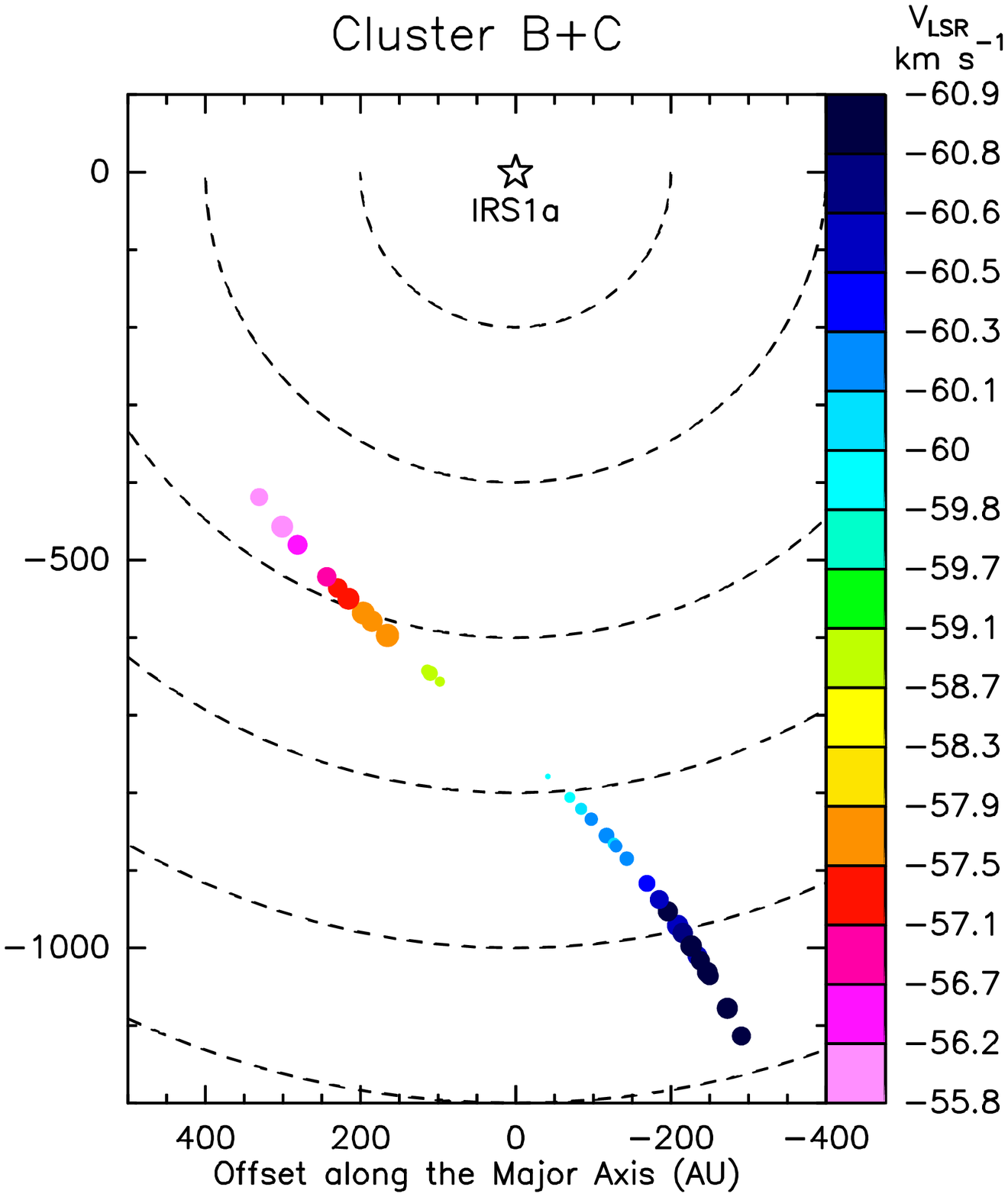}
\caption{Locations of maser features as derived from the edge-on disk model described in Sect.~\ref{mas_kin_dis}, assuming that the masers emerge from the near-side of the disk. 
{\it Colored dots} give the position for features of cluster(s)~"A" ({\it left panel}) and "B"+"C" ({\it right panel}).
For each cluster, only the maser features closely aligned along the cluster major axis are considered.
Symbols and colors have the same meaning as in Fig.~\ref{vlsr_pos}.
The labeled {\it star} marks the YSO position at the disk centre.
{\it Dashed arcs} indicate circular orbits at steps in radial distance of 500 and 200~AU, for the plot 
of cluster(s)~"A" and "B"+"C", respectively.  
}
\label{disk_pl}
\end{figure*}

Fig.~\ref{ac_pm_fit} presents the comparison between the measured and the best-fit accelerations and velocities.
Finally, Fig.~\ref{disk_pl} shows the modeled positions for maser features of cluster(s) "A" and "B"+"C"
on the near-side of the edge-on disk.

\section{Nature of the YSOs Exciting the 6.7~GHz Masers}

\label{phy_sen}

We propose that the three radio continuum peaks, associated with clusters 
"A", "B"+"C", and "E" of 6.7~GHz \meth\ masers, 
could mark the position of three distinct YSOs in the region, responsible for the maser excitation. 
Hereafter, we refer to these YSOs with the names \ IRS1b, IRS1a, and IRS1c. 
In the following, we describe their physical properties, based on our \meth\ and \nh3\ measurements 
as well as the edge-on disk model presented in Sect.~\ref{mas_kin}.

\subsection{IRS1a: the high-mass YSO Associated with the Maser Clusters~"B"+"C"}
\label{nat_b+c}

Figure~\ref{nh3_pvI_mas} reveals that the strongest absorption in all the \nh3\ lines  occurs at the centre of the maser clusters~"B"+"C", indicating that dense and warm molecular gas is associated with this cluster.
Towards the same position, 
\citet{Beu13} detected emission from warm dust and typical hot-core tracers using the PdBI. 
In particular, the 843~$\mu$m dust emission peaks just in correspondence of the maser cluster~"B" \citep[][see their Fig.~1]{Beu13}, where the map brightness temperature reaches 219~K. Towards the dust continuum peak, 
they derive a maximum value of  \ $n_{\rm H_2}$~density of \ 10$^{9}$~cm$^{-3}$. Observing \NGC1\ with the SMA, \citet{Zhu13}~and~\citet{Qiu11}  detected molecular lines from several dense gas tracers  (e.g., CH$_3$CN and CH$_3$OH) and derived gas temperatures of \ $\approx$250~K.
Therefore, we conclude that the 6.7~GHz masers of the clusters~"B"+"C"
are associated with an hot-core excited by a {\em massive} YSO, that we call IRS1a, which is surrounded by a rotating disk.

In the edge-on disk model proposed in Sect.~\ref{mas_kin_fit} for the clusters~"B"+"C", 
we only poorly constrain the value of the parameter \ $q = 0.8 \pm 0.8$, 
but this would be consistent with a Keplerian rotation if $q$ is actually close to zero.
Then, most of the \ $\approx$25~M$_{\sun}$ \ predicted within a radius of \ $\approx$740~AU \ would actually 
constitute the mass of the high-mass YSO at the disk centre. 
A YSO of \ $\approx$25~M$_{\sun}$ could account for most of the
far-IR luminosity of \NGC1, \ $\sim$10$^{5}$~L$_{\sun}$ \citep[see, e.g.,][]{Dav11}. 

The quasi-Keplerianity of the disk around IRS1a is also supported by the properties of the thermal emission from warm dust and hot gas probed by highly excited molecular lines. 
For example, the peak value of the \ 843~$\mu$m emission imaged by \citet{Beu13}, converts to a gas mass contribution over the \ $\sim$0\farcs2 beam ($\approx$500~AU) of only \ $\sim$1~M$_{\sun}$, much lower than the total mass of  25~M$_{\sun}$. 
More compelling, in multi-transition studies of hot-cores  it has been demonstrated that
molecular lines of higher excitation emerge from warmer gas at smaller radii from the YSO (see for ex. the study of CH$_3$CN transitions in IRAS20126+4104 by \citealt{Ces99}). 
The same result seems to hold also for the highly-excited \nh3\ lines in \NGC1 (Goddi et al. in prep.). 
In this context, the steepening of the slope (from \ $4 \times10^{-3}$~\kmo\ to \ 10$^{-2}$~\kmo) of the position--velocity plots with the excitation energy of the \nh3\ line 
(Fig.~\ref{grad_B+C}, lower panel) is consistent with (quasi-)Keplerian rotation, where the angular
velocity is expected to increase at smaller radii from the central mass. 
Unfortunately, the angular resolution of the \nh3\ observations does not permit a more quantitative analysis on
the relation between \Vlsr\ gradients and sizes of the emitting regions. 
As a matter of fact, the measured \Vlsr\ gradients are likely
only lower limits and no reliable estimate of the size of the absorption region for individual \nh3\ lines is possible.

Since IRS1a is the dominant YSO in the region and it is surrounded by a rotating disk, we could ask the question if this YSO is responsible for driving some of the outflows identified in the region. 
 It is evident that the direction perpendicular to the disk plane (PA $=-19^\circ$, the expected outflow direction) is significantly misaligned from the axes of the bipolar CO outflow (PA $-50^\circ$, eyeball value) and the inner core    ($\la$0\pas5)  of the radio continuum imaged with the VLA A-array (Fig.~\ref{vlsr_pos}), orientated mainly N-S (PA $\sim $ $0^\circ$). It is however consistent with the observed NIR fan-shaped region around IRS1 probed by H$_2$ emission (PA $\sim $ $-20^\circ$; see Fig. 3 by \citealt{Krau06}) and the bending of the outer core of the radio continuum (0\pas5-1\arcsec) towards the West at  PA $\sim $ $-25^\circ$ \citep[e.g.,][]{Cam84,Sand09}.

\subsection{IRS1b: the YSO Associated with the Maser Cluster~"A"}
\label{nat_a}

Comparing with the maser clusters~"B"+"C", the cluster~"A" presents a more regular distribution of maser 3-D velocities
and l.o.s. accelerations, which in turns allows a more accurate determination of the parameters of the edge-on disk model (see Sect.~\ref{mas_kin_fit}). The value of \ $q = 1.9$ \ derived for the cluster~"A" implies 
\ $ M(R) \propto R^{1.9} $, i.e., the mass within a given radius increases approximately with the square of the radius. 
Clearly that can be verified {\it only if the disk mass dominates the YSO mass}.
Since the model determines that a mass  \ $M_0 = 16$~M$_{\sun}$ \ is contained 
within a radius \ $R_0 = 550$~AU, the   mass of IRS1b can be at most of a few~M$_{\sun}$.  
Assuming an open disk geometry with the disk height \ $H(R) = 2 \, \tan(\alpha) \, R $ \; ($\alpha$ denoting the
semi-opening angle of the disk), the molecular (number) density at radius \ $R$ \ is:
\begin{equation}
\label{nh2_a}
 n_{\rm H_2} \approx 2.5 \, 10^{9} \; \frac{1}{\tan(\alpha)} \; \left[\frac{R_0}{R}\right]^{1.1} \; \; {\rm cm}^{-3}
\end{equation} 

The spatial distribution and \Vlsr\ of the weak, most redshifted \nh3\ absorption 
in Fig.~\ref{nh3_pvI_mas} is consistent with the edge-on disk model based on the 6.7~GHz \meth\ masers.
The peaks of the most redshifted \nh3\ absorption distribute close to the SE extension of the major axis of the cluster~"A" maser disk. 
Their projected separations from the estimated position of the YSO IRS1b \ fall in the range \ 0\farcs1 -- 0\farcs15, and their \Vlsr\ vary over the interval \ [$-$53, $-$48]~\kms. Using the quadratic polynomial fitted to the change of maser
\Vlsr\ with the projected separation, $s$, from the star 
(see Sect.~\ref{6.7_vlsr_regu} and Eq.~\ref{Q_cf_A}), one would expect a \Vlsr\ variation in the interval \ [$-$52, $-$48.4]~\kms\ for \ $s$ \ varying 
across \ 0\farcs1 -- 0\farcs15. The good match with the observed range of \Vlsr\ for the reddest \nh3\
absorption features suggests that \nh3\ and 6.7~GHz masers trace the same kinematical structure.
The relatively coarse, positional accuracy ($\ga$30~mas) of the weak, reddest \nh3\ peaks prevents 
a more accurate comparison with the edge-on disk model.  

Although we detect \nh3\ absorption  only from the most redshifted SE end of the disk, 
we cannot exclude that absorption may originate from the whole disk structure, 
including the blueshifted NW end. 
However, since the  angular resolution of the \nh3\ observations does not allow to resolve the disk around IRS1b as well as IRS1a from IRS1b, we expect the  \nh3\ absorption at more negative \Vlsr\ to be dominated by the hot-core excited by the high-mass YSO IRS1a, hiding any potential contribution from IRS1b.

Since the brightness of the radio continuum towards the cluster~"A" and "B"+"C"
is comparable, the relative weakness (by a factor of \ $\approx$10) of the \nh3\ absorption (see Fig.~\ref{nh3_pvI_mas}) 
suggests that towards the cluster~"A" 
either the \nh3\ column density is scarce or there are less favourable excitation conditions.
Equation~\ref{nh2_a} indicates that the gas density traced by the masers 
in the cluster~"A"  is comparable or even higher (if the disk is thin) than that in the clusters~"B"+"C".  
The weakness of the \nh3\ absorption lines can then be explained with a difference in temperature between IRS1a and IRS1b.  
 This hypothesis is also supported by the properties of the mm dust continuum.  
Examining the 843~$\mu$m map of dust emission of \citet[][see Fig.~1]{Beu13}, we see that at 
the position of the maser cluster~"A" (crossed by the \ 5$\sigma$ \ contour level) 
 the map intensity is about one third of the peak value, in the direction 
of the maser cluster~"B". If the gas density is comparable between the two clusters, 
the decrease of intensity from dust emission could reflect a lowering in temperature from cluster~"B" to cluster~"A". 
Therefore, we argue that the gas around IRS1b should be at lower temperature than towards IRS1a (i.e. T$<$250~K). 
Coupled with the results of the edge-on disk model, we also expect that the YSO IRS1b 
 is not so massive to significantly ionize (and heat up) the surrounding gas.

Since IRS1b is also a YSO surrounded by a rotating disk, 
we could ask a similar question as done for IRS1a if also this YSO is responsible for driving any of the outflows identified in the region.
The PdBI observations of the HCO$^+$(4--3) line (beam FWHM $\approx$ 0\farcs2) by \citet[][see their Fig.~2, right panel]{Beu13},  reveal a collimated bipolar outflow, whose axis 
is oriented NE--SW and intersects the dust emission a few tenths of arcsec to the NE of the 843~$\mu$m continuum peak (aligned
with the maser cluster~"B"). 
These two findings are consistent with an outflow driven by IRS1b and collimated by its rotating disk. 
We also argue that the both the velocity distribution and the collimation degree indicate an outflow in the plane of the sky.  
\citet{Beu13} noticed that the HCO$^+$ spectrum extracted towards the direction of the blue-shifted (red-shifted) lobe contained also weak red-shifted (blue-shifted) emission, 
and suggested that the outflow should be oriented close the l.o.s.. 
However, the detection of both blue- and red-shifted emission
from a single outflow lobe can occur also in the case that the outflow axis is close to the plane of sky, 
if the flow has a significant aperture angle 
(see for ex. the well-studied case of IRAS~20126+4104 imaged in the HCO$^+$ 1--0 line; see, e.g., \citealt{Ces99}). 
Considered also the high degree of collimation, we think it is more likely that the outflow is oriented close to the plane of the sky rather than along the l.o.s.. 

We conclude that IRS1a is surrounded by an edge-on disk and drives an outflow  on the plane of the sky.

\subsection{IRS1c: the YSO associated with the maser cluster~"E"} 
\label{nat_e}

Comparing with the other \meth\ maser clusters, cluster~"E" presents a more scattered, quasi-spherical distribution of masers (see Fig.~\ref{vlsr_pos}). The overall pattern of proper motions is also quite scattered (Fig.~\ref{prmot}, bottom panel), with a hint at an average motion of the whole cluster (with respect to the clusters~"A" and
"B"+"C") towards East. The internal motions of the masers of cluster~"E" are of a few kilometer per seconds
both along the l.o.s. and across the sky plane.
The feature l.o.s. accelerations are all zero within the measurement errors
(see Table~\ref{tab_acc}). Towards this cluster, \nh3\ lines of different excitation, from the (6,6) to the (9,9)
line (see Fig.~\ref{nh3_mom1}), show absorption at similar \Vlsr$\approx$ $-$60~\kms 
(close to the systemic velocity of \ $-$59.4~\kms), indicating that layers of molecular gas with different temperatures and densities (and likely distances from the ionized gas), have no significant relative motions (at least, along the l.o.s.). 
The \nh3\ \Vlsr\ is slightly more negative than that of the 6.7~GHz masers 
($-$58.5~\kms). 
We conclude that the circumstellar gas associated with IRS1c 
 is less structured and more quiescent than that in the other YSOs in the region.

 Again, the highly excited thermal lines help us estimating qualitatively the physical conditions of the molecular gas around IRS1c. 
Towards the cluster~"E", the \nh3\ inversion lines of highest excitation (the (10,10) and (13,13) lines) are 
not observed in absorption (see Fig.~\ref{nh3_mom1}). 
A possible explanation is that the gas
temperature is not high enough to excite the population levels of the highest excitation
\nh3\ transitions, suggesting that the temperature in IRS1c may be lower than that of IRS1b, 
where absorption, although weak, is detected in all the \nh3\ lines.
Accordingly, the 843~$\mu$m map of \citet[][see their Fig.~1]{Beu13} shows only a weak ($\le$3$\sigma$) level of 
dust emission in correspondence of the southern maser cluster~"E", 
which again points out to lower temperatures  (and densities)  than towards the cluster~"B", where the dust emission peaks. 
The presence of a (weak) radio continuum peak (see Fig.~\ref{vlsr_pos}) 
 suggests that the YSO IRS1c associated with the cluster~"E" may also be a massive YSO, 
if the free-free emission is coming from photoionized gas. 
If this is the case, IRS1c is likely less massive than the dominant object in the region, IRS1a.
If the 6.7~GHz masers of cluster~"E" traced molecular gas at the edge of a slow expanding, ultra-compact \HII\ region, that could account for both the sparse spatial
distribution and the slow and scattered proper motions of the masers. Assuming that the expansion velocity increases with
the radius, the \nh3\ absorption could trace less dense gas at larger radii than the 6.7~GHz masers,
which could explain why the \Vlsr\ of the \nh3\ lines is more negative (more blueshifted) than that of the masers.

The depicted scenario for the maser cluster~"E" is clearly very speculative, and more sensitive, 
high angular resolution data are needed to put it on a firm observational ground.

\section{Discussion} 
\label{discu}

\subsection{General Implications of the Model for \NGC1}
In previous section, we described the main properties of the individual YSOs and their surrounding disks, as derived from  our \meth\ and \nh3\ measurements as well as our edge-on disk model.
We discuss here general implications of this model. 

We start discussing the structure, geometry, and velocity field of the circumstellar gas associated with individual YSOs with respect to the natal molecular clump.
Looking at Fig.~\ref{nh3_pvI_mas}, three regions at different \Vlsr\ can be identified in the \nh3\ absorption:
\ 1)~the most redshifted (weak) absorption to the SE of the maser elongation of cluster~"A";  
\ 2)~the (strong) absorption near the systemic velocity ($-$59.4~\kms) towards the centre of the maser clusters~"B"+"C"; 
\ 3)~absorption in between the two maser clusters or YSOs IRS1a and IRS1b, where the gas \Vlsr\ steadily increases, going from the centre of the clusters~"B"+"C" towards the cluster~"A".  
One should distinguish here between gas motion {\it between} the two YSOs and {\it around} a given YSO. 
Owing to limited angular resolution, close to the cluster~"A"
the two contributions to the gas motion are not resolved and the direction of the \Vlsr\ gradient bends from close to N--S (as it is
at the middle point between the two clusters) to SW--NE. 
Blending of the two different motions accounts also for the
orientation of the \Vlsr\ gradient visible in the \nh3\ first-moment maps (see Figure~\ref{nh3_mom1}), which is
directed NE--SW at PA = 30\degree -- 40\degree. 
Taking into account the bias introduced by the blending of the two motions,
Fig.~\ref{nh3_pvI_mas} then indicates that the \Vlsr\ gradient in the gas  {\it between} the two YSOs IRS1a and IRS1b  
is directed close to N--S and extends from \ $\approx$ $-$59~\kms (at the centre of the clusters~"B"+"C")
to \ $\approx$ $-$56~\kms (close to the cluster~"A"). 
This finding suggests that the observed \Vlsr\ gradient between the two maser clusters
could reflect original gas motions inside the molecular core out of which the YSOs IRS1a and IRS1b have formed. 

Since we model the maser clusters~"A" and "B"+"C" as two disks close to edge-on, separated by less than 500~AU, we could ask which are the chances to observe two nearby edge-on disks oriented at different PA in the plane of the sky. 
A possible explanation could be maser excitation effects. In fact, an edge-on geometry implies large column densities of masing molecules which in turn enable to strongly amplify the maser emission. In principle, there could be many YSOs in the region but only those with  edge-on disks would be traced by the methanol masers. In practice, we think that this is not the case. 
In fact, the small separation between IRS1a and IRS1b (i.e., $<$500~AU) would imply a Jeans length for the collapsing and fragmenting core of at most a few hundreds of AU.
Even  assuming an average density as high as \ $n_{\rm H_2}$ = 10$^9$~cm$^{-3}$ (i.e., the peak value measured towards IRS1a by \citealt{Beu13}), the mass of individual fragments would correspond to only a few hundredths of solar mass. 
We conclude therefore that the presence of other undetected massive YSOs  in the region, besides the ones identified by the methanol maser clusters, is unrealistic. 
An alternative and more plausible explanation could be a physical mechanism producing an edge-on alignment. 
\citet{Sur11} measured the linear polarized emission of all \meth\,masers in the region, finding that the magnetic field 
is aligned close (within 20\degr) to the plane of the sky and preferentially oriented perpendicularly to the elongation axes of clusters "A", "B", and "C", that is perpendicular to the disks around IRS1a and IRS1b (see their Table~3). 
This finding agrees with predictions of star formation models which include  magnetic fields \citep[e.g., ][]{Ban07,Sei11}. In these models, the collapsing core flattens preferentially along a direction perpendicular to that of the magnetic field lines. These predictions are consistent with the general result of a polarimetric survey of  \meth\,masers towards high-mass star forming regions  \citep{Sur13} and suggest that the magnetic field may be dynamically important in the mass-accretion and mass-loss regulating the formation of high-mass YSOs.  
In this framework, an edge-on geometry for protostellar disks could be favoured 
if the magnetic field in the natal core (before collapse)  were orientated
near the plane of the sky. 
Although this qualitative argument explains the edge-on geometry of the two disks around IRS1a and IRS1b,   it does not explain their different PA in the sky.

One could now ask whether the disks around IRS1a and IRS1b are
stable or transient structures. The Toomre stability parameter for disks, $Q$, is given by the
expression \ $ Q = (2 \, \Omega \, \Delta V) / (\sqrt{8 \ln2} \, \pi \, G \, \Sigma) $, with \ $\Omega$ \ the angular velocity,
 $\Sigma$ \ the surface density, and \ $ \Delta V$ \ the FWHM linewidth. 
 For the maser cluster~"A", the disk model requires that at \ $R = R_0 = 550$~AU, $\Omega = \beta = 0.025$~\kmo, with the disk/envelope mass equal
to \ $M_0 = 16$~M$_{\sun}$ (see Equation~\ref{fit_clA}). Using these parameters, we derive
\ $Q_{clA} = 0.18 \, \left[\frac{\Delta V}{{\rm km \, s}^{-1}}\right] $. 
The condition for instability,\ $Q_{clA} \le 1$, is then satisfied if \ $\Delta V \le $5.6~\kms.
From our \nh3\ lines, we measure FWHM linewidths of 7--10 \kms, where however the two YSOs IRS1a and IRS1b, with a velocity separation of 3 \kms\ (see Sect.~\ref{mas_kin}), are blended together. 
We then expect \ $Q_{clA} \le 1 $, that is the massive disk/envelope around IRS1b is probably not gravitationally stable and could break up into fragments.  
For IRS1a, if a large fraction of the \ 25~M$_{\sun}$ \ inside a radius \ $R_0$ = 740~AU (see Equation~\ref{fit_clBC}) actually constitutes the YSO mass, then the disk surface density should be about one order of magnitude lower than for the disk/envelope
around IRS1b. Since the value of \ $\Omega = 0.02$~\kmo\ is comparable, the condition for instability for IRS1a would require 
\ $\Delta V \le $1~\kms. The measured values of FWHM linewidths are definitively larger than this threshold,
indicating that the disk surrounding the high-mass YSO IRS1a is actually gravitationally stable. 

In the rest of this section, we will discuss excitation effects and implications on the physical structure of the observed disks.
Excitation models predict that the 6.7~GHz \meth\ masers, radiatively pumped by IR radiation, 
are strongly inverted over an extended
range of gas density,  10$^4$~cm$^{-3}$ $\le n_{\rm H_2} \le$ 10$^{9}$~cm$^{-3}$, and kinetic temperature,
 25~K $\le T_K \le$ 250~K \citep{Cra05}. The IR radiation can be produced by warm dust nearby
 the methanol masing molecules, with an allowed range of dust temperature for efficient maser pumping \ 100~K $\le T_d \le$ 300~K.
 Following the discussion in Sect.~\ref{phy_sen}, the gas temperature towards the 6.7~GHz maser clusters in \NGC1\ should vary from the lowest value in IRS1c to the highest value  (250~K) in IRS1a. 
 Therefore, the gas temperature measured and/or expected towards each maser cluster is within the range predicted 
 by the model of \citet{Cra05} for strong maser action. 
The value of gas density \ $n_{\rm H_2} \sim$10$^{9}$~cm$^{-3}$ \ towards IRS1a derived
from the SMA observations by \citet{Beu13}, is close to the upper end of the maser excitation range, but still
 consistent with the models of \citet{Cra05}. 
 
 A more quantitative comparison between the physical conditions deduced from our kinematical model and
 those required for maser excitation can be done for IRS1b and the cluster~"A".
Looking at the spatial distribution of the maser features onto the edge-on disk 
 (see Fig.~\ref{disk_pl}, left panel), one can see that the maser
emission abruptly decreases in intensity and vanishes at radii between \ 500 and 300~AU.
As indicated by \ Equation.~\ref{nh2_a}, at radii smaller than \ $R_0 = 550$~AU \ the local gas density 
rapidly increases to values \ $n_{\rm H_2}$ > 10$^9$~cm$^{-3}$, and the masers are probably thermally quenched,
in line with the model predictions of \citet{Cra05}. The edge-on disk model for cluster~"A"
predicts a maximum radius of 6.7~GHz maser emission of \ $\approx$2500~AU, which can also be explained in terms of the maser excitation model
of \citet{Cra05} if the IR radiation flux at this radius falls below the threshold for efficient maser pumping. Assuming that the dust
absorbs the stellar photons  and re-emits them at IR wavelengths like a  black-body, one derives an inverse square-root dependence
for the dust temperature on the distance from the star. 
If $T_d \approx$300~K (i.e., the maximum value expected for efficient
pumping) at the closest maser radii of \ $\approx$300~AU, 
at the largest maser radius of \ $\approx$2500~AU \ $T_d \approx$100~K, 
that is  the value indicated by \citet{Cra05} as the lower limit for efficient excitation of the 6.7~GHz masers.

From   \ Equation~\ref{nh2_a} it is also clear that
the opening angle \ $\alpha$ \ of the disk around IRS1b has to be large enough ($\alpha$ $\approx$ 45\degree) to ensure that the gas density does not exceed the threshold for the thermal quenching 
even at \ $R \ge R_0$. Such a geometry reminds the profiles of the rotationally flattened, circumstellar envelopes 
observed towards low-mass YSOs at millimeter wavelengths \citep{Koe95} and in the near-IR \citep{Cot01}.
If high-mass star formation proceeds as a scaled-up version of the low-mass case, rotationally flattened, thick disks/envelopes could characterize the earliest protostellar phase, when only a minor fraction of the available circumstellar material
has been accreted onto the protostar.

Finally, it is worth discussing more thoroughly the distribution of the 6.7~GHz masers around IRS1a and IRS1b as predicted by the  edge-on disk model (see Fig.~\ref{disk_pl}). 
We note two main differences between the two maser distributions:  
1)~the maser pattern in the IRS1b disk is more elongated along the l.o.s and more compact in the perpendicular direction than the one in IRS1a;
\ 2)~approaching the line-of-sight to the star, the maser intensity increases in the IRS1b disk and decreases in the IRS1a disk.
The analysis of \citet[][see Table~3]{Sur11} shows that most of the 6.7~GHz masers in \NGC1\ are unsaturated, so that their intensity depends linearly on the
background radiation. Therefore, a l.o.s. elongated maser pattern can naturally result if  masers amplify a compact continuum source (like a hypercompact \HII\ region or an ionized jet)
around the star, in such a fashion that only the l.o.s. intercepting sufficiently strong continuum background radiation would result in maser emission.
Regarding point 1), since IRS1a is more massive and evolved than IRS1b, 
it should have a more developed and extended continuum, which naturally explains the observation of the 6.7~GHz features at larger projected separation from the star 
(up to \ $\approx$400~AU compared with \ $\approx$200~AU for IRS1b).  
Regarding point~2), the opposite trend observed in IRS1a and  IRS1b 
can be explained with slightly different disk geometries for the two YSOs.  
For a perfectly edge-on disk, as in the case of IRS1b, 
any l.o.s. offset from the star will intercept  less column density than the l.o.s. along the star. 
The increasing maser intensity approaching the l.o.s. to IRS1b can then result from longer velocity-coherent paths, 
since, for a pure rotation pattern, velocities at all radii have equally null 
projection along the l.o.s. to the centre of rotation. 
If the disk is however not perfectly edge-on, as it is probably the case for IRS1a (see discussion in Sect.~\ref{mas_kin_qua}),  
this argument does not hold anymore. 
In this case one key element to consider is instead the role played by circumstellar dusty disks 
in shielding the masing molecules against the disruptive action of the UV stellar radiation.
 The dissociation of methanol molecules follows the temperature distribution, 
 which, for an hydrostatic Keplerian disk, 
 depends on both the radial distance on the disk plane and the vertical height above/below the mid-plane. 
 In the following, we show that any l.o.s. offset from the star provides better shielding  
 enabling the methanol molecules to more easily survive.  
Let us consider a point above the disk midplane, which we assume (for simplicity) to lie on the plane 
containing the disk major axis and the l.o.s.. 
Indicating with \ $i_d$ \ the angle between the disk plane and the l.o.s., 
with \ $s$ \ and \ $z$ \ the offset from the star along the major axis and 
the distance along the l.o.s. to the the major axis, respectively, 
the height \ $H$ \ from the disk mid-plane of the selected point is given by \ $ H = z \sin(i_d) $. 
The projection of the point onto the disk mid-plane is at a radial distance 
\ $R$ \ from the star given by \ $  R = \sqrt{ z^2 \cos^2(i_d) + s^2 }=\sqrt{ H^2 / \tan^2(i_d) + s^2 }$.
This expression shows that at a given height \ $H$, 
the l.o.s. to the star ($s=0$) intersects the disk atmosphere  at a radial distance from the star 
{\em smaller} than that of the l.o.s. offset by \ $s$ \ along the major axis. 
Similarly, it is possible to show that at a given radial distance $R$, 
the l.o.s. to the star ($s=0$) intersects the disk atmosphere  at a  height from the disk plane 
{\em larger} than that of the l.o.s. offset by \ $s$ \ along the major axis. 
Consequently, l.o.s. with smaller and smaller offset \ $s$ \ from the star cross {\em increasingly warmer} portions of the disk atmosphere,
where methanol molecules can be more easily dissociated and maser emission reduced. 
As a result, (strong) maser emission can be observed only along l.o.s. sufficiently 
offset from the star to intersect better shielded disk positions. 
This interpretation can readily account for the 6.7~GHz maser emission fading away towards the l.o.s. to IRS1a, 
and indirectly  proves the reliability of the position of IRS1a derived by our model. 
The opposite trend observed in IRS1b can be used to conclude that its disk has to be very close to edge-on.

Since here we propose kinematical models, we do not attempt to explain 
the physical nature of the ordered maser patterns, which would require a full analysis of the physical and kinematical conditions of the gas, coupled with a consistent treatment of the
radiative transport in the methanol maser line.

\subsection{Comparison with Previous Models} 

For completeness, here we discuss briefly some of the alternative models previously proposed to explain NGC7538 IRS1. 

As mentioned in the Introduction, \citet{Pes04} modelled the linear distribution of 6.7 and 12.2 GHz \meth\ masers of cluster "A" as an edge-on Keplerian disk, rotating around a 30 \ms \,protostar. 
While the geometrical properties of this modeled disk are qualitatively compatible with the model described in Sect.~\ref{mas_kin} (e.g., PA and radius), that model unlike ours does not set constraints on the mass/density distribution in the disk  and the central mass, which is given in input  to the model (assuming that the YSO exciting the maser cluster "A" accounts for 
the region's bolometric luminosity). 
Another limitation is the assumption of Keplerian rotation, despite an evident bend in the position-velocity diagram, which can be naturally explained if the rotation velocity increases with the radius.  
Our model includes as constraints l.o.s. accelerations and proper motions (besides l.o.s. velocities and positions),  and allows us to constrain the mass/density distribution in the disk  and the central mass, as well as  to establish the 
radial profile of the rotation velocity. 

An alternative model to the edge-on disk model was proposed by \citet{DeBui05}, based on a mid-IR study. They find that the circumstellar dust associated with IRS1 is extended both northwest-southeast (on scales $\sim4000$ AU, PA$\sim-45$\degree) and northeast-southwest (on small scales $\sim400$ AU, PA$\sim30$\degree). 
Since the large-scale mid-IR emission is extended along a position angle similar to that of the CO outflow, they suggest that it is coming from dust heated on the walls of the outflow cavities near the star, as opposed to trace a circumstellar disk. 
They also propose that the small-scale elongation seen in the mid-IR, nearly perpendicular to the axis of the CO outflow (and the linearly distributed methanol masers), is a circumstellar disk. 
While this "outflow" model explains the main properties of the mid-IR emission, individual maser features in cluster "A" would be either clumps in the cavity or recent ejecta from the outflow, but both options are inconsistent with our  measurements of velocities and accelerations of \meth\ masers (see arguments at the end of Sect.~\ref{mas_kin_qua} ).  

In an attempt to explain the different orientations of the elongated structures of cluster "A" of \meth\ masers and of the near/mid-IR emission  as well as the CO outflow axis, \citet{Krau06} proposed that the edge-on disk modeled by \citet{Pes04} is driving a precessing jet.
This model requires as the most plausible explanation, the presence of a relatively tight binary (with an orbital separation of tens of AU and an orbital period of tens of years). While we cannot exclude it, 
there is no evidence of a tight binary in \NGC1, 
and we believe that the presence of multiple outflows from individual YSOs provides a more natural explanation.

Aiming for a more complete picture, \citet{Sur11} also tried to incorporate the presence of methanol clusters B, C, D, and E (besides A) in a "torus" model alternative to the  edge-on Keplerian disk model for feature A. 
Evidence for a large torus (of several thousands of AU) is provided by the measurements of velocity gradients in molecular lines approximately perpendicular to the  large-scale CO bipolar outflow \citep{Kla09,Qiu11,Beu13} 
They proposed that all the observed \meth\ maser clusters should mark the interface between an infalling envelope and this large-scale torus, oriented perpendicularly to the elongated mid-IR emission observed by \citet{DeBui05}. 
This scenario has the merit of incorporating all maser clusters (and not just cluster "A" as in previous models), but it does not explain the changing of the position angle of the outflow based on radio continuum imaging \citep[e.g., ][]{Cam84}. 
Nevertheless, this simplified and qualitative model where all the observed maser clusters are associated with one large-scale torus of several thousands of AU, does not exclude the presence of smaller accretion disks, like the ones probed individually by clusters "A" and "B"+"C" in our edge-on disk model.

\section{Summary}
\label{summary}

We have used four epochs of EVN archival 6.7~GHz \meth\ maser observations and JVLA B-Array \nh3\ observations in
several inversion lines, (6,6), (7,7), (9,9), (10,10) and (13,13), to study the  3D kinematics and dynamics of the molecular gas towards the high-mass star forming region \NGC1\ on linear scales of 10--1500~AU. 
\NGC1\ has been extensively studied in the past, and several outflows with different orientation have been identified as well as  claims for accretion disks with different orientations have been made, which led to competing models and some controversy in the interpretation of this region. 
Previous observations have detected three compact 1.3~cm continuum
sources and five 6.7~GHz maser clusters distributed over a region extending N--S across \ $\approx$1500~AU (labeled from "A" to "E"). 
Our measurements provide compelling evidence that \NGC1\ is not simply a single massive YSO but 
its core is actually forming a multiple system of massive YSOs within 1500 AU. %The YSOs are massive and surrounded by disks. 
In particular, we propose that the three radio continuum peaks, associated with clusters "A", "B"+"C", and "E" of 6.7~GHz \meth\ masers, could mark the position of three distinct high-mass YSOs in the region, responsible for the maser excitation:  \ IRS1b, IRS1a, and IRS1c, respectively. IRS1a and IRS1b are surrounded by disks and may be driving outflows in the region. We argue that the presence of a multiple stellar system solves the controversy aroused in previous works. 

The evidence for accretion disks comes from accurate measurements of positions, \Vlsr, proper motions, and l.o.s. accelerations of \meth\ maser features in individual clusters. %allowed by the multi-epoch dataset / study over time 
 In particular, we find that the 6.7~GHz masers distribute along a line with a regular variation of \Vlsr\ with projected position both inside the maser cluster~"A" and 
the combined clusters~"B"+"C". Interestingly, the variation of \Vlsr\ with the projected position is not linear but quadratic for both clusters. 
Towards the maser clusters~"A" and "B"+"C", a \Vlsr\ gradient is also detected in the \nh3\ inversion lines, although shallower than that measured with the 6.7~GHz masers (owing to lower angular resolution).
We measured proper motions for 33 persistent maser features with an average (1-$\sigma$) accuracy of 0.6~\kms. 
For the clusters~"A", "B", "C" and "D", most of the measured proper motions are oriented parallel 
to the major axis of the maser clusters and the average amplitude is \ $\approx$5~\kms. Masers belonging to the
cluster~"E" move slower, with an average speed of only \ $\approx$3~\kms, and present a more scattered pattern of proper motions. 
By studying the time variation of the emission spectrum, we derive also l.o.s. accelerations 
for 30 maser features. Features in cluster~"A" have similar values of l.o.s. acceleration around \ 0.01~\kmsy, 
while in clusters "B" and "C" there is a larger scatter, with values from \ $-$0.019 to 0.016~\kmsy. Finally, in cluster~"E" most of the measurements are compatible with a null value of maser l.o.s. acceleration.

We built a simple model of  edge-on disk in centrifugal equilibrium to explain individually the linear maser distributions in clusters~"A" and "B"+"C", with the regular pattern of maser and \nh3\ \Vlsr, maser proper motions and l.o.s. accelerations. 
Masers of clusters~"B"+"C" may trace a quasi-Keplerian $\sim$1~M$_{\sun}$, thin disk, orbiting a high-mass YSO, IRS1a, of up to $\approx$25~M$_{\sun}$. IRS1a should be responsible for most of the bolometric luminosity of the region,\ $\sim$10$^5$~L$_{\sun}$,   the intense radio continuum emission,
and  the excitation of the hot-core detected towards the maser clusters~"B"+"C" by subarcsecond observations in
several dense gas molecular tracers. 
The disk around IRS1b, traced by the masers of cluster~"A", is both massive ($\la$16~M$_{\sun}$ inside a
radius of $\approx$500~AU) and thick (opening angle $\approx$ 45\degree), and the mass of the central YSO
is constrained to be at most of a few M$_{\sun}$.  
Towards the cluster~"E", both \nh3\ and 6.7~GHz masers appear to trace less structured
and more quiescent dynamics than for the other clusters.
The presence of a relatively weak radio continuum peak could  suggest that the YSO IRS1c associated with the cluster~"E" is also an ionizing and massive YSO, and we speculate that a slow expanding, ultra-compact \HII\ region 
could account for both the sparse spatial distribution and the slow and scattered proper motions of the 6.7~GHz masers.

Our \nh3\ images provide evidence that besides the circumstellar gas associated with individual YSOs, there is also hot and dense molecular material between the YSOs, which probably constitutes the parental core from which the YSOs are forming. 
Interestingly, the \nh3\ inversion lines identify a clear \Vlsr\ gradient 
in the region between the YSOs IRS1a and IRS1b, which may reflect the global rotation of the parental massive core. 
Based on intensity-weighted velocity  (or first moment) maps, this gradient appears to be oriented NE--SW (PA = 30\degree -- 40\degree, depending on the line excitation). Since the angular resolution of the \nh3\ images (0\pas2) is comparable with the separation between IRS1a and IRS1b, the first moment maps may be affected by the blending of the gas motions 
{\it around} a given YSO and {\it between} the two YSOs (which have a separation in the systemic velocity  of 3~\kms). 
Using the \nh3\ absorption peak positions in individual spectral channels  (fitted with Gaussians), we find that the \Vlsr\ gradient in the gas between the two YSOs is close to N--S.   
We conclude that the correct orientation of the \nh3\ \Vlsr\ gradient may be between NE-SW and N--S, not too far from the relative displacement between IRS1a and IRS1b. 

\begin{acknowledgements}
      We are grateful to Dr. Q. Zhang for fuitful discussion on the kinematics traced by thermal molecular lines in the hot-core. We are also grateful to Dr. G. Surcis for useful discussion on the magnetic field traced by 6.7 GHz methanol masers in the region. 
\end{acknowledgements}

% for the bibliography, at the end
\bibliographystyle{aa} % style aa.bst
\bibliography{biblio} % your references Yourfile.bib

\Online

\begin{appendix} %First online appendix

\section{Derivation of Star Position and \Vlsr}
\label{appe}

This section describes our method to determine the star position, $s^{\prime}_{\star}$, 
and star (systemic) \Vlsr, $V_{\star}$,
for the clusters "A" and "B"+"C". 
Indicating with \ $V_z$ \ the maser l.o.s. velocity (with respect to the star), 
for edge-on rotation we can write:
\begin{equation}
\label{ome_s}
V_{\rm LSR} = V_z + V_{\star} = \Omega s + V_{\star}
\end{equation}
where \ $\Omega$ \ and \ $s$ \ are respectively the angular velocity and the positional offset
from the star projected along the major-axis, $\bf{D}$, of the maser distribution.

From our observations we know the maser \ \Vlsr, the $\bf{D}$-axis projected offset, $s^{\prime}$, 
{\em evaluated from the reference maser spot} (feature \#1 of cluster~"A", see Table~\ref{tab_6.7}), 
and the derivative, $dV_{\rm LSR}/ds^{\prime} $, of \Vlsr\ with respect to \ $s^{\prime}$.
We determine the values of \ $s^{\prime}_{\star}$ \ and \ $V_{\star}$, in two steps.
Since from the fit of \ \Vlsr\  versus \ $s^{\prime}$ (see Sect.~\ref{vlsr_regu})
the value of the second order coefficient \ $\alpha$ \ is always found to be small 
compared to the first order coefficient \ $\beta$, in the first step we take 
\ $\Omega \approx dV_{\rm LSR}/ds^{\prime} $,
which would be an exact relation only if \ $\alpha = 0$ \ and \  $\Omega$ \ were constant.
Basing on the Equation~\ref{ome_s}, we look for the best values of \ $V_{\star}$ \ and
 \ $s^{\prime}_{\star}$ \ 
% ($s = s^{\prime} -s^{\prime}_{\star}$)   
 which minimize the \ $\chi^{2}$ \ expression:
\begin{equation}
\label{chi}
\chi^2 = \displaystyle\sum_{i} \left( \frac{(V_{\rm LSR}^{i}-V_{\star})}{dV_{\rm LSR}^{i}/ds^{\prime}_{i}} - 
(s^{\prime}_{i}-s^{\prime}_{\star}) \right)^{2}
\end{equation}
where the sum of index \ $i$ \ is performed over all the features belonging to the maser cluster(s).
$s^{\prime}_{\star}$ \ and \ $V_{\star}$ \ are searched over an interval of values comparable
with the maximum maser offset and the global \Vlsr\ extent ($\pm$100~mas and \ $\pm$5~\kms, respectively),
in steps of \ 5~mas \ and \ 0.1~\kms, respectively.

The first step allows us to approximate the maser projected offset, $s$, and  l.o.s. velocity, $V_z$, in the reference
system of the star, using the equations:
\begin{align}
\label{spp}
s \approx s^{\prime\prime} &= s^{\prime} - s^{\prime}_{\star} \\
\label{vpp}
V_z \approx V_z^{\prime} &= V_{\rm LSR} - V_{\star}
\end{align}

In the second step we refine our determination of the star parameters, taking advantage of the relation:
\begin{equation}
\label{O_dV}
\Omega = \alpha s + \beta = dV_{\rm LSR}/ds - \alpha s
\end{equation}
which derives directly from Equations~\ref{Ome}~and~\ref{Q_cf}. Using \ $s^{\prime\prime}$ \ to express \ $s$,
the Equation~\ref{O_dV}  allows us to improve our estimate of the angular velocity \ $\Omega$.
We look for small corrections of the star position, $s^{\prime\prime}_{\star}$, and \Vlsr,
$\delta V_{\star}$, by minimizing the \ $\chi^{2}$ \ expression:
\begin{equation}
\label{chi2}
\chi^2 = \displaystyle\sum_{i} \left( \frac{(V_z^{\prime i}-\delta V_{\star})}{(dV_{\rm LSR}^{i}/ds^{\prime\prime}_{i} 
- \alpha s^{\prime\prime}_{i})} - 
(s^{\prime\prime}_{i}-s^{\prime\prime}_{\star}) \right)^{2}
\end{equation}
which is more precise of the first-step \ $\chi^2$ (see Equation~\ref{chi}), since it 
employs the more accurate evaluation of the maser angular velocity \ $\Omega$.
The parameters  \ $s^{\prime\prime}_{\star}$ \ and \ $\delta V_{\star}$ \ are looked for in small steps
of \ 1~mas \ and \ 0.02~\kms, respectively. For both  maser clusters "A" and "B"+"C", the second step 
derives corrections for the star position and \Vlsr\ of only a few mas and a few tenths of kilometer per second,
respectively. 

\begin{figure}
\includegraphics[angle=0.0,width=0.45\textwidth]{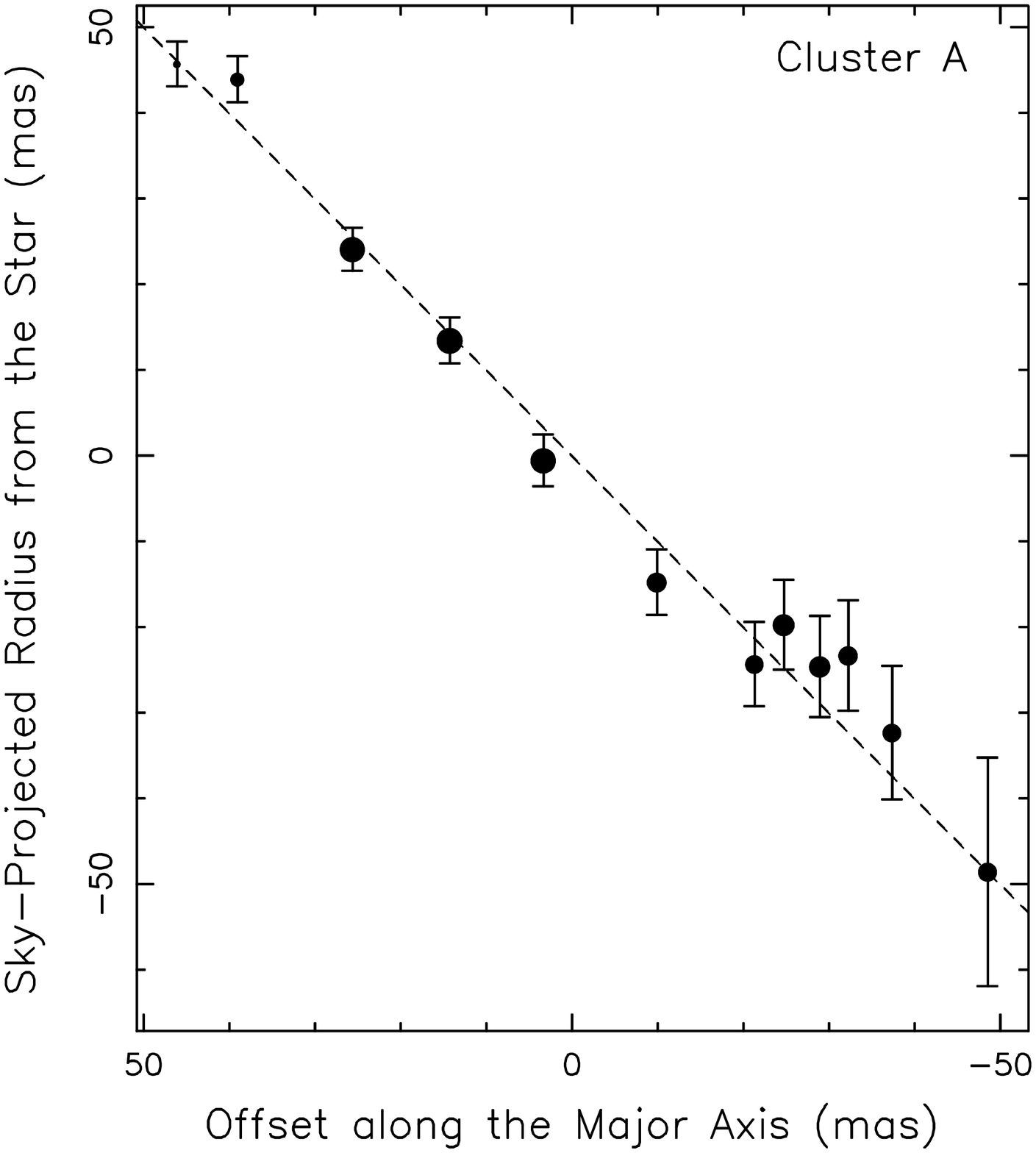}

\vspace{0.5cm}

\includegraphics[angle=0.0,width=0.45\textwidth]{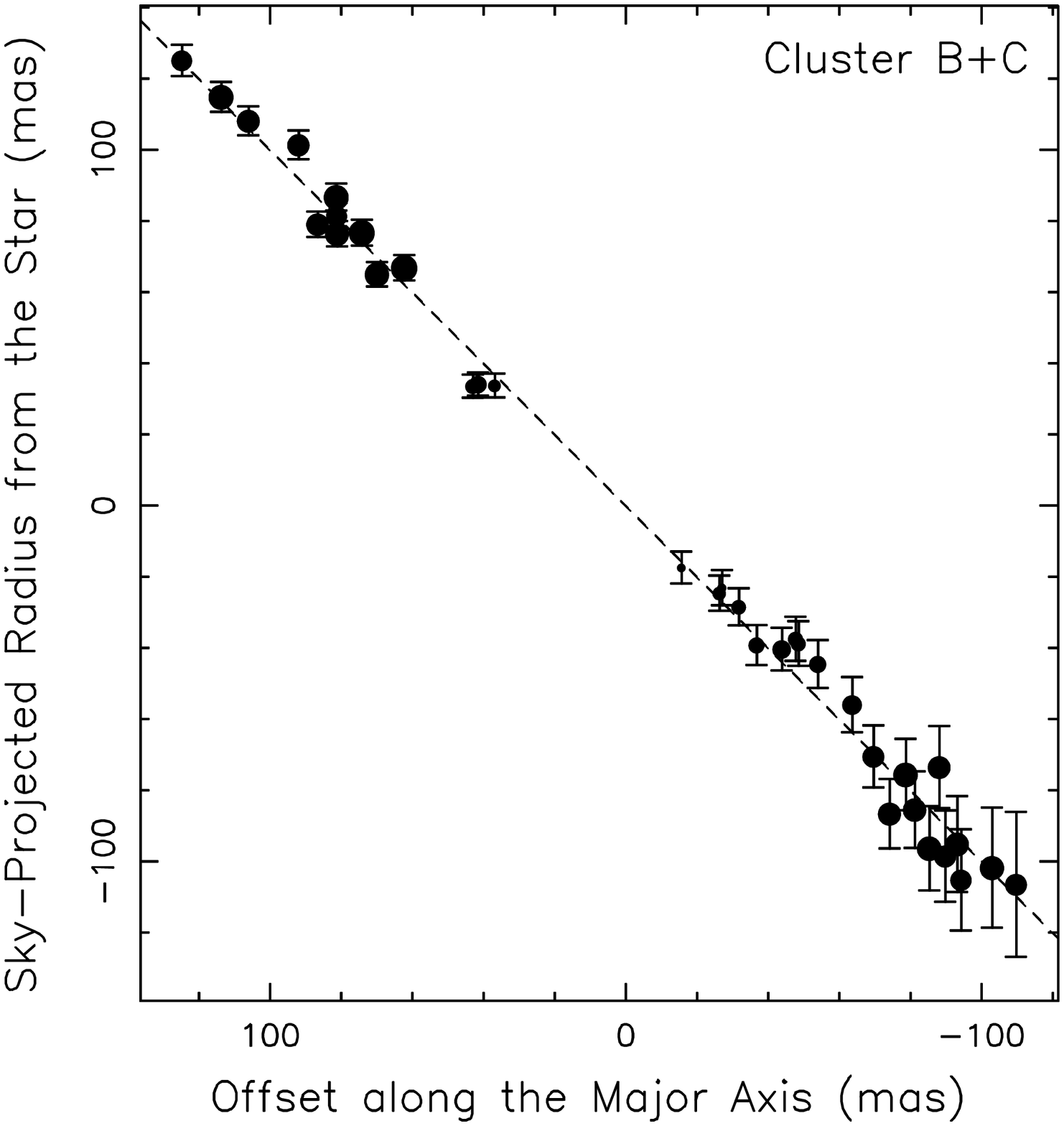}
\caption{Plot of sky-projected maser radii, derived from the ratio
of the maser l.o.s. velocity with the angular velocity  (see Sect.~\ref{appe}),
versus position projected along the major axis of the spatial distribution 
for the maser cluster~"A" {\it (upper panel)} and the combined clusters "B"+"C" {\it (lower panel)}.
Dot size is proportional to the logarithm of the maser intensity and vertical errorbars give
the estimated errors of the maser radii. The {\it black dashed line} denotes the \ "Y = X" \ axis.}
\label{rp_off}
\end{figure}

Finally, maser projected offset, $s$, and  l.o.s. velocity, $V_z$, in the reference
system of the star, are evaluated using the relations:
\begin{align}
\label{spp2}
s &\approx  s^{\prime} - (s^{\prime}_{\star}+s^{\prime\prime}_{\star}) \\
\label{vpp2}
V_z &\approx V_{\rm LSR} - (V_{\star}+\delta V_{\star})
\end{align}
To appreciate the quality of the final result, Fig.~\ref{rp_off} compares the estimate of \ $s$ \ 
(reported on the "X" axis) with the ratio \ $V_z/\Omega$ \ (along the "Y" axis), 
where \ $\Omega$ \ is derived from Equation~\ref{O_dV}.
 
\end{appendix}

\end{document}